\documentclass[aps, prx,amsmath,amssymb,longbibliography,showpacs,letterpaper,twocolumn,nofootinbib,floatfix,superscriptaddress]{revtex4-1}

\usepackage{CJKutf8}
\usepackage{tikz}

\usepackage{amsmath}
\usepackage{graphicx}
\usepackage{fullpage}
\usepackage[colorlinks=true,allcolors=blue]{hyperref}
\usepackage[capitalise]{cleveref}
\usepackage[utf8]{inputenc}
\usepackage{siunitx}

\usepackage{url}
\usepackage{rotating}
\usepackage{epsfig,graphics,rotate}
\usepackage{wrapfig}
\usepackage[caption=false]{subfig}

\usepackage{bm}
\usepackage{amssymb}
\usepackage{amsmath}
\usepackage{amsfonts}

\usepackage{booktabs}
\usepackage{microtype}
\usepackage{multirow, makecell}
\usepackage{lipsum}

\usepackage{xspace}
\usepackage{comment}
\usepackage{floatrow}
\usepackage{ragged2e}

\usepackage{nicefrac, xfrac}
\usepackage{mathtools} % for 'pmatrix' environment
\usepackage{relsize}   % for '\larger' macro

\usepackage{float}

\newcommand{\nsum}[1][1.44]{\mathop{\vcenter{\hbox{%   % 1.44=(1.2)^2
   \scalebox{#1}{$\displaystyle\sum$}}}}}

\definecolor{lime}{HTML}{A6CE39}
\DeclareRobustCommand{\orcidicon}{
	\begin{tikzpicture}
	\draw[lime, fill=lime] (0,0) 
	circle [radius=0.16] 
	node[white] {{\fontfamily{qag}\selectfont \tiny ID}};
	\draw[white, fill=white] (-0.0665,0.095) 
	circle [radius=0.005];
	\end{tikzpicture}
	\hspace{-2mm}
}

\foreach \x in {A, ..., Z}{\expandafter\xdef\csname orcid\x\endcsname{\noexpand\href{https://orcid.org/\csname orcidauthor\x\endcsname}
			{\noexpand\orcidicon}}
   }

\DeclareSIUnit \s {\second}
\DeclareSIUnit \ns {\nano\second}
\DeclareSIUnit \mus {\micro\second}
\DeclareSIUnit \ms {\milli\second}
\DeclareSIUnit \MB {\mega\byte}
\DeclareSIUnit \GB {\giga\byte}
\DeclareSIUnit \TB {\tera\byte}
\DeclareSIUnit \PB {\peta\byte}
\DeclareSIUnit \Mbps {\mega\bit/\s}
\DeclareSIUnit \Gbps {\giga\bit/\s}
\DeclareSIUnit \Tbps {\tera\bit/\s}
\DeclareSIUnit \Pbps {\peta\bit/\s}
\DeclareSIUnit \kton {\kilo\tonne} % changed  back to kton
\DeclareSIUnit \kt {\kilo\tonne}
\DeclareSIUnit \Mt {\mega\tonne}
\DeclareSIUnit \eV {\electronvolt}
\DeclareSIUnit \keV {\kilo\electronvolt}
\DeclareSIUnit \MeV {\mega\electronvolt}
\DeclareSIUnit \GeV {\giga\electronvolt}
\DeclareSIUnit \TeV {\tera\electronvolt}
\DeclareSIUnit \PeV {\peta\electronvolt}
\DeclareSIUnit \EeV {\exa\electronvolt}
\DeclareSIUnit \m {\meter}
\DeclareSIUnit \cm {\centi\meter}
\DeclareSIUnit \in {\inchcommand}
\DeclareSIUnit \km {\kilo\meter}
\DeclareSIUnit \kV {\kilo\volt}
\DeclareSIUnit \kW {\kilo\watt}
\DeclareSIUnit \MW {\mega\watt}
\DeclareSIUnit \MHz {\mega\hertz}
\DeclareSIUnit \mrad {\milli\radian}
\DeclareSIUnit \year {years}
\DeclareSIUnit \POT {POT}
\DeclareSIUnit \sig {$\sigma$}
\DeclareSIUnit\parsec{pc}
\DeclareSIUnit\lightyear{ly}
\DeclareSIUnit\foot{ft}
\DeclareSIUnit\ft{ft}
\DeclareSIUnit \ppb{ppb}
\DeclareSIUnit \ppt{ppt}
\DeclareSIUnit \samples{S}
\DeclareSIUnit \pe{PE}

\newcommand{\enu}{\E_\enu}

\begin{document}
\begin{CJK*}{UTF8}{gbsn}

\title{Measuring Oscillations with a Million Atmospheric Neutrinos}

\begin{abstract}
After two decades of measurements, neutrino physics is now advancing into the precision era. With the long-baseline experiments designed to tackle current open questions, a new query arises: can atmospheric neutrino experiments also play a role? 
To that end, we analyze the expected sensitivity of current and near-future water(ice)-Cherenkov atmospheric neutrino experiments in the context of standard three-flavor neutrino oscillations.
In this first in-depth combined atmospheric neutrino analysis, we analyze the current shared systematic uncertainties arising from the common flux and neutrino-water interactions.
We then implement the systematic uncertainties of each experiment in detail and develop the atmospheric neutrino simulations for Super-Kamiokande, with and without neutron-tagging capabilities, IceCube Upgrade, ORCA, and Hyper-Kamiokande detectors.
We carefully review the synergies and features of these experiments to examine the potential of a joint analysis of these atmospheric neutrino data in resolving the $\theta_{23}$ octant at 99\%~confidence level, and determining the neutrino mass ordering above 5$\sigma$ by 2030.
Additionally, we assess the capability to constrain $\theta_{13}$ and the $CP$-violating phase ($\delta_{CP}$) in the leptonic sector independently from reactor and accelerator neutrino data.
A combination of the atmospheric neutrino measurements will enhance the sensitivity to a greater extent than the simple sum of individual experiment results reaching more than 3$\sigma$ for some values of $\delta_{CP}$. 
These results will provide vital information for next-generation accelerator neutrino oscillation experiments such as DUNE and Hyper-Kamiokande.
\end{abstract}

\author{C.~A.~Arg{\"u}elles\orcidA{}}
\email{carguelles@fas.harvard.edu}
\affiliation{Department of Physics \& Laboratory for Particle Physics and Cosmology, Harvard University, Cambridge, MA 02138, USA}

\author{P.~Fern\'andez\orcidB{}}
\email{pablo.fernandez@dipc.org}
\affiliation{University of Liverpool, Department of Physics, Liverpool, United Kingdom}
\affiliation{Donostia International Physics Center DIPC, San Sebasti\'an/Donostia, E-20018, Spain}

\author{I.~Mart\'inez-Soler\orcidC{}}
\email{imartinezsoler@fas.harvard.edu}
\affiliation{Department of Physics \& Laboratory for Particle Physics and Cosmology, Harvard University, Cambridge, MA 02138, USA}

\author{M.~Jin (靳淼辰)\orcidD{}}
\email{miaochenjin@g.harvard.edu}
\affiliation{Department of Physics \& Laboratory for Particle Physics and Cosmology, Harvard University, Cambridge, MA 02138, USA}

\maketitle
\end{CJK*}

%%%%%%%%%%%%%%%%%%%%%%%%%%%%%%%%%%%%%%%%%%%%%%%
\section{Introduction}\label{sec:intro}
\begin{figure}[t]
\centering
\includegraphics[width=\columnwidth]{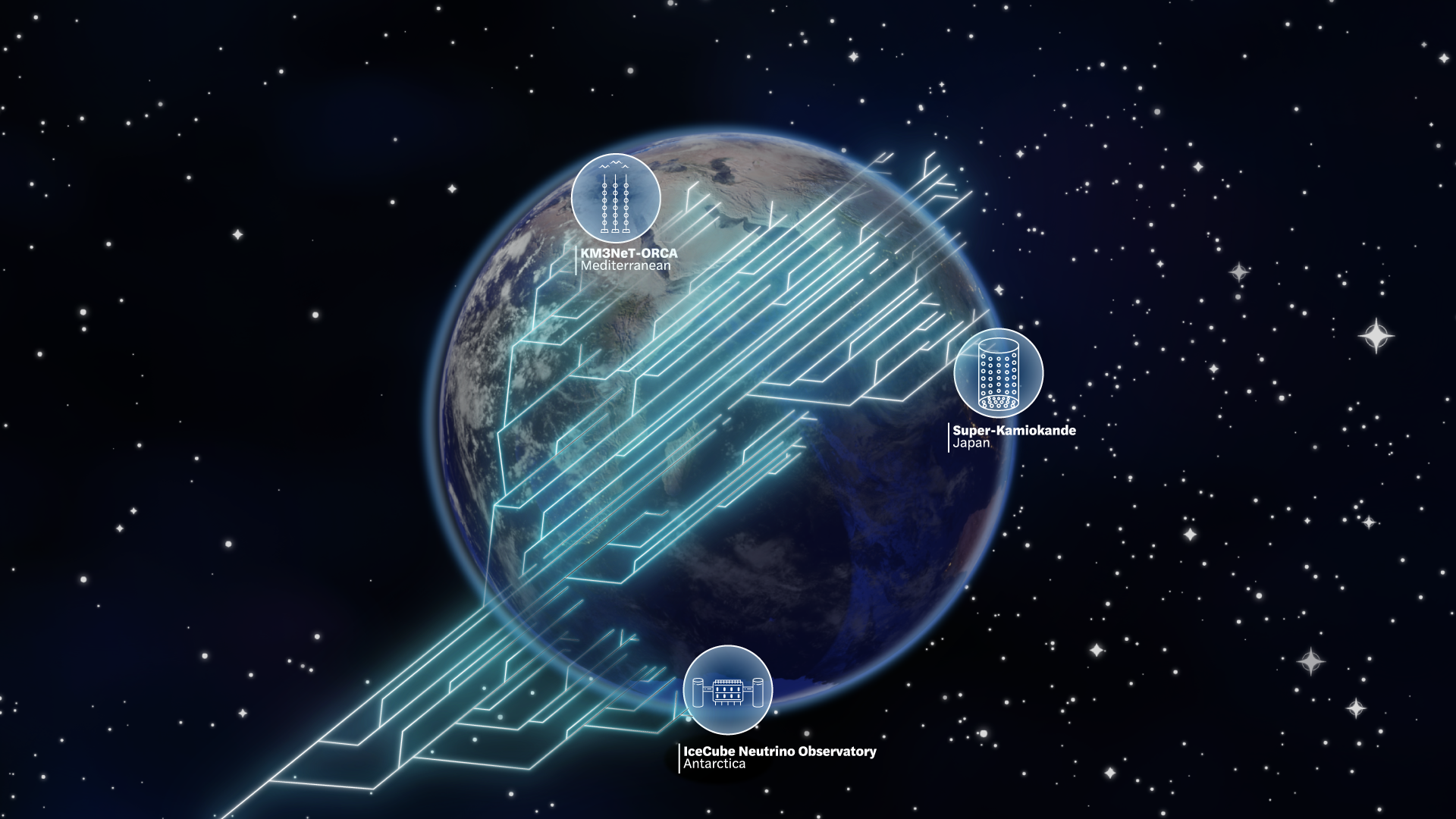}
\caption{\textbf{\textit{Illustration of this analysis.}}
Locations of experiments used in this work are shown. Note that Hyper-Kamiokande has roughly the same location as Super-Kamiokande.
\label{fig:illustration}}
\end{figure}

Atmospheric neutrinos have played a key role in discovering and understanding neutrino oscillations~\cite{Lipari:2018qqz,Koshio:2020yex}.
The first hint of resilient signatures of deviations from the Standard Model in neutrinos came from a deficit of muon neutrinos in IMB~\cite{PhysRevLett.69.1010} and Kamiokande~\cite{Kamiokande-II:1992hns}, which were later confirmed by Super-Kamiokande (SuperK)~\cite{Super-Kamiokande:2004orf}.
These anomalies are now known to be due to neutrinos having non-zero, small masses, and the flavor states and mass states being misaligned~\cite{deGouvea:2016qpx}.
This misalignment produces a characteristic neutrino flavor change, often called oscillations due to its periodic behavior in vacuum, and this flavor change depends on the ratio of the distance from source to detector, $L$ -- called baseline -- and the neutrino energy, $E_\nu$. 

One of the reasons why atmospheric neutrinos are good discovery experiments is that they cover ten orders of magnitude in the ratio of baseline to energy, $L/E_\nu$.
The baseline $L$ covers ranges from $15$~${\rm km}$ to $12,700$~${\rm km}$ and the neutrino energy ranges from $\mathcal{O}(10^{-2})$~${\rm GeV}$ to $\mathcal{O}(10^5)$~${\rm GeV}$.
This results in a coverage of $L/E_\nu$ from $\mathcal{O}(10^{-4})$ ${\rm km/GeV}$ to $\mathcal{O}(10^6)$ ${\rm km/GeV}$.
This broad range of $L/E_\nu$ and the fact that atmospheric neutrinos go through the largest amounts of matter~\cite{Wolfenstein:1977ue} have allowed them to be used for measuring neutrino oscillation parameters~\cite{Esteban:2018azc,deSalas:2020pgw} and placing stringent bounds on new neutrino states~\cite{IceCube:2020tka,IceCube:2020phf}, non-standard neutrino interactions~\cite{Fornengo:2001pm,Esmaili:2013fva,Salvado:2016uqu,Gonzalez-Garcia:2016gpq,IceCube:2017zcu, IceCubeCollaboration:2021euf,IceCube:2022ubv,Proceedings:2019qno, Demidov:2019okm}, space-time symmetries~\cite{IceCube:2017qyp}, and other physics beyond the Standard Model~\cite{Gonzalez-Garcia:2004pka,Koshio:2022zip}; see~\Cref{fig:illustration} for an artistic illustration of atmospheric neutrinos and the detectors used to observe them.

\begin{figure}[hbt!]
\centering
\includegraphics[width=\textwidth]{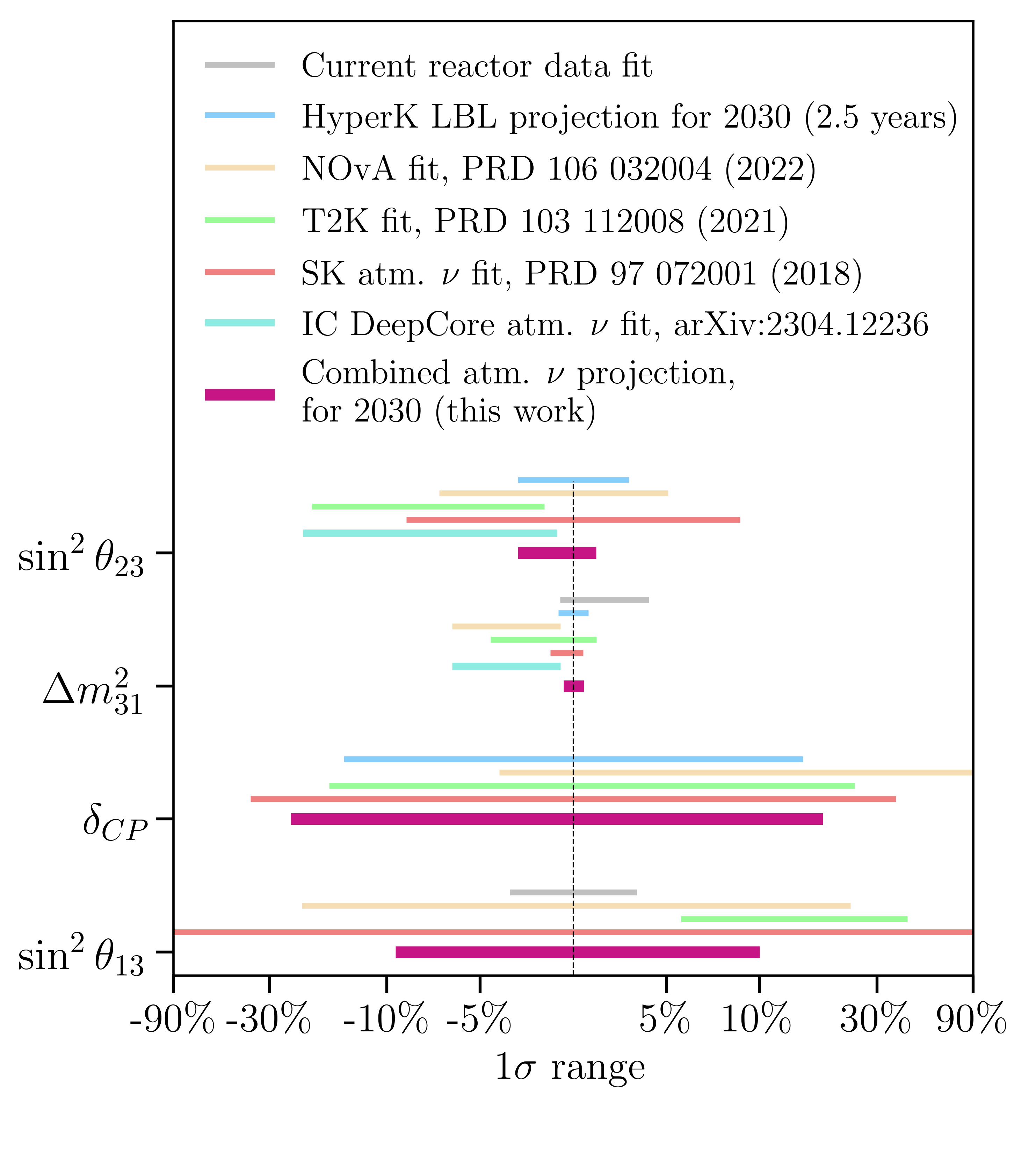}
\caption{\textbf{\textit{Comparison between the present and future projected sensitivities for the oscillation parameters.}} 
This figure showcases the power of atmospheric combined neutrino experiments, as they have smaller or comparable errors on these parameters than accelerator neutrino experiments also shown in this figure.
Oscillation parameters that can be measured by atmospheric experiments are arranged in the vertical axis, while their measured precision is quantified in the horizontal axis.
The present 1-sigma region allowed by fit to data from T2K~\cite{T2K:2021xwb} (green), NOvA~\cite{NOvA:2021nfi} (light orange), SuperK atmospheric neutrinos~\cite{Super-Kamiokande:2017yvm} (light red), Deep Core atmospheric neutrinos~\cite{icecubecollaboration2023measurement} (turquoise), the reactor experiments (grey) and the projected sensitivity of Hyper-Kamiokande's accelerator program (blue) for 2030~\cite{hk_tdr} is compared with the expected 1-sigma region from a combined atmospheric neutrino analysis (this work in red-violet).
The sensitivity from DUNE is excluded due to the low significance achievable by this experiment for 2030 assuming a 2029 starting date~\cite{DUNE:2021mtg}.
}
\label{fig:present}
\end{figure}

In the last two decades, the neutrino community has developed a vast program using accelerator, reactor, and solar neutrinos to measure their evolution, leading to the conclusion that atmospheric measurements will not contribute to precision neutrino physics.
In this article, we change that paradigm, showing that atmospheric neutrinos will improve our knowledge of some of the largest unknowns describing the evolution of neutrino states.
See~\Cref{fig:present} for a comparison between our results, the present status, and the future predictions for the next generation of neutrino experiments. 
By combining the atmospheric measurements, we aim to achieve the most accurate determination of atmospheric parameters ($\Delta m^2_{31}$ and $\theta_{23}$).
Additionally, this combined approach will yield valuable insights into $\theta_{13}$ due to the Earth matter effect, as we will elaborate upon later.
Lastly, the atmospheric measurements are expected to offer a level of precision comparable to current results in determining $\delta_{cp}$.

\begin{figure}[hbt!]
\centering
\includegraphics[width=\textwidth]{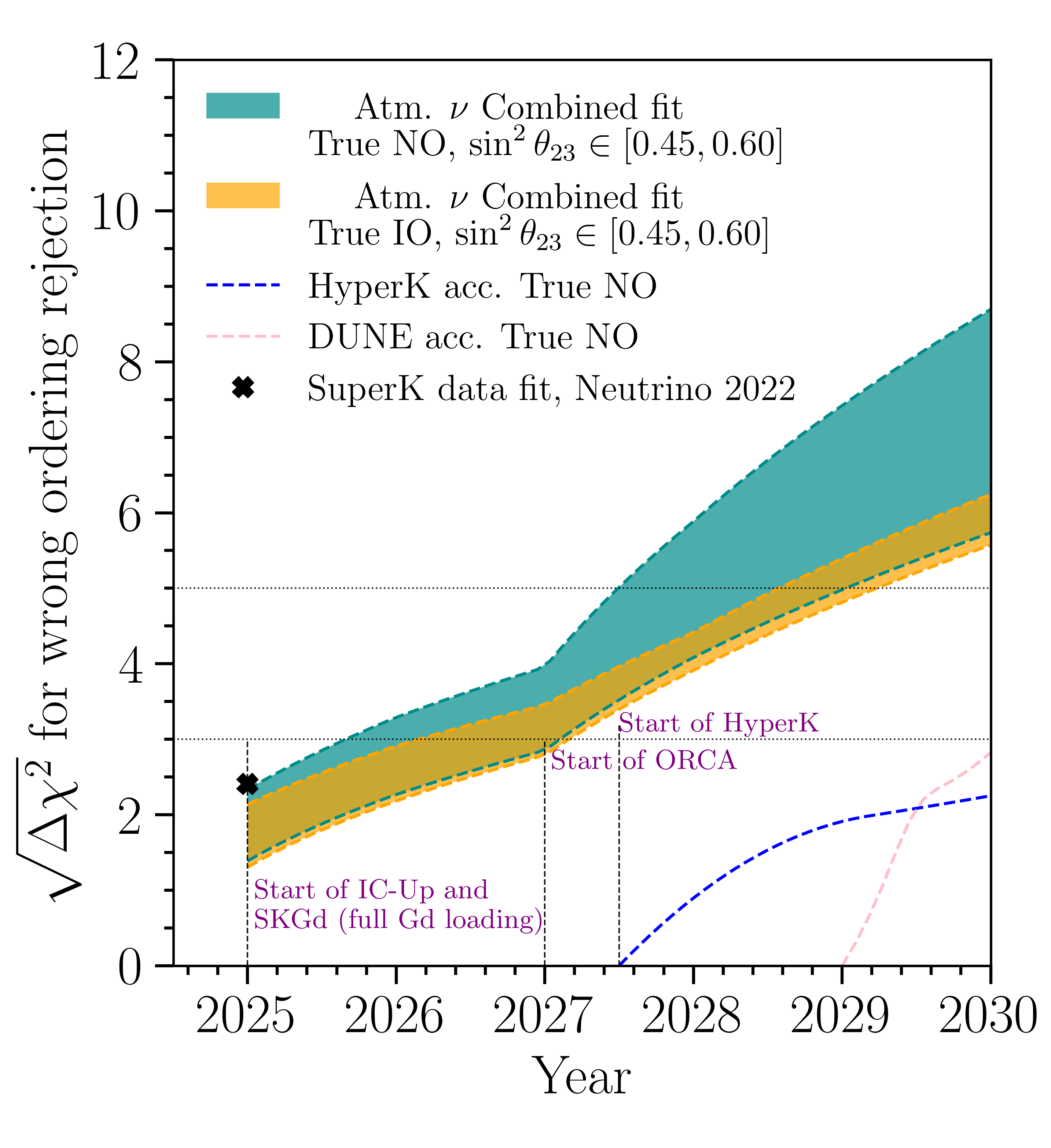}
\caption{\textbf{\textit{Neutrino mass ordering sensitivity as a function of years in operation.}} The cyan (orange) band shows the sensitivity for rejecting the wrong ordering hypothesis for true normal (inverted) ordering, assuming fixed $\sin^2\theta_{13}=0.022$.
The width of the bands covers the allowed values for $\sin^2\theta_{23}$ from 0.45 to 0.6. The black dot corresponds to the last reported SuperK neutrino mass ordering analysis~\cite{linyan_wan_2022_6694761}.
For comparison, we also include the prediction of the next-generation long-baseline neutrino experiments that are supposed to start in mid 2027 in the case of HyperK and 2029 in the case of DUNE~\cite{https://doi.org/10.48550/arxiv.2209.06872}.}
\label{fig:ordering}
\end{figure}

As we will demonstrate in this article, data from current and soon-to-operate atmospheric neutrino experiments can be combined to address some of the most pressing questions in neutrino physics.
The questions that we will discuss in this article can be organized into three categories: determining the neutrino oscillation parameters, establishing the neutrino mass spectra, and measuring the CP-phase in the lepton sector. 
The neutrino oscillation parameters are encoded in the so-called Pontecorvo-Maki-Nakagawa-Sakata (PMNS) matrix, which relates the neutrino weak and mass eigenstates~\cite{a_footnote}.
The precise determination of the lepton mixing parameters is crucial for understanding the neutrino evolution, and it can also be the first indication of a hidden flavor symmetry~\cite{Altarelli:2010gt, Ishimori:2010au}.
Measuring a large $CP$-violation in the neutrino sector can also be an explanation to the baryon asymmetry of the early universe via a sphaleron process~\cite{Fukugita:1986hr}.
Finally, the determination of the neutrino mass spectrum in the next few years will significantly impact experiments whose goal is to determine the absolute scale of the neutrino masses~\cite{KATRIN:2021uub}, to discriminate between the Dirac or Majorana nature of the neutrino masses~\cite{GERDA:2018pmc,KamLAND-Zen:2016pfg}, and even to understand the evolution of the universe~\cite{DiValentino:2021hoh}.

In the rest of the section, we will provide a concise summary of the main results from this study.
In atmospheric neutrino oscillations, the dominant mixing angle is $\theta_{23}$, and the relevant mass-squared-difference is $\Delta m^2_{31}$.
The atmospheric mixing angle is currently the least precisely determined of the mixing angles and is known to be close to maximal, i.e., $\sin^22\theta_{23} \approx 1$.
However, current experimental data points towards deviations from maximality, but is unable to resolve the octant, i.e., whether $\theta_{23}$ is smaller or greater than $\pi/4$.
In terms of the neutrino flavor structure, the maximality and octant question can be rephrased as understanding the relative contribution of tau and muon flavor in the second mass state, where a maximal angle implies equal amounts, the first octant more muon, and the second octant less muon.
Therefore, the large muon neutrino component of the atmospheric flux can provide significant knowledge on that parameter.
In this article, we will demonstrate that by combining the SuperK, IceCube Upgrade, ORCA and Hyper-Kamiokande (HyperK) measurements (see~\cref{sec:exp}), we will reach a half percent-level precision on $\Delta m^2_{31}$ and $\sim 2\%$ in the case of $\sin^2\theta_{23}$: see~\Cref{fig:present}.
Furthermore, the atmospheric measurements will allow us to discriminate the wrong octant at more than $3\sigma$ by 2030 assuming the current best-fit value of $\theta_{23}$: see~\Cref{{fig:t23vdm231}}.
As we will discuss in~\cref{sec:disc}, those measurements are primarily limited by statistics.
In~\Cref{sec:TH23} we checked that our main results and conclusions are independent of the benchmark scenario considered. 

The neutrino mass spectra enter the neutrino oscillation expressions through the sign of the mass-square differences.
The neutrino mass states are conventionally labeled by their decreasing number of electron-neutrino components, where $\nu_1$ is the state that has the most electron neutrino and $\nu_3$ the least; in symbols, $|U_{e1}| > |U_{e2}| > |U_{e3}|$.
From solar neutrino oscillations it is known that the second mass state, $\nu_2$ is heavier than the first one, $\nu_1$; however, its not known if $\nu_3$ is heavier (normal ordering, NO) or lighter (abnormal or inverted ordering, IO) than the other two states.
This is referred to as the neutrino ordering problem and is one of the main objectives of JUNO~\cite{JUNO_ORCA} and next-generation neutrino experiments DUNE~\cite{DUNE:2020ypp} and Hyper-Kamiokande~\cite{hk_tdr}.
Currently, the combination of neutrino data favors normal ordering mildly, although no conclusive measurement of this has been achieved to date.
In this article, we show that the combination of neutrino experiments discussed in this work will determine the neutrino ordering at more than $5\sigma$~\cite{b_footnote}.
Atmospheric neutrino experiments can reach $4\sigma$ sensitivity before the next generation of neutrino accelerator experiments starts taking data,~\Cref{fig:ordering}.
The sensitivity to this parameter is mainly affected by statistics and the miss-reconstruction of $\nu_{\tau}$ interaction represented by the $\nu_{\tau}$ cross section as discussed in~\cref{sec:disc}.
In~\cref{sec:xsec}, we summarize the present status of the neutrino cross section measurements along with the most relevant experimental measurement that will happen in the next year and will contribute to increasing the sensitivity over that parameter.

Finally, we turn to the question of $CP$-violation in the leptonic sector, which has been previously discussed in the context of atmospheric neutrino experiments~\cite{Razzaque:2014vba}.
The combination of the experiments considered in this work will provide a measurement of the $CP$-violating phase that will shrink the allowed region in a factor five (\Cref{fig:present}), assuming the preferred value of T2K~\cite{T2K:2019bcf} and SuperK~\cite{linyan_wan_2022_6694761,Abe_2018} measurements, being able to exclude more than half of the allowed parameter space at more than $90\%$~confidence level for any value of $\delta_{CP}$: see~\Cref{sec:results}.
The capacity to measure the $CP$-phase is dominated by SuperK and HyperK due to their large low-energy neutrino efficiency and neutrino-antineutrino separation ability.
The improved neutrino-antineutrino separation is due to a recent upgrade of SuperK, where the detector is doped with gadolinium (SKGd).
See~\cref{sec:exp} for a detailed description of the detector response and the simulation used in this analysis.
The combination of a precision measurement of the oscillation parameters by the neutrino telescopes and an improve SuperK detector allows us to bring new, and complementary information on the $CP$-phase compared to that obtained by long-baseline experiments.
This is of particular interest since measurements by current long-baseline experiments in the continental United States and Japan are in mild tension.
This work shows, for the first time, that atmospheric experiments have the potential to weigh in on this tension.
Finally, our work implies that before the operation of the next-generation neutrino detectors --- DUNE, Hyper-Kamiokande, and IceCube-Gen2 --- we will have two independent measurements of the $CP$-phase: one from the combination of accelerator neutrinos (i.e., T2K and NOvA), and another one from the combination of atmospheric neutrinos.

The combination of the  SuperK, IceCube Upgrade, ORCA, and HyperK has a twofold purpose of solving open questions in neutrino physics and providing initial input for the next generation of neutrino experiments.
In this work, we have developed for the first time the necessary tools to perform such a combined analysis along with the most realistic publicly available simulations for each experiment involved.
All of this allows us to make the first in-depth analysis of those three experiments taking into account a detailed description and implementation of the detector responses, and their common systematic uncertainties.
The work performed here is well-beyond what has currently been done in any prior global analysis of atmospheric neutrinos.
The result is the realistic projection of the sensitivity of atmospheric neutrinos to the remaining mixing parameters and the identification of the uncertainties limiting these measurements, thus establishing a competitive and independent approach from accelerators to the neutrino physics precision era.

The rest of this article is organized as follows.
In~\Cref{sec:flux}, we outline the primary characteristics of the atmospheric neutrino flux and discuss associated uncertainties. \Cref{sec:xsec} delves into the interaction of these neutrinos with water, while \Cref{sec:WC} focuses on key aspects of reconstructing these interactions. Atmospheric neutrinos needs to cross the Earth before to reach the detector, the main aspect of the neutrino evolution are summarized in~\Cref{sec:OSC}. In this analysis, we have incorporated four experiments: Super-Kamiokande, IceCube Upgrade, ORCA and Hyper-Kamiokande. The scope of measurements conducted by each of these experiments are elucidated in~\Cref{sec:exp}. \Cref{sec:results} contains a description of the main results of this analysis, while \Cref{sec:disc} engages in a comprehensive discussion of their implications. We present out conclusion in~\Cref{sec:concl}.

%%%%%%%%%%%%%%%%%%%%%%%%%%%%%%%%%%%%%%%%%%%%%%%
\section{Atmospheric Neutrino Fluxes}\label{sec:flux}
\begin{figure}[hbt!]
\centering
\includegraphics[width=1.\textwidth]{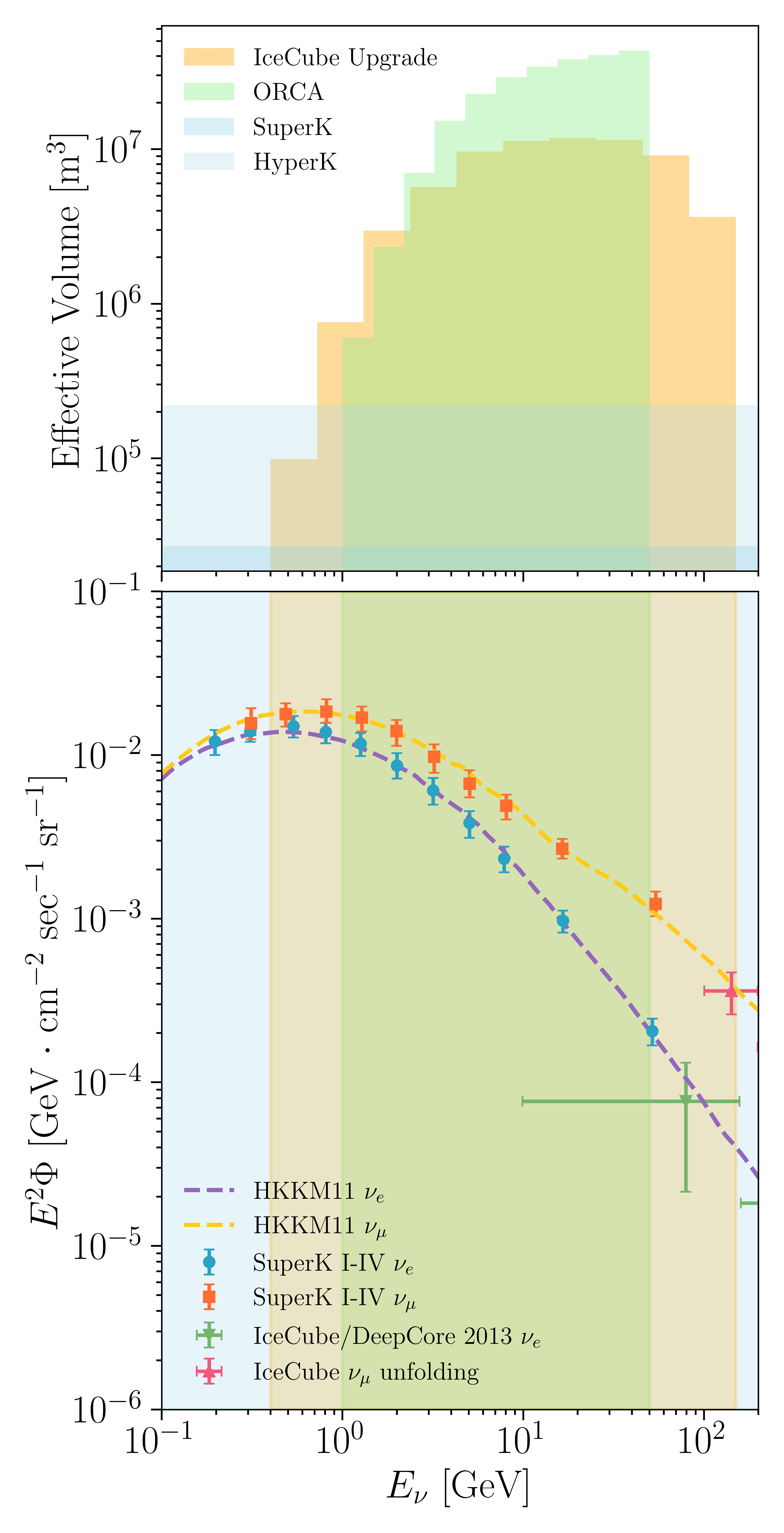}
\hfill
\caption{\textbf{\textit{Measurements of the atmospheric neutrino flux as a function of the energy.}} 
The total neutrino flux measured by different experiments~\cite{IceCube:2010whx,IceCube:2012jwm,Super-Kamiokande:2015qek} together with the energy range covered by the four experiments (SuperK, IceCube Upgrade, ORCA and HyperK) considered in this analysis is shown in the lower panel. On that figure, we have also included the flux prediction from HKKM2014 model~\cite{Honda:2015fha}. On the top panel, we have the effective volume for the three experiments as a function of the neutrino energy.}
\label{fig:fluxmeasured}
\end{figure}

In this section, we will elucidate the key elements of the atmospheric neutrino flux that play a pivotal role in determining oscillation parameters. 
Additionally, we will examine the primary sources of uncertainty influencing the measurement of this flux and  propose a parametrization that incorporates them.

Atmospheric neutrinos are produced by the collision of cosmic rays with Earth's atmosphere.
The primary spectrum of cosmic rays spans from MeV to EeV energies~\cite{Gaisser:2002jj,Dembinski:2017zsh}, and is composed of free protons ($\sim 80\%$) and bound nuclei ($\sim 20\%$).
Their interaction with nuclei in the atmosphere initiates hadronic showers on average about 20~km above the surface, producing copious amounts of mesons.
Neutrinos are produced predominantly from the decay of muons, pions, and kaons, which dominate the muon-neutrino flux below 10, 100, and $10^6$ ~${\rm GeV}$, respectively~\cite{Fedynitch:2018cbl,Fedynitch:2022vty}.
With $\nu_{\mu}$, $\nu_{e}$ are also produced in the atmosphere, and above the TeV scale there also exists a $\nu_{\tau}$ flux from the charmed mesons decays~\cite{Enberg:2008te,Gauld:2015kvh}.  
At the lowest energies, when decay of the mesons is prompt, the spectrum follows the cosmic-ray spectrum, but it softens by approximately one unit in spectral index as the mesons start interacting in the air~\cite{Gaisser:2016uoy}. 
In~\Cref{fig:fluxmeasured} (lower panel), we show the measurement of the total neutrino flux carried out by different experiments from $\sim 100$~MeV to $\sim 100$~GeV.
We added the prediction from Honda et al.~\cite{Honda:2015fha}, which is used in this work as a benchmark scenario for the neutrino flux.
We use the \texttt{NuFlux}~\cite{nuflux} package to interpolate those tables for the energy and directions relevant to this analysis.
The three experiments considered in our analysis measure the flux at energies below $\sim \SI{100}\GeV$.
The effective volume for SuperK, IceCube Upgrade, and ORCA as a function of the neutrino energy is shown in~\Cref{fig:fluxmeasured} (top panel).

\begin{figure*}[t]
\centering
\includegraphics[width=0.45\textwidth]{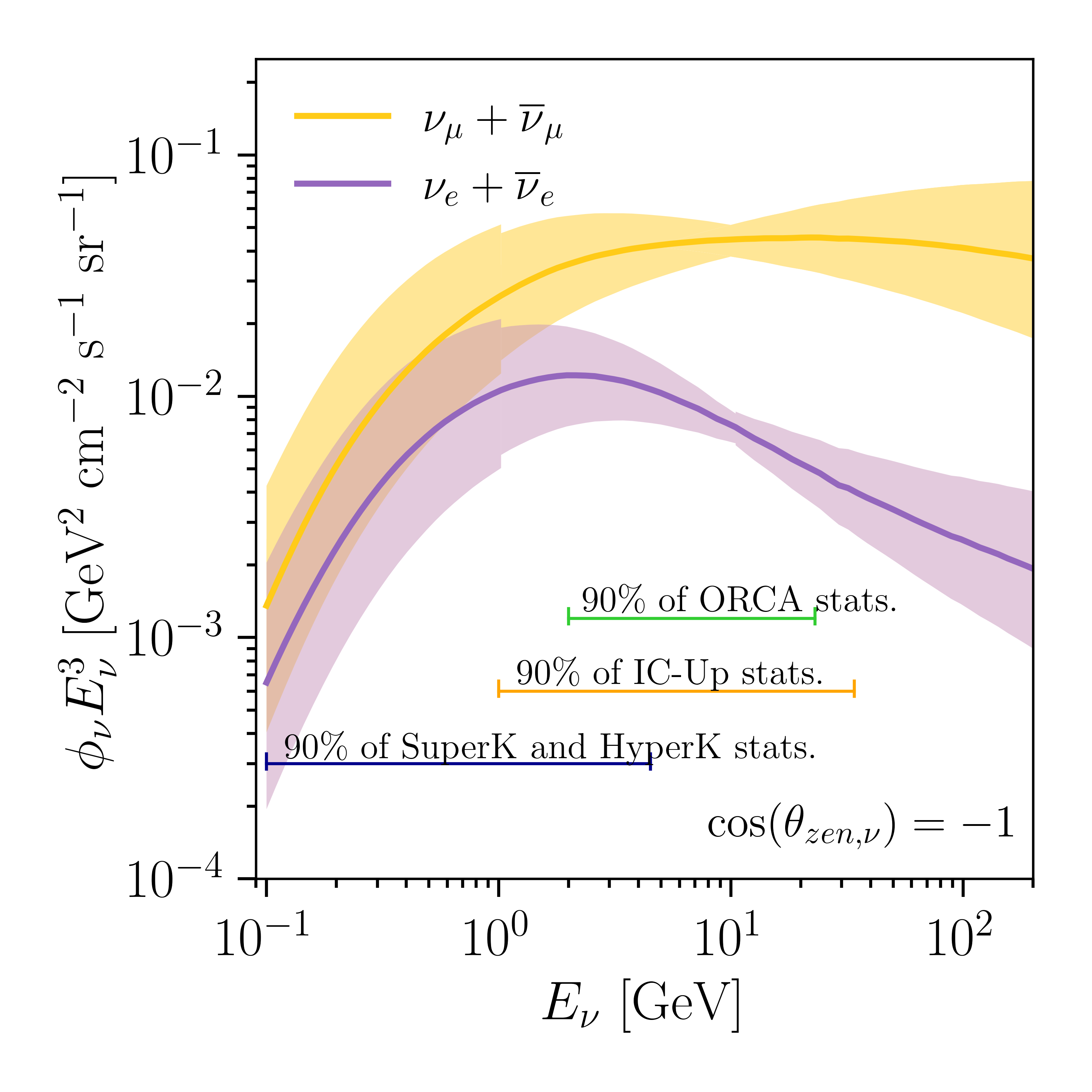}
\hfill
\includegraphics[width=0.45\textwidth]{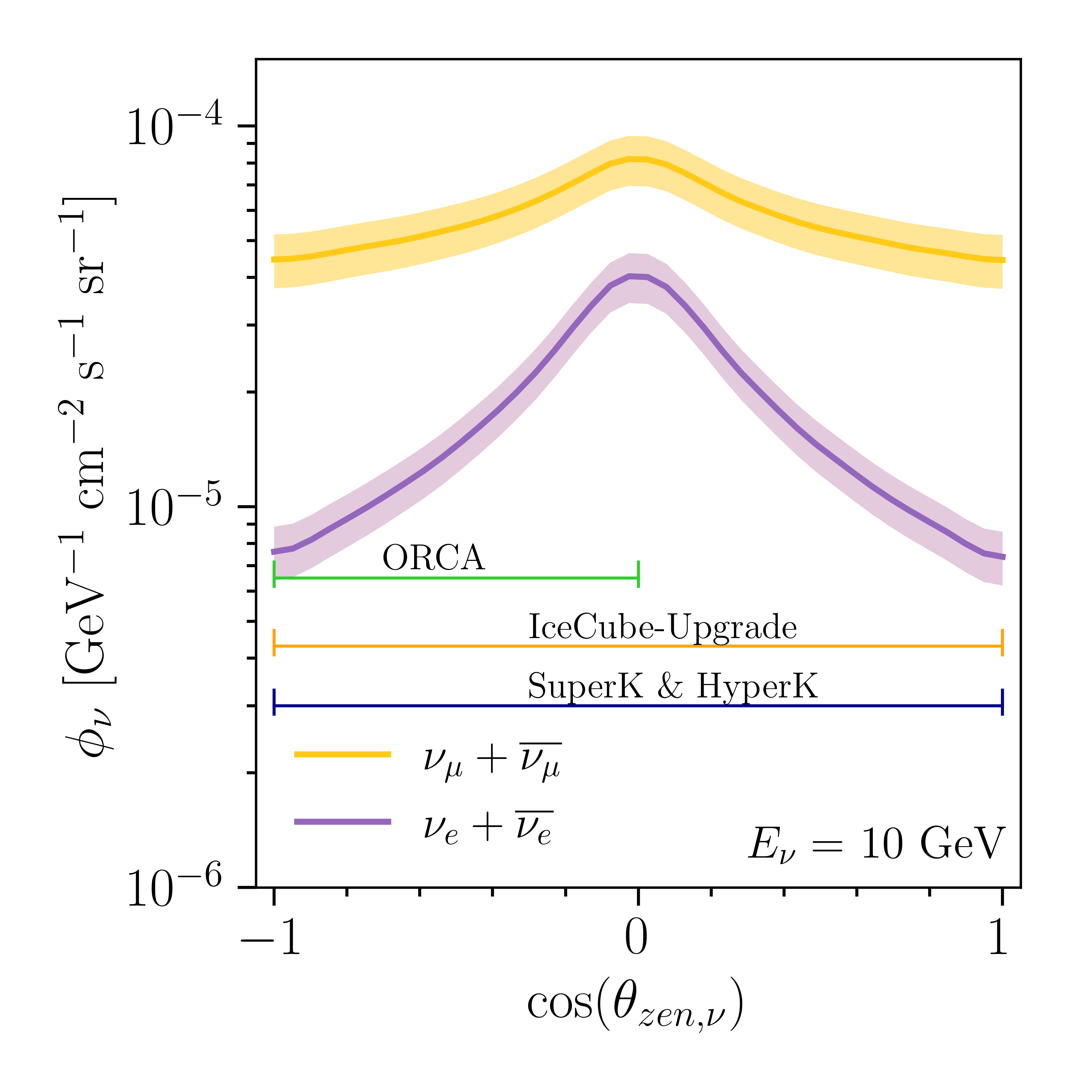}
% \vspace{2cm}
\caption{\textbf{\textit{Energy (left) and zenith (right) distribution of the atmospheric neutrino flux.}}
The flux is based on Honda et al.~\cite{Honda:2015fha} calculation.
Around the prediction for each flavor component of the flux at the detector, we show the $1\protect\sigma$ uncertainty band.
}
\label{fig:flux}
\end{figure*}

The zenith distribution of the neutrino flux at the detector shows an enhancement for the horizontal directions due to the longer paths that mesons have to travel before hitting Earth.
An example of this effect is shown in~\Cref{fig:flux} (right) for $\cos (\theta_{zen}) = 0$ for both flavor components of the flux at $E=\SI{10}\GeV$, although the same effect also happens at larger energies.
For energies where the mesons and muons have decayed before reaching Earth, there is still a horizontal enhancement due to the spherical geometry~\cite{Lipari:2000wu} of the volume where the neutrinos are produced.
The absorption of the mesons and muons by the Earth also contributes to modifying the flavor composition of the initial flux.
If all the parent particles are able to decay, we can expect that $(\nu_{e} + \overline{\nu}_{e}) / (\nu_{\mu} + \overline{\nu}_{\mu}) \sim 1/2$.
As the energy increases and muons hit the Earth, losing their energy, this ratio decreases in the meantime. 

In the sub-GeV energy range, the flux shows additional anisotropies due to the interaction of the charged mesons with Earth's magnetic field.
At low momentum, the mesons produced in the interaction of the cosmic rays get trapped by the magnetic field and can decay into neutrinos, enhancing the flux at lower energies.
The trajectory of the primary cosmic rays that reach Earth also gets modified by the magnetic effects inducing an east-west asymmetry~\cite{Super-Kamiokande:1999mpf} and a dependence of the flux with the location of the Earth~\cite{Honda:2015fha}.
Further, the interaction of the meson flux with Earth's magnetic field modifies the neutrino to antineutrino ratio.
At lower energies, the multiple scatterings of the mesons wash out the differences between both fluxes.
As the energy increases, the geomagnetic effect becomes less important on the meson fluxes, and the neutrino production is dominated by the primary flux mesons.

The uncertainties in the calculation of the atmospheric neutrino flux from cosmic-ray interactions described above arise from four factors: incident cosmic-ray flux, the hadronic interaction model, the atmospheric air density profile, and the magnetic effect at low energies.
Precision measurements of the Earth's atmospheric density have been performed by the NASA Atmospheric Infrared Sounder (AIRS) on board the Aqua satellite~\cite{ImprovingGlobalAnalysisandForecastingwithAIRS}.
This data has been used to compute the atmospheric neutrino fluxes and study the effects of seasonal variations ~\cite{2018PhDT.......184C,AtmosUncertaintiesCollin,euan2015}.
The effects of seasonal variations on the atmospheric muon neutrino flux are less than 10\% for neutrinos below a TeV~\cite{AtmosUncertaintiesCollin} and average out significantly for data sets that span multiple years.
Additionally, cosmic-ray spectral measurements in the relevant energy range have been recently performed by AMS.
Hadronic interaction models play a significant role in the uncertainty of atmospheric neutrino calculations; see~\cite{Barr:2006it,Fedynitch:2022vty} for reviews on this topic.
The energy range relevant for our analyses is well covered by accelerator data predominantly by the NA49 and NA61 experiments at CERN, though measurements by HARP, PHENIX and STAR are relevant on the lower and higher energy bands considered in this work~\cite{Barr:2006it,Fedynitch:2022vty}.
These sources of uncertainty can be translated to bands on the atmospheric neutrino fluxes. 
As recently estimated~\cite{Fedynitch:2022vty}, at below $\approx\SI{100}\GeV$, the normalization uncertainty of $\nu_\mu + \bar\nu_\mu$ and $\nu_e + \bar\nu_e$ is below 10\% , while at energies  below $\SI{10}\GeV$, the precision on the ratio of muon-to-electron neutrinos is known with a at an error below 2\%.
Additionally, at energies below $\approx\SI{10}\GeV$, the ratio of neutrinos to antineutrinos for muons is known with a precision between 1\% and 5\%, while for electrons this ratio is known to a precision of 10\%.

To include all the uncertainties mentioned before, we use a similar parametrization of the flux as in Ref.~\cite{Barr:2006it,Kelly:2021jfs}
\begin{equation}
 \Phi_{\alpha}(E,\cos\zeta) =  f_{\alpha}(E,\cos\zeta)\Phi_{0}\left(\frac{E}{E_{0}}\right)^{\delta} \eta(\cos\zeta),
\end{equation}
where $f_{\alpha}(E,\cos\zeta)$ are the Honda table's values interpolated by \texttt{NuFlux}~\cite{nuflux}.
The symbol $\Phi_{0}$ describes the uncertainty over the normalization of the flux; $(E/E_{0})^{\delta}$ modifies the energy dependence of the flux; and $\eta(\cos\zeta)_{\text{u,d}} = 1 - C_{\text{u,d}}\tanh(\cos\zeta)^{2}$ describes the relative uncertainty between horizontal and up-going or down-going directions. \Cref{table:FluxSysts} summarizes the atmospheric flux systematic uncertainties of our analysis. For easy comparison, we use a systematic uncertainty budget that is consistent with recent SuperK~\cite{Super-Kamiokande:2017yvm} and IceCube~\cite{IceCube:2019dqi} analyses.
However, the choice of uncertainty parameterization used in this article is conservative given the discussion above, and is expected to be further improved by further measurements.
In~\Cref{fig:flux}, we show the $1\sigma$ range of the energy (left) and zenith (right) uncertainties used in our analysis.

\begin{table}
\begin{tabular}{l|c}
Systematic source & 1$\sigma$-range\\
\hline\hline
Norm. $E_\nu<$1~GeV		&	25\%	\\
Norm. $E_\nu>$1~GeV   &	15\%	\\
Spectral index  & 20\%         \\
Flux $\nu/\overline{\nu}$  & 2\%           \\
Flux $\nu_e/\nu_\mu$  & 2\%           \\
Up and Horizontal  & 2\%           \\
Down and Horizontal  & 2\%           \\
\end{tabular}
\caption{Summary of atmospheric neutrino flux systematic uncertainties used in this work.}
\label{table:FluxSysts}
\end{table}

%%%%%%%%%%%%%%%%%%%%%%%%%%%%%%%%%%%%%%%%%%%%%%%
\section{Neutrino Water Cross Section}\label{sec:xsec}

The experiments in this work share, in addition to the neutrino flux, the neutrinos cross sections in water.
These experiments cover a wide range of neutrino energies in which neutrinos may interact via the exchange of charged or neutral currents.
The relevance of distinct interaction channels differs from one experiment to another, since they measure the atmospheric flux at distinct energy scales.
Therefore, we need to study the different interaction channels as they may affect the determination of oscillation parameters in a different way in each experiment. 
In this section, we will provide a concise overview of the essential aspects of the neutrino cross-section that bear relevance to the determination of oscillation parameters.

\begin{figure}
\centering
\includegraphics[width=\textwidth]{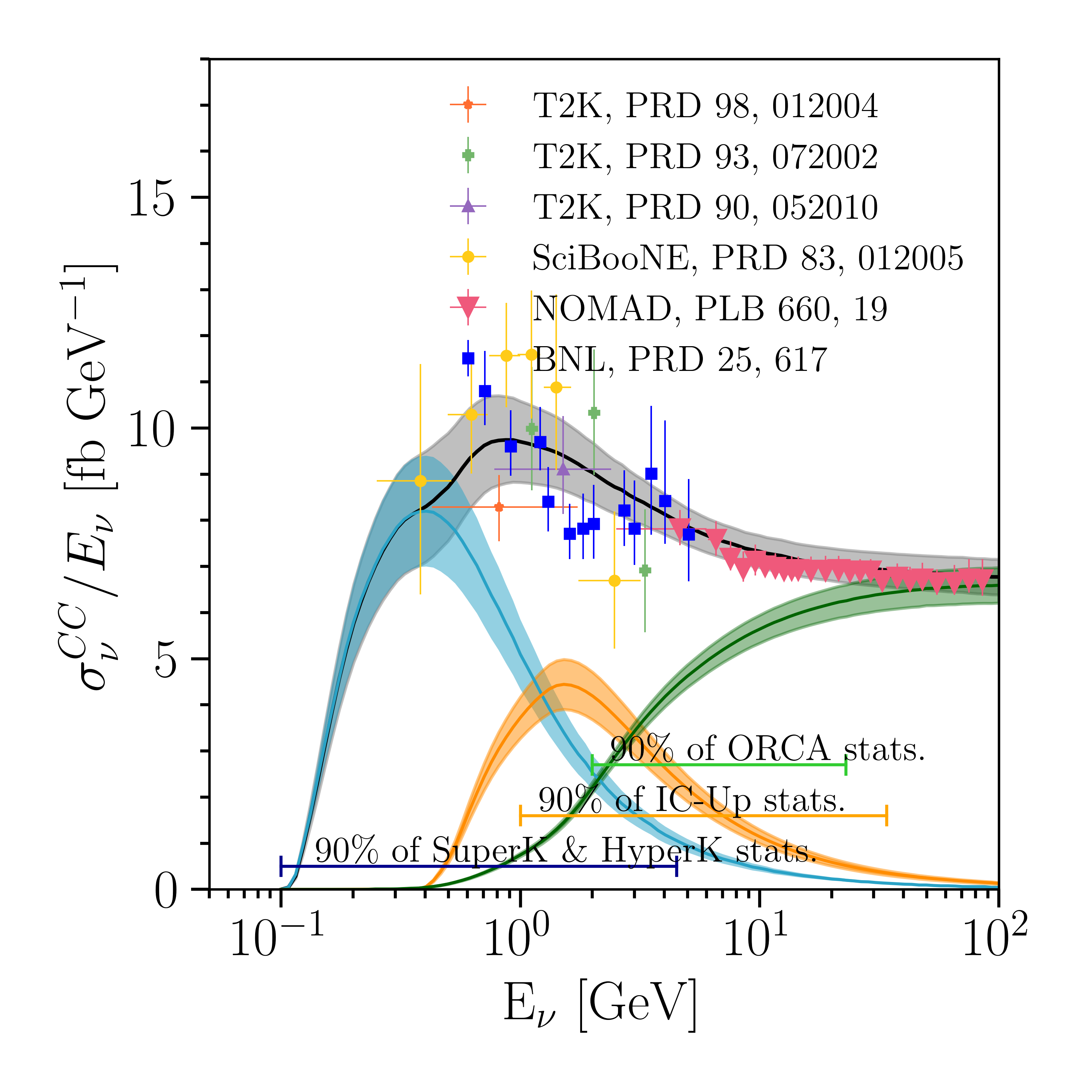}
\caption{\textbf{\textit{Charged-current $\nu_\mu$ cross section per nucleon as a function of energy split by its main contributions, QE, RES, and DIS.}}
The cross section models used are from GENIE pre-computed splines, shaded regions correspond to the 1$\sigma$ priors assumed in~\Cref{table:XSecSysts}, and data points correspond to the most relevant inclusive cross section measurements for this work.}
\label{fig:xsec}
\end{figure}

Charged-current interactions produce a charged lepton that shares flavor with the original neutrino and are divided into three major contributions shown in~\Cref{fig:xsec}. 
\begin{itemize}
\item{\textit{Charged current quasi-elastic (CCQE):}}
These interactions dominate in the lower energy region, below $\SI{2}\GeV$, and scatter off one of the bound nucleons, exchanging a W$^\pm$ boson, emitting the charged lepton partner of the interacting neutrino.
The outgoing nucleon is either a proton, for neutrinos, or a neutron, for antineutrinos.
These interactions are most relevant in the sub-GeV region of the SuperK atmospheric neutrino dataset and the lowest energy bins for IceCube Upgrade and ORCA, where the sensitivity to the $CP$-phase resides.
Additionally, this channel provides a clear link between the matter-antimatter character of the incoming neutrino and the presence of a neutron in the final state, which becomes relevant when introducing neutron-tagging capabilities in SuperK.

\item{\textit{Resonance production (CC RES):}}
At slightly higher energies up to $\SI{4}\GeV$, neutrinos can excite an entire nucleon, producing a baryon resonance that in turn, quickly decays into a nucleon and single or multiple mesons.\\
The $\Delta(1232)$ baryon resonance dominates this channel, producing a single pion in the final state.
All three experiments are sensitive to this channel.\\
Similarly to CCQE, in these interactions, antineutrinos tend to produce more neutrons than protons in the final state.
Further, in single-pion production, neutrinos are linked to $\pi^+$ and antineutrinos to $\pi^-$.
In turn, oxygen atoms in water quickly absorb $\pi^-$ before decaying, providing a potential signature to separate neutrinos from antineutrinos from the reconstruction of $\pi^+$ decay products, namely Michel electrons.
The SuperK detector is sensitive enough to use these features, which in this energy region, provide sensitivity to the neutrino mass ordering from the core and mantle resonances between $\SI{2}\GeV$ and $\SI{10}\GeV$ from MSW effect of neutrinos propagating through Earth (\Cref{fig:Pmue}).\\
Beyond resonance production, single pion final states can also be produced from neutrinos coherently scattering the whole nucleus (Coh~$\pi$).

\item{\textit{Deep inelastic scattering (CC DIS):}}
At energies above $\SI{4}\GeV$, neutrinos can scatter off a single quark inside the nucleon, producing the corresponding charged lepton plus a hadronic shower in the final state.
In this type of event, the flavor reconstruction may get confused unless the outgoing charged lepton has sufficient momentum to be distinguished from the hadronic showers. \\
Given the large mass of the $\tau$ lepton, this is the only channel allowed for charged-current tau neutrino interactions with a threshold of $\SI{3.5}\GeV$.\\
This channel dominates the neutrino interactions measured at IceCube Upgrade and ORCA, and thus the sensitivity to the neutrino mass ordering through the second Earth's matter-effect resonance and to the atmospheric squared mass difference.
\end{itemize}

On the other hand, neutral-current interactions do not produce an accompanying charged lepton but a neutrino and, therefore, do not carry any information about the flavor of the incoming neutrino in water-Cherenkov detectors. 
%Still, they can be used to separate between neutrinos and anti-neutrinos if a relativistic proton is produced after a quasielastic interaction~\cite{Beacom:2003zu}, but the expected event rate is very small.
Thus, these interactions are a background for the neutrino flavor oscillation analyses.
For neutral currents, only analog resonance and deep-inelastic scattering channels can be reconstructed in water-Cherenkov detectors if the produced mesons or hadronic showers are sufficiently energetic, or if the products decay to other detectable particles.
An example of these interactions is the production of a $\pi^0$ decaying to a pair of photons, which are a source of background for charged-current electron-like (e-like) events below $\SI{1}\GeV$ in SuperK.

Besides, some products of primary interactions undergo secondary interactions within the nuclear media, making the final state of the neutrino interaction more complex and difficult to reconstruct and classify.
Secondary interactions become more prominent at higher energies and obscure some distinct signatures expected from primary interactions, like the content of pions and neutrons, weakening the flavor identification as to the separation between neutrinos and antineutrinos in the case of SuperK.

Over the last four decades, several experiments have measured neutrino cross sections with different targets, and over a wide range of energies and channels relevant to this work.
The first measurements were done in hydrogen and deuterium bubble chambers~\cite{PhysRevD.25.617, BARANOV1979258, ERRIQUES1977383}.
These experiments provided precise measurements but lacked the complexity and features of heavier target nuclei required in current and future neutrino oscillation experiments.
The structure and kinematics of nucleons as well as the re-scattering within the nucleus can potentially alter the neutrino interaction cross section and outcome, affecting the measurement of the oscillation parameters~\cite{Ankowski:2015jya, Benhar:2015wva, Coyle:2022bwa}.
As neutrino physics developed, neutrino cross section measurements on the relevant targets were necessary to account for these nuclear effects.

Both charged- and neutral-current neutrino interactions on water, carbon, and hydrocarbon targets are considerably well measured and theoretically understood~\cite{Ankowski:2015lma,Rocco:2016ejr,Simons:2022ltq}.
Most of the existing measurements are for muon neutrinos and antineutrinos, and cover the previously sketched channels, ranging from $\SI{0.4}\GeV$ to $\mathcal{O}(10^2)$~GeV and performed by Miner$\nu$a~\cite{Ren_2017, Lu_2021, minerva1, Coplowe_2020, Mislivec_2018, Altinok_2017}, K2K~\cite{Rodriguez_2008, Gran_2006, Hasegawa_2005, Nakayama_2005}, NOMAD~\cite{nomad, Wu_2008, Lyubushkin_2009, Kullenberg_2012}, SciBooNE~\cite{PhysRevD.83.012005}, and T2K~\cite{Abe_2014t2kprd90, Abe_2017t2k, PhysRevD.97.012001, Abe_2016t2k, Abe_2019t2k32, Abe_2021t2k43, Abe_2020t2k64, Abe_2020t2k456, T2K:2016jor, T2K:2018lnf}.
Additionally, T2K has also performed measurements for sub-GeV electron neutrinos~\cite{Abe_2014t2k, t2k_2020}.

One of the remaining issues is to achieve the same precision for electron neutrinos and antineutrinos, for which the data is much scarcer since the beams are mainly made of muon neutrinos.

In addition to the aforementioned Miner$\nu$a and T2K, there are other experiments with water targets in operation, such as ANNIE~\cite{annie} and NINJA~\cite{Hiramoto_2020, Fukuda_2017}.
These experiments, together with near future detectors like Hyper-Kamiokande's Intermediate Water-Cherenkov Detector (IWCD)~\cite{hk_tdr} and the upgrade of the current T2K's near detector ND280~\cite{nd280up}, will play a crucial role in reducing the uncertainties of both, muon and electron neutrino and antineutrino cross sections as well as differential cross sections at energies from hundreds of~MeV to a few GeV on a water target.
Additionally, the upgrade of ND280 will measure the ratio between carbon and oxygen to better precision, enabling a more precise extrapolation of cross section measurements on carbon and hydrocarbon targets.
Moreover, the tagging of neutrons from neutrino interactions is proven to be an effective tool for separating neutrinos from antineutrinos.
It is expected that the ANNIE and IWCD experiments will be Gd-doped detectors exposed to high-intensity neutrino beams, thus providing valuable input to reduce the current neutron production uncertainties.

In addition to more experimental measurements, the development of more precise nuclear and interaction models is crucial.
In this context, two topics will be of great importance in the coming years for neutrino physics.
First, the model-independent reassessment of bubble chamber data and, second, the deeper study of nucleon form factors and resonance production from updated models and lattice QCD calculations~\cite{Meyer_2022, xsec_theory_kat, xsec_theory}.
The latter work would be especially relevant for the CCQE and RES channels.

For charged-current tau neutrino data is much more limited; it comes primarily from the DONuT~\cite{donut} and OPERA experiments~\cite{OPERA:2021xtu} and, more recently, from tau appearance measurements in the atmospheric neutrino flux from IceCube and SuperK~\cite{Aartsen_2019,Super-Kamiokande:2017edb}.
These measurements found the cross section to be in good agreement with the expectations within 10\% of the expected value.
Tau appearance does not play a significant role in extracting the parameters of interest in this article and thus we do not discuss them further.

For a detailed review of neutrino cross sections and available data, the reader is referred to~\cite{Formaggio_2012,https://doi.org/10.48550/arxiv.2209.06872,Lovato:2020kba,CLAS:2021neh, Benhar_2017, Alvarez_Ruso_2018, Mahn_2018}.

The systematic uncertainties assumed in this work are summarized in~\Cref{table:XSecSysts} and follow the ranges assumed by the official oscillation analyses from each of the experiments considered in this work.
During the upcoming data-taking period of IceCube Upgrade, ORCA, and SuperK, it is expected that the neutrino cross section uncertainties will further improve with new data from experiments in operation and theoretical development.
These measurements and theoretical developments are motivated by the upcoming next-generation neutrino detectors, DUNE and Hyper-Kamiokande.

\begin{table}[ht]
\begin{tabular}{l|c}
Systematic source & 1$\sigma$-range \\
\hline\hline
CCQE            &       10\%    \\
CCQE $\nu/\overline{\nu}$   &   10\%    \\
CCQE $e/\mu$  & 10\%         \\
CC1$\pi$ production  & 10\%           \\
CC1$\pi$ $\pi^0/\pi^\pm$  & 40\%           \\
CC1$\pi$ $\nu_e/\overline{\nu_e}$  & 10\%           \\
CC1$\pi$ $\nu_\mu/\overline{\nu_\mu}$  & 10\%           \\
Coh. $\pi$ production  & 100\%           \\
Axial Mass ($M_A$)  & 10\%           \\
CC DIS  & 5\%           \\
NC hadron prod.  & 10\%           \\
NC over CC  & 20\%           \\
$\nu_\tau$  & 25\%           \\
Neutron prod. (SuperK only)~\cite{c_footnote}  & 15\%           \\
\end{tabular}
\caption{Summary of neutrino-water interactions systematic uncertainties used in this work.}
\label{table:XSecSysts}
\end{table}

%%%%%%%%%%%%%%%%%%%%%%%%%%%%%%%%%%%%%%%%%%%%%%%
\section{Physics of Water Cherenkov Detectors}\label{sec:WC}
Due to their large active and instrumented volumes, water- or ice-Cherenkov detectors have proven to be a very effective and successful technology for measuring the properties of neutrinos.

In these kinds of detectors, neutrinos interact with the nuclei in water molecules producing several particles depending on the primary interaction channel and subsequent secondary interactions within the nuclear media.
Charged particles with momenta above a given threshold determined by the refractive index of the medium emit Cherenkov radiation detected with photomultiplier tubes (PMTs).
Ice and water have sligtly different refractive indexes~\cite{Kravchenko2004InSI}, and thus, Cherenkov energy thresholds; namely $\SI{0.13}\GeV$ and $\SI{1.2}\GeV$ in Antartic ice and $\SI{0.16}\GeV$ and $\SI{1.4}\GeV$ in water for muons and protons, respectively.

Furthermore, both media exhibit well-understood light-propagating properties, enabling reliable event reconstruction.
This is clearly shown by all three experiments considered in this work.
On the one hand, thanks to its large photo-coverage and the exhaustive control of the water conditions, SuperK can reconstruct very low-energy events (MeV-scale), which gives a comprehensive reconstruction of atmospheric neutrino events at energies below $\SI{1}\GeV$.
On the other hand, IceCube Upgrade and ORCA experiments are capable of precisely parametrizing the optical properties of Antarctic ice and Mediterranean seawater, so it is possible to control such large volumes of unprocessed media and use them as particle physics detectors with energies as small as $\SI{1}\GeV$.
In this context, the ORCA experiment provides a better directional reconstruction than IceCube where there is more scattering of photons due to the presence of air bubbles trapped in the ice.\\
As charged particles propagate through a medium, they polarize the medium around them.
When these particles move faster than light in that medium, photons from the polarization, unable to keep up with the particles' pace, form a characteristic cone of light around the direction of motion, with a characteristic opening angle depending on the medium's refractive index.
The detection of this radiation allows the reconstruction of the direction and production vertex of the charged particle.
Additionally, being proportional to the number of photons emitted, the momentum of the particle is inferred from the charge collected by the PMTs.

In the context of neutrinos, this means WC detectors can reliably reconstruct the promptly charged leptons that originate from a neutrino charged-current interaction, namely $e^\pm$ and $\mu^\pm$ from electron and muon (anti)neutrinos, respectively. 
Moreover, in large-photo-coverage experiments, some particles are identifies based on the ring patterns; electrons produce diffuse ring patterns due to the electromagnetic showers produced as they propagate through water ($e$-like or showers), whereas muons, being more massive, produce rings with sharper edges ($\mu$-like or tracks).
Other particles such as high-energy photons and charged pions can be detected and reconstructed similarly, the former with a pattern practically indistinguishable from showers and the latter from tracks. 
All these features extend the capabilities of neutrino WC detectors, allowing for a complete reconstruction of the particles in the final state of the neutrino interaction as well as the decay products from heavier short-lived particles and other charged particles produced in secondary interactions.

There are, however, differences in the reconstruction among the detectors considered.
In the case of SuperK, the Cherenkov cone gets projected to the densely instrumented inner surface, which registers all the signals produced in the event.
In the case of IceCube and ORCA, strings of PMTs are scattered throughout the volume, enabling the measurement of the energy loss of particles as they travel through the ice (IceCube) and seawater (ORCA) at each step of the track.
See Refs.~\cite{IceCube:2013dkx,Abbasi:2021ryj,Garcia-Mendez:2021vts} for recent discussions on reconstruction on these type of detectors.

An additional yet relevant point is the detection of neutrons in WC detectors.
Currently, in the context of atmospheric neutrinos, this only applies to SuperK.
Despite having a null electric charge, neutrons get captured by hydrogen atoms of water, forming an excited state of deuterium.
Its de-excitation emits a $\SI{2.2}\MeV$ photon which is detected with an efficiency of $\sim$20\%.
Despite the modest tagging efficiency, this capability adds more information to the reconstructed final state of the interaction, improving the oscillation analysis sensitivity, which we will discuss in~\Cref{sec:SKntag}.

\begin{figure}
\centering
\includegraphics[width=0.75\textwidth]{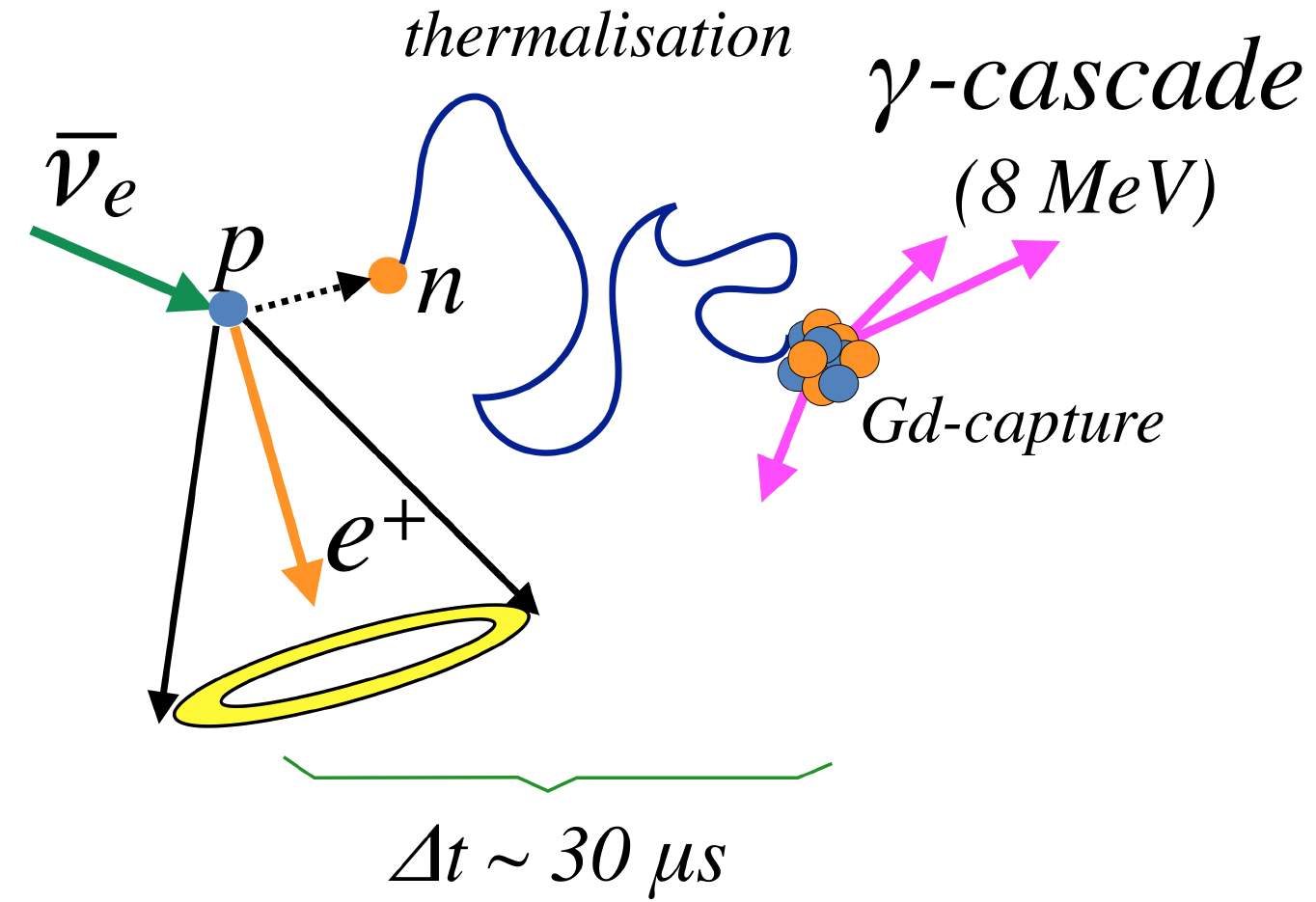}
\caption{\textbf{\textit{Diagram showing neutron tagging on gadolinium in an inverse-$\beta$ interaction.}}}
\label{fig:ntag}
\end{figure}

Additionally, given the proven relevant of neutron tagging~\cite{Fernandez:2017uko}, SuperK is being upgraded to dissolve a gadolinium (Gd) salt in the water to improve the detection of neutrons through their capture on this element~\cite{Beacom_2004}.
Gadolinium is the stable isotope with the largest thermal neutron absorption cross section, which, together with the emission of an $\SI{8}\MeV$ cascade of photons, as shown in~\Cref{fig:ntag}, enhances the neutron detection efficiency up to $\sim$80\%.

%%%%%%%%%%%%%%%%%%%%%%%%%%%%%%%%%%%%%%%%%%%%%%%
\section{Neutrino oscillations}\label{sec:OSC}
\begin{figure*}
 \includegraphics[width=0.92\textwidth]{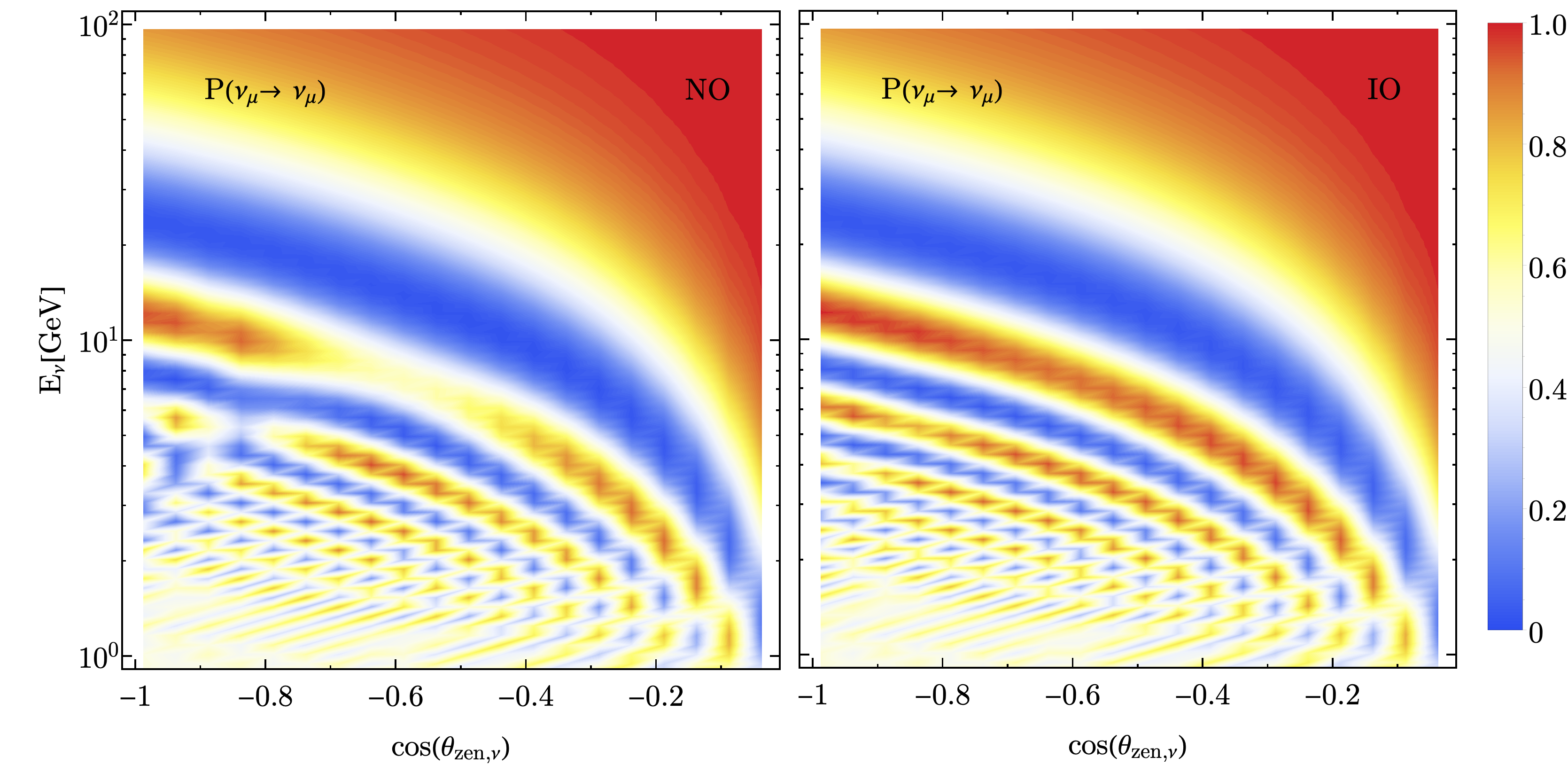}
\centering
\caption{\textbf{Muon-disappearance probability}. 
For energies above $\SI{1}\GeV$ and all the trajectories crossing the Earth  ($-1 < cos(\theta_{zen}) < 0.$), we have computed $P(\nu_{\mu}\rightarrow\nu_{\mu})$ for normal (left) and inverted (right) mass ordering.
We have considered the PREM for the Earth matter distribution. }
\label{fig:Pmumu}
\end{figure*}

\begin{figure*}
 \includegraphics[width=0.92\textwidth]{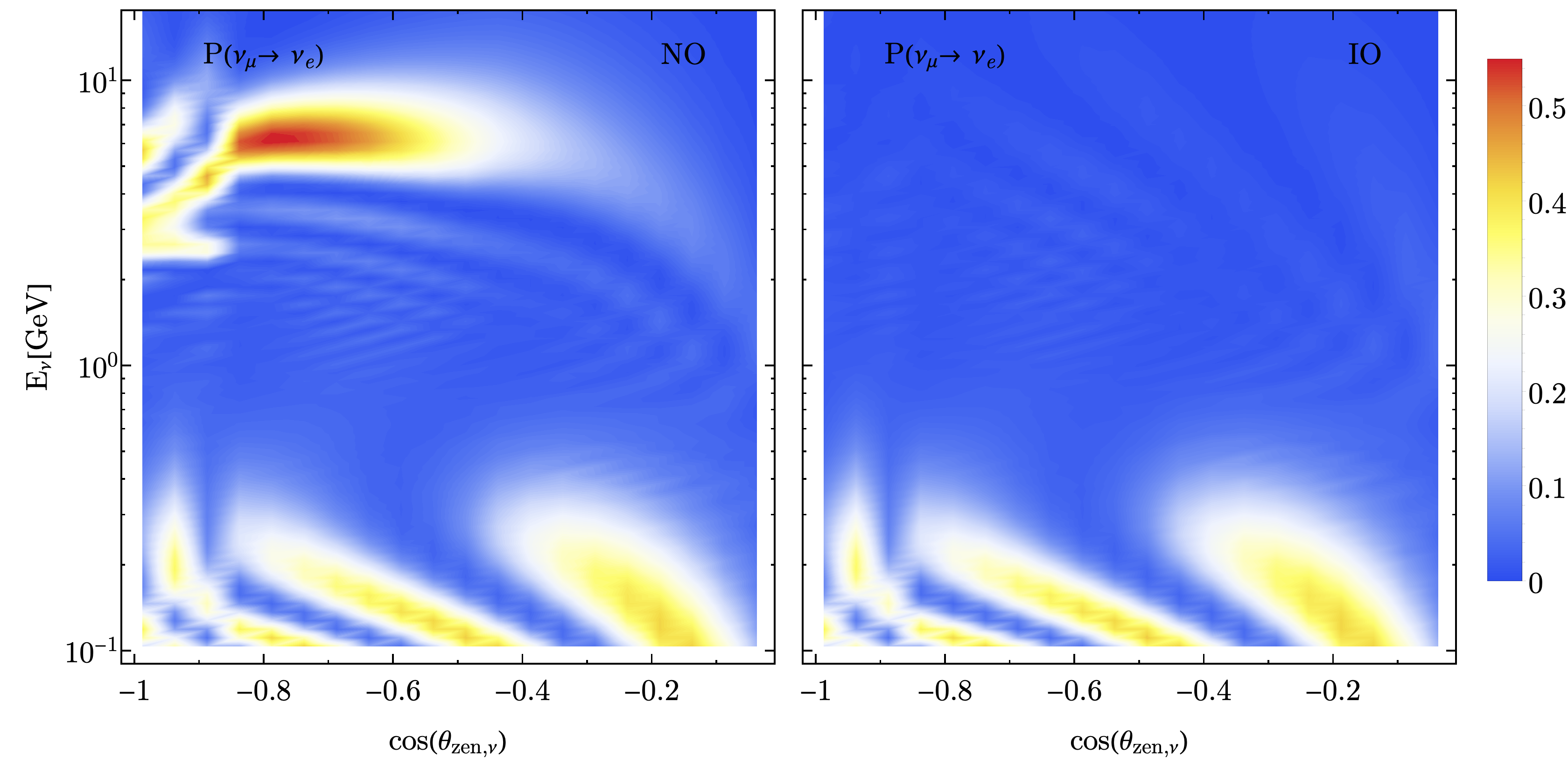}
\centering
\caption{\textbf{Electron-appearance probability}. Similar to~\Cref{fig:Pmumu}, we have computed $P(\nu_{\mu}\rightarrow\nu_{e})$ for both mass orderings, normal (left) and inverted (right). In this case, we have considered neutrinos with energies between $\SI{0.1}\GeV$ and $\SI{15}\GeV$.}
\label{fig:Pmue}
\end{figure*}

In the $3\nu$ scenario, neutrino evolution is described by six parameters: two mass-squared differences ($\Delta m^2_{31}$ and $\Delta m^2_{21}$), three mixing angles ($\theta_{12}$, $\theta_{13}$, and $\theta_{23}$), and a complex phase that parameterizes the violation of the $CP$-symmetry in the lepton sector.
The so-called solar parameters ($\Delta m^2_{21}$ and $\theta_{12}$) have been measured by solar experiments and KamLAND~\cite{SNO:2011hxd,PhysRevD.94.052010,BOREXINO:2020aww,KamLAND:2013rgu} looking for the disappearance of the electron neutrinos or anti-neutrinos.
Reactor experiments with a baseline of $\mathcal{O}(10^0~\text{km})$ and using a configuration with a near and a far detector has determined $\theta_{13}$, becoming the most precisely measured parameter to date~\cite{DoubleChooz:2019qbj,RENO:2018dro,DayaBay:2016ssb}.
Finally, the atmospheric parameters ($\Delta m^2_{31}$ and $\theta_{23}$) and the $CP$-violation phase require high-energy beams and, therefore are constrained by long-baseline experiments~\cite{T2K:2021xwb,NOvA:2019cyt,MINOS:2013utc,IceCube:2017lak,ANTARES:2012tem}.
Using a $\nu_{\mu}$ flux with energies at or near the GeV scale, those experiments search for the disappearance of muon neutrinos ($\nu_{\mu}\rightarrow\nu_{\mu}$) and for the appearance of electron neutrinos ($\nu_{\mu}\rightarrow\nu_{e}$).
In the case of muon-disappearance channel, since the matter effects are suppressed by $\sin^4\theta_{23}$~\cite{Akhmedov:2006hb}, the oscillation probability at leading order~\cite{Nunokawa:2005nx} can be written as
\begin{equation}\label{eq:dis}
    P_{\mu\mu} \approx 1 - 4 |U_{\mu 3}|^2 (1 - |U_{\mu 3}|^2)\sin^2\Delta m^2_{\mu\mu},
\end{equation}
where
\begin{equation}
\begin{split}
    \Delta m^2_{\mu\mu} = \sin^2\theta_{12} \Delta m^2_{31} + \cos^2\theta_{12} \Delta m^2_{32} \\
    + \cos\delta_{CP} \sin\theta_{13}\sin 2\theta_{12}\tan\theta_{23}\Delta m^2_{21},
\end{split}
\end{equation}
showing that we can constraint $\Delta m^2_{\mu\mu}$ and $\sin^2 2\theta_{23}$ by looking at the muon distribution in the detector.

In the appearance channel the matter effects are more important.
To describe long-baseline experiments (LBL) sensitivity using this channel, we use an expression valid in a constant matter neutrino evolution~\cite{Elevant:2015ska, Cervera:2000kp,Freund:2001pn} given by
\begin{align}
\label{eq:app}
    P_{\mu e} &\approx 4\sin^2\theta_{13}\sin^2\theta_{23}\frac{\sin^2 \Delta_{31}(1-a s A)}{(1-asA)^2}\\\nonumber
    &+s \frac{\Delta m^2_{21}}{|\Delta m^2_{31}|}\sin 2\theta_{13} \sin 2\theta_{12} \sin 2\theta_{23}\\\nonumber
    &\cos(s\Delta_{31} + a\delta_{CP}) \frac{\sin \Delta_{31} A}{A}\frac{\sin \Delta (1 - asA)}{1-asA},
\end{align}
where $a=1$ for neutrinos and $a=-1$ for anti-neutrinos, $s=\text{sign}(\Delta m^2_{31})$, and $\Delta_{ij} = \Delta m^2_{ij}L/4E$.
The matter effects in this channel are introduced via the term $A= 2EV/\Delta m^2_{31}$, where $V$ is the matter potential.
The dependence of this channel on the mass ordering are proportional to the matter effects.
The $\nu_{e}$ appearance also depends on $\sin^2\theta_{23}$ bringing the possibility to resolve between both $\theta_{23}$ octants.
Also, as we see from equation~\Cref{eq:app}, using this channel we have a dependence on $\theta_{13}$. 
Moreover, the possibility of long-baseline experiments in running in neutrino and anti-neutrino modes (or in the atmospheric case, having a mix beam of neutrinos and antineutrinos) enables these experiments to resolve the differences in the oscillation of both modes.
The comparison of the neutrino and antineutrino oscillation patterns can be translated into a measurement of $\delta_{CP}$. 

The latest results of the global analyses indicate that some of those parameters can be constrained to the percent level~\cite{deSalas:2020pgw,Esteban:2020cvm,Capozzi:2021fjo}, although there are still several sizeable uncertainties in $3\nu$ mixing scenario.
Among the less constrained parameters, we have $\theta_{23}$, for which values above and below maximal mixing are allowed within $1\sigma$; for the mass ordering, we have a $2\sigma$ preference for normal ordering (NO)~\cite{d_footnote}; and $\delta_{CP}$, for which just a small region around $\pi/2$ is excluded by T2K and SuperK at $3\sigma$~CL. 
Atmospheric neutrinos can contribute to narrowing down those uncertainties.

\subsection{Neutrino evolution through the Earth}

\begin{figure}
 \includegraphics[width=\textwidth]{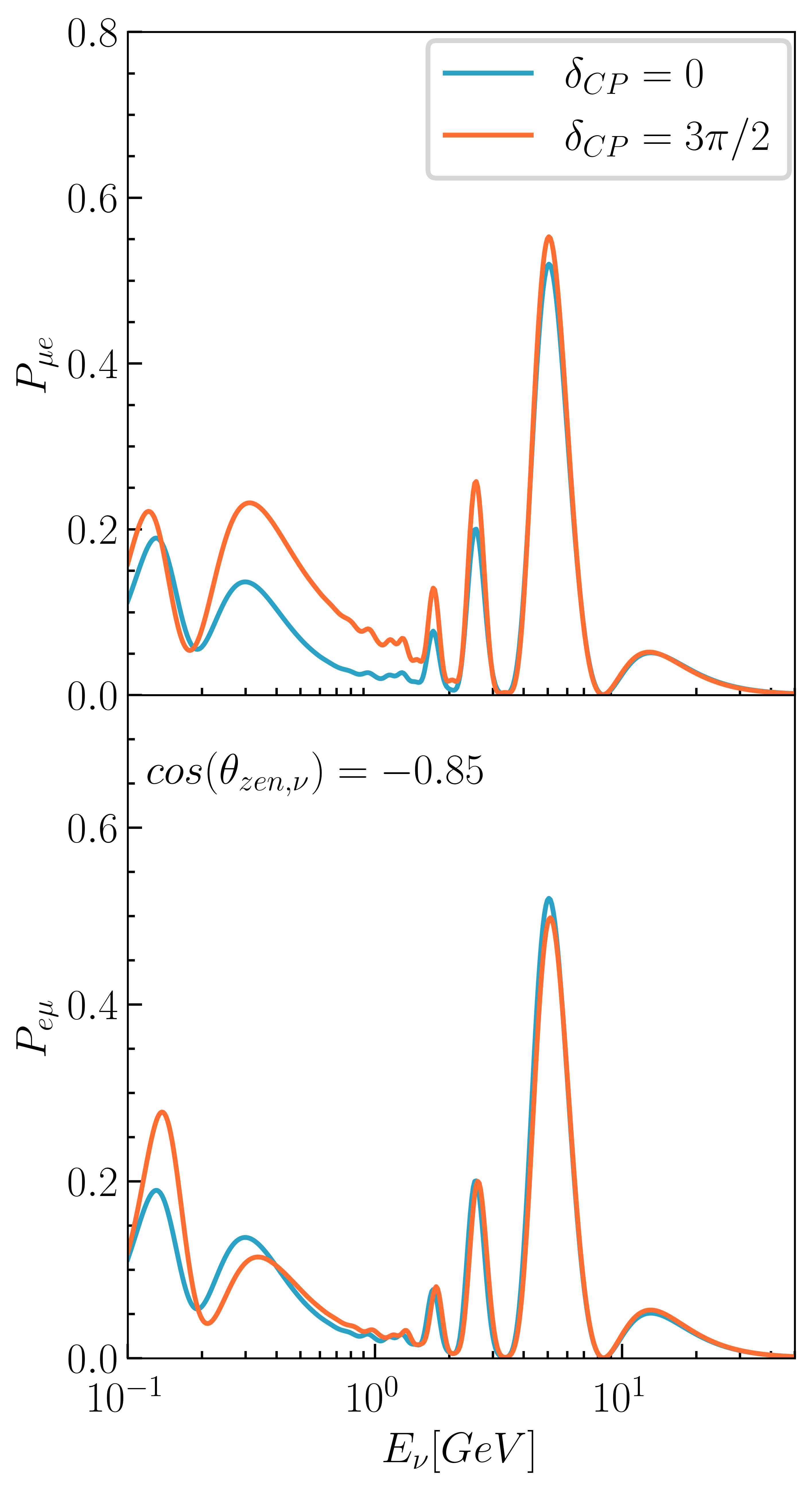}
\centering
\caption{\textit{\textbf{Electron (top) and muon (bottom) appearance probabilities for $\cos(\theta_{zen}) = -0.85$.}}
We show the impact of $\delta_{CP}$ in both oscillation channels.
The fast oscillations have been smeared assuming a Gaussian uncertainty of $5\%~$E/GeV.}
\label{fig:ProbCP}
\end{figure}

The large range of baselines covered by atmospheric neutrinos that extend from $\mathcal{O}(10~\text{km})$ to $\mathcal{O}(10^4~\text{km})$, and the vast energy range where the flux can be measured from $\mathcal{O}(10^{-2}~\text{GeV})$ to $\mathcal{O}(10^5~\text{GeV})$ ensures the access to a vast neutrino oscillation phenomenology, as we can see in~\Cref{fig:Pmumu,fig:Pmue}.

In the sub-GeV region, and for baselines $\sim\SI{1000}\km$, the neutrino evolution is dominated by $\Delta m^2_{21}$. The finite energy resolution makes the experiments inaccessible to the oscillation driven by $\Delta m^2_{31}$ and $\Delta m^2_{32}$. The average over those two oscillatory terms enhances the effects of $\delta_{CP}$ over neutrino evolution~\cite{Kelly:2019itm,Martinez-Soler:2019nhb}. The asymmetry between neutrino and anti-neutrino oscillation comes through the Jarlskog invariant~\cite{Jarlskog:1985ht,Jarlskog:1985cw} defined as $J = \Im[U_{\alpha i}U^{\ast}_{\alpha j}U^{\ast}_{\beta i}U_{\beta j}] = J_{r}\sin\delta_{CP}$, where we have factorized the dependence with $\delta_{CP}$. In vacuum, the $CP$-violation term is given by the product of the three oscillation wavelengths~\cite{Denton:2016wmg} and the Jarlskog invariant; namely
\begin{equation}
P_{CP} = -8 J_r \sin\delta_{CP} \sin\Delta_{21} \sin\Delta_{31} \sin\Delta_{32},
\end{equation}
where $\Delta_{ij} = \delta m^2_{ij}L/4E$. The average over the two largest mass-splittings suppresses the $P_{CP}$ term by $\sim 1/2$~\cite{e_footnote}.
\Cref{fig:ProbCP} shows the electron- and muon-appearance probability for two values of $\delta_{CP} = 0$ and $\pi$. In the case of $CP$-conservation, both probabilities are the same, as seen in the top and bottom panels of the figure. As $\delta_{CP}$ takes a different value, both appearance channels separate, increasing the sensitivity over that parameter~\cite{f_footnote}.
The $CP$-violation results results in a different normalization of the probability and a shift in the oscillation phase: these effects are quite broad in energy and more relevant specifically in the sub-GeV energy range. The matter effects depend on the neutrino trajectory along the Earth and create a dependence on the $CP$-effects with the neutrino direction. 

\begin{figure}
\centering
 \includegraphics[width=\linewidth]{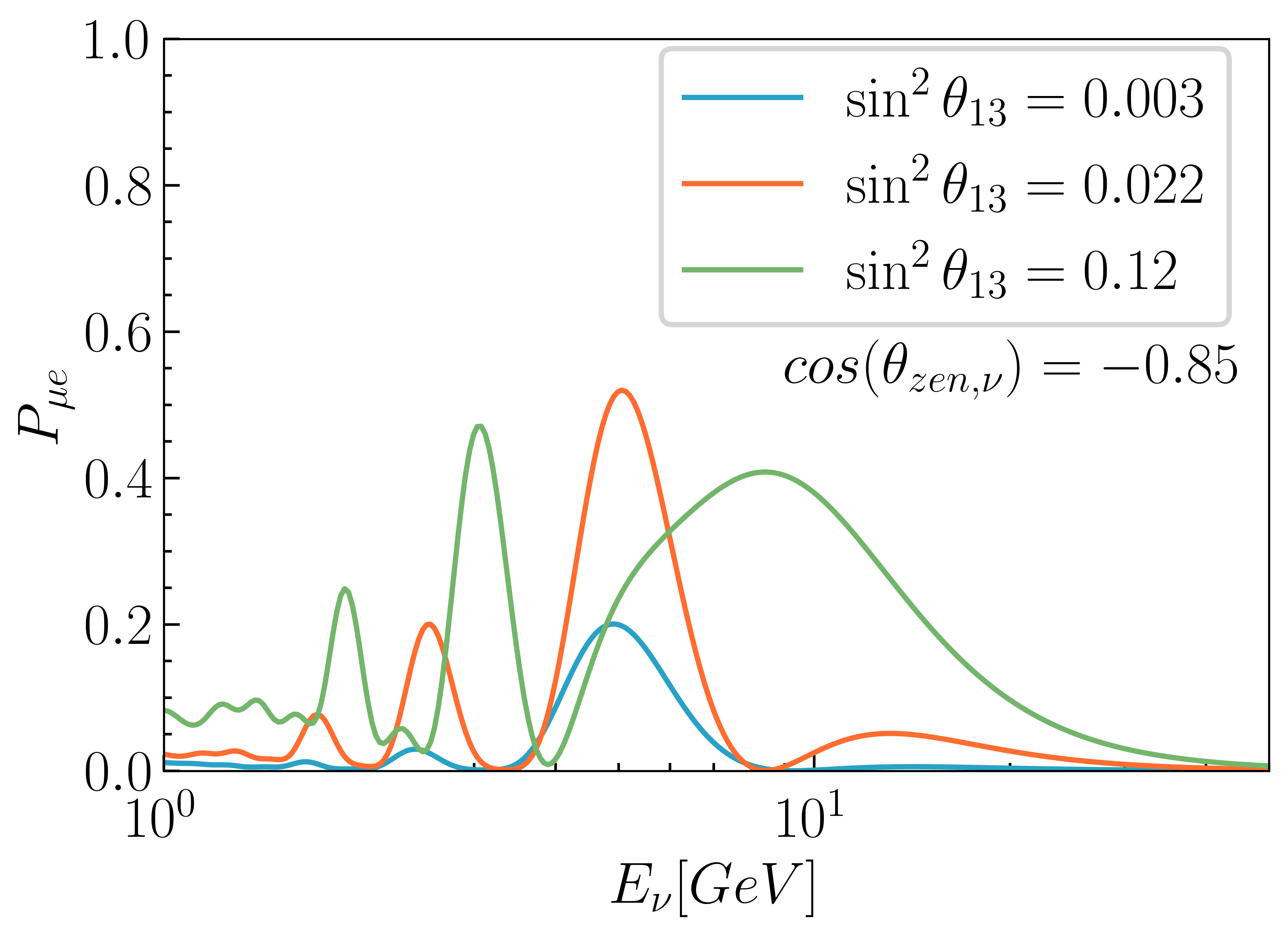}
\caption{\textit{\textbf{Electron appearance probability for different values of $\sin^2\theta_{13}$ }}.The neutrino direction is fixed to $\cos(\theta_{zen})=-0.85$. The fast oscillations that happen for $E\sim \text{GeV}$ have been smeared assuming a Gaussian uncertainty of $5\%~$E/GeV.}
\label{fig:ProbT13}
\end{figure}

As the neutrino energy rises, the matter effects become more important. 
At the GeV energy scale and for trajectories crossing the mantle, there is an enhancement of the effective mixing angle $\tilde{\theta}_{13}$ due to the coherent-forward elastic scattering of the neutrinos with the electrons in the Earth, the so-called MSW effect~\cite{Wolfenstein:1977ue,Mikheyev:1985zog}.
The matter-modified mixing angle is given by
\begin{equation}
    \sin 2\tilde{\theta}_{13}=
    \frac{\sin 2\theta_{13}}{\sqrt{(\cos\theta_{13} - 2 E V/\Delta m^2_{31})^2 + \sin^2 2\theta_{13}}}.
\end{equation}
For energies around $\SI{6}\GeV$ and densities around $5 \text{g/cm}^3$, $\sin 2\tilde\theta_{13}$ becomes maximal, giving rise to an enhancement of the flavor conversion: see in~\Cref{fig:ProbT13}.
The location of the resonance is controlled by $\sin^2\theta_{13}$ for a given value of $\Delta m^2_{31}$, providing sensitivity to this angle.
As $\sin^2\theta_{13}$ becomes larger, the resonance moves to lower energies and densities.
The opposite happens if $\sin^2\theta_{13}$ becomes smaller, we will need larger energies and densities to meet the resonant condition.
For very small values of $\sin^2\theta_{13}$, the oscillation length at the resonance becomes larger than the size of the Earth~\cite{Akhmedov:2006hb}, making impossible to get a large flavor conversion with atmospheric neutrinos, as is the case with the blue line in~\Cref{fig:ProbT13}. 

Beyond the MSW effect, at energies around $\SI{1}\GeV$, the oscillation is enhanced for some trajectories crossing Earth's core and mantle due to the matter distribution, the so-called parametric resonance~\cite{Akhmedov:1988kd,Petcov:1998su, Akhmedov:1998ui, Akhmedov:1998xq, Chizhov:1998ug,Petcov:1998su, Liu:1998nb, Chizhov:1999az, Chizhov:1999he, Ohlsson:1999um, Palomares-Ruiz:2004cmm, Akhmedov:2005yj, Gandhi:2004bj, Kelly:2021jfs}.
Both types of resonance happen for neutrinos if the mass ordering is normal and for anti-neutrinos in the case of inverted ordering, as shown in~\Cref{fig:Pmumu,fig:Pmue}.
The differences in the flux and the cross section for both fermions bring the possibility to measure the neutrino mass ordering using atmospheric neutrinos. 

\begin{figure*}
 \includegraphics[width=\textwidth]{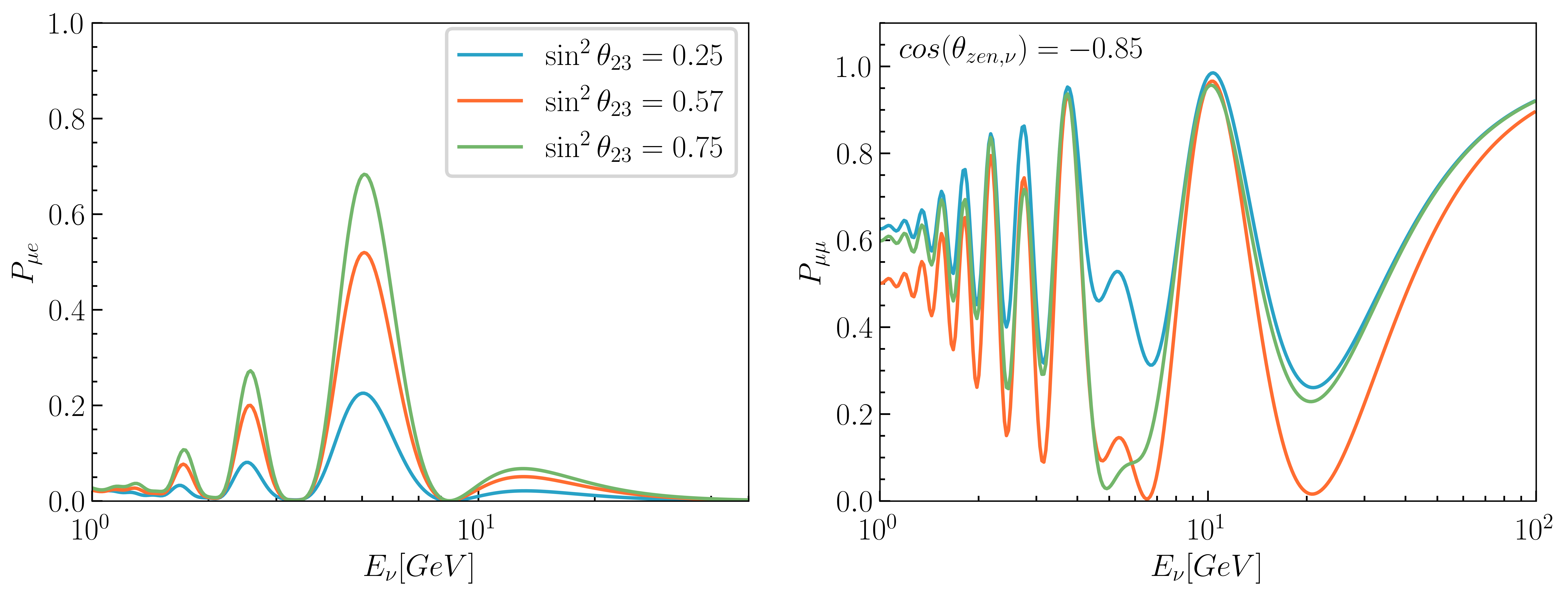}
\centering
\caption{\textit{\textbf{Electron appearance (left) and muon disappearace (right) probabilities for $\cos(\theta_{zen}) = -0.85$.}} We show impact of $\sin^2\theta_{13}$ in both oscillation channels. The fast oscillations have been smeared assuming a Gaussian uncertainty of $5\%~$E/GeV.}
\label{fig:ProbT23}
\end{figure*}

In the multi-GeV scale, the neutrino oscillation length gets longer, and the neutrino oscillation is dominated by $\Delta m^2_{31}$ and $\theta_{23}$.
The first oscillation minimum for $P(\nu_{\mu}\rightarrow\nu_{\mu})$ happens at $E\sim \SI{20}\GeV$ for baselines that cross the Earth: see in~\Cref{fig:Pmumu}.
The energies where that oscillation minimum happens depend on $|\Delta m^2_{31}|$, and the oscillation amplitude is controlled by $\sin^2 2\theta_{23}$, as shown in~\Cref{fig:ProbT23}~\cite{g_footnote}.
It is important to notice that the octant of $\theta_{23}$ can be measured with atmospheric neutrinos in two ways as shown in~\Cref{fig:ProbT23}: similarly to LBL experiments through the electron appearance channel as it is proportional to $\sin^2\theta_{23}$,~\Cref{eq:app}, and through muon disappearance as matter effects break its dependence with $\sin 2 \theta_{23}$ due to the enhancement of $\sin 2\tilde{\theta}_{13}$.

In this analysis, we numerically solve the neutrino evolution using \texttt{nuSQuIDS}~\cite{Arguelles:2021twb}. For the Earth matter density, we used the Preliminary Reference Earth Model~\cite{Dziewonski:1981xy} (PREM), which divides the Earth into eleven concentric layers, and for each of the layers, the density is given by a polynomial function that depends on its distance to the center of the Earth. As a benchmark scenario for the mixing parameters, we followed the Nu-Fit results~\cite{Esteban:2020cvm}. For the solar parameters, we used $\Delta m^2_{21} = 7.42\times 10^{-5}\text{eV}^2$ and $\sin^2\theta_{12} = 0.304$ that are fixed in the whole analysis. The reactor angle is fixed at $\sin^2\theta_{13} = 0.022$ unless otherwise specified. For the atmospheric parameters, we assumed $\Delta m^2_{31} = 2.5\times 10^{-3}\text{eV}^2$ and $\sin^2\theta_{23} = 0.572$, and for the $CP$-violation phase $\delta_{CP} = 234^o$. This is summarized in~\Cref{table:OscPar}.

\begin{table}
\begin{tabular}{l|c|c}
Parameter & True value & Constraints \\
\hline\hline
$\sin^2\theta_{12}$ &  $0.304$    & fixed \\
$\sin^2\theta_{13}$ &  $0.022$     & free  \\
$\sin^2\theta_{23}$ &  $0.572$     & free   \\
$\delta_{CP}$       &  $4.082$     & free \\
$\Delta m^2_{21}$ [eV$^2$]  &  $7.42\times 10^{-5}$     & fixed \\
$\Delta m^2_{31}$ [eV$^2$]  &  $2.50\times 10^{-3}$    & free \\
Ordering            & Normal  & free
\end{tabular}
\caption{\textbf{\textit{Summary of the true values assumed for 3-flavor neutrino oscillation parameters and their treatment in the oscillation analysis.}}}
\label{table:OscPar}
\end{table}

%%%%%%%%%%%%%%%%%%%%%%%%%%%%%%%%%%%%%%%%%%%%%%%
\section{Experimental Analyses}\label{sec:exp}
To analyze the sensitivity to the neutrino oscillation parameters, we produce an Asimov dataset for each experiment, and perform a combined fit to the Monte Carlo simulation assuming different values of the oscillation parameters sampled from two 4-dimensional ($\Delta m^2_{31}$, $\theta_{23}$, $\theta_{13}$, and $\delta_{CP}$) grid of points, one for each neutrino ordering.
To reduce the impact of stochastic uncertainties from the simulations, we produce large-exposure Monte Carlo datasets for each experiment.

We bin the events in observable quantities, which differ from experiment to experiment and are discussed in the following subsection.
We then construct the following test statistic, $\chi^2$, to compare data with prediction. 
This is given by \Cref{eq:chi2},
\begin{widetext}
\begin{equation}
    \chi^2 = 2 \nsum_{Exp.} \nsum_{i\in Bins} \Bigg( \mu_i\Bigg(1+\sum_{j\in Syst.}
    \eta_j f_{ij}\Bigg)-O_i+O_i\cdot \log\Bigg( \frac{O_i}{\mu_i\big(1+\sum_{j\in Syst.}
    \eta_j f_{ij}\big)}\Bigg) \Bigg) + \nsum_{j\in Syst.} \Bigg(
    \frac{\tilde\eta_j-\eta_j}{\sigma_j} \Bigg)^2,
    \label{eq:chi2}
\end{equation}
\end{widetext}
where $\mu_i$ and $O_i$ are the expected number of events in bin $i^{th}$ respectively, and $f_{ij}$ is the fractional change in the number of events in bin $i^{th}$ due to the $j^{th}$ systematic source of uncertainty which takes value $\eta_j$ and has $\tilde\eta_j$ nominal value with $\sigma_j$ error size.

The first term in~\Cref{eq:chi2} is the negative times two of the log-likelihood ratio between the Asimov dataset and the MC at a given point in the oscillation parameter space assuming Poisson statistics, and the last term takes into account the penalty from each source of systematic uncertainty, for which gaussianity is assumed.
The Asimov dataset is fitted against the MC using a binned $\chi^2$ method assuming the number of events per bin follows a Poisson statistic.
For introducing the effect of systematic uncertainties, the number of entries in each bin is re-weighted accordingly and an additional penalty term is introduced in~\Cref{eq:chi2}, see Ref.~\cite{Fogli_2002} for details.

Minimizing~\Cref{eq:chi2} requires solving the system of equations defined by $\nabla\chi^2 = \vec{0}$ over all systematic uncertainties.
This procedure brings the Asimov dataset and MC into the best agreement allowed by the size of systematic uncertainties.
Usually, this process requires extensive CPU resources as it relies on the numerical computation of partial derivatives with respect to all systematic sources, which has to be repeated for all points in the defined grid of oscillation parameters to be tested while managing large simulation files.
This problem can be alleviated by analytically computing the Jacobian of~\Cref{eq:chi2}, at the cost of implementing in the code the partial derivatives of the expected number of events in each bin for each systematic source, as shown in~\Cref{eq:chi2_jac}.
Such implementation improves the convergence time to a minimum by almost two orders of magnitude in this analysis, and provides more robust minimization.\\
For the minimization of $\chi^2$ we use the \texttt{SciPy} \texttt{Python} package~\cite{2020SciPy-NMeth}.

\begin{widetext}
\begin{equation}
    \nabla_j \chi^2 = 2~\nsum_{Exp.} \nsum_{i\in Bins} \Bigg(\Bigg( 1 - \frac{O_i}{\mu'_i}\Bigg)
    \frac{\partial \mu'_i}{\partial \eta_j}\Bigg)
    +2~\Bigg(\frac{\tilde\eta_j-\eta_j}{\sigma_j^2}\Bigg)
    \text{~, ~where~  }
    \frac{\partial \mu'_i}{\partial \eta_j} = \mu_i\cdot f_{ij}.
    \label{eq:chi2_jac}
\end{equation}
\end{widetext}

The events are divided into the samples according to the definitions of each experiment and binned by their reconstructed zenith angle and energy as explained in~\Cref{sec:SK},~\Cref{sec:SKntag},~\Cref{sec:IC}, and~\Cref{sec:ORCA}.

The $f_{ij}$ are introduced by computing them on run-time before the fit on an event-by-event basis or read-out from publicly available tables.
Unlike prior global analyses of atmospheric neutrinos, flux and cross section systematics are common to all experiments considered, while detector systematics are only applied to each experiment.

As mentioned earlier, in this work, we consider four experiments: SuperK, IceCube Upgrade, ORCA, and HyperK.
Additionally, we consider the three distinct phases of operation within SuperK: SuperK without H-neutron tagging (SuperK), SuperK with H-neutron tagging (SK-Htag), and SuperK with gadolinium (SKGd).
The phases of SuperK are treated in the analysis as three independent data-taking experiments, with uncorrelated detector systematics.
The first phase of SuperK covers the first three runs of the experiment, from SuperK-I to SuperK-III; SK-Htag covers the detector from SuperK-IV to SuperK-V with neutron tagging capabilities on hydrogen; finally, SKGd covers the projected running time of SuperK with gadolinium enabling improved neutron tagging.
In terms of exposure, we assume the full data-taking periods through SuperK-I to SuperK-V as reported in~\cite{linyan_wan_2022_6694761}.
For SKGd, we project five years of operation with the final concentration of Gd dissolved in water, 0.2\%, starting in 2025.

For the soon-to-be-deployed IceCube Upgrade and ORCA, we conservatively foresee five and three years of operation starting in 2025 and 2027, respectively.
For HyperK, we assume 2.5 years of operation starting in mid 2027 as foreseen by the collaboration.

Thus, the combined analysis in the report assumes the running of SuperK until the last reported exposure time, the SKGd and IceCube Upgrade data-taking period extending from 2025 until 2030, and ORCA from 2027 to 2030.
Additionally, HyperK is projected to be completed by mid-2027, ensuring a large amount of statistics by 2030, and thus its included in the fit.
The situation is less certain for DUNE, as its final volume has not been defined, for this reason it is not included in the fit.
Nonetheless, its expected performance is described in~\cref{sec:DUNE}. 
According to the current status of construction, this fit conservatively considers the neutrino oscillation measurement picture these experiments will provide by the end of the decade.

The combined fit is then performed over a total of 3595 bins (SuperK-I to SuperK-III: 459, SuperK-IV to SuperK-V: 539, SKGd: 539, IceCube Upgrade: 800, ORCA: 180, and HyperK: 1078) from events classified in 75 samples (SuperK-I to SuperK-III: 16, SuperK-IV to SuperK-V: 18, SKGd: 18, IceCube Upgrade: 2, ORCA: 3, and HyperK: 18). 
A total of 103 systematic uncertainties are considered, 20 of which come from flux and cross section and are common to all experiments, and the rest are related to each detector (SuperK-I to SuperK-III: 16, SuperK-IV to SuperK-V: 17, SKGd: 17, IceCube Upgrade: 6, ORCA: 10, and HyperK: 17).

\subsection{Super-Kamiokande}\label{sec:SK}
SuperK is a 50-kton cylindrical water-Cherenkov detector located in Kamioka, Japan.
The detector is instrumented with more than 11,000 20-inch PMTs in the inner surface and facing inwards.
The mountain above shields the experiment from most of the cosmic-ray muons and the large photo-coverage enables the measurement of low-energy atmospheric neutrinos up to $\SI{100}\MeV$~\cite{FUKUDA2003418}.

SuperK started its operation in 1996 and has largely contributed to the current knowledge of neutrinos; particularly, it contributed to the discovery that neutrinos have a definite mass by measuring oscillations in atmospheric neutrinos~\cite{Super-Kamiokande:1998kpq}.
Over the years, the experiment went through various phases, from SuperK-I to the current SuperK-VI~\cite{linyan_wan_2022_6694761}.
From 2020, the experiment is undergoing a major upgrade going from an ultra-pure to a Gd-doped water Cherenkov detector~\cite{Abe_2022}, which will improve its capabilities by enabling highly efficient neutron tagging on gadolinium.

After more than 25 years of nearly continuous operation, SuperK data provides a very comprehensive picture of neutrino oscillations measuring solar~\cite{PhysRevD.94.052010, Abe_2011, Renshaw_2014}, accelerator --- serving as K2K's~\cite{Ahn_2006, Suzuki_2000, NITTA_2004} and T2K's~\cite{Abe_2011zz, https://doi.org/10.48550/arxiv.hep-ex/0106019, T2K:2019bcf, Abe_2021, Abe_2021b} far detector---, and atmospheric fluxes~\cite{Wendell_2010, PhysRevD.91.052003, Abe_2015, Super-Kamiokande:2015qek, Super-Kamiokande:2019gzr, Super-Kamiokande:2017edb}.
From the latter and according to to~\cite{Abe_2018}, SuperK has been able to constrain the values of the atmospheric parameters, $\Delta m^2_{31} = 2.50^{+0.13}_{-0.20}\times 10^{-3}$~eV$^2$ and $\sin^2\theta_{23} = 0.588^{+0.031}_{-0.064}$, and of the $CP$-phase, $\delta_{CP}=4.18^{+1.41}_{-1.61}$, assuming $\theta_{13}$ constrained by reactor experiments.
Results also show a preference for normal ordering at $\sim2~\sigma$ level.

Recently, due to the knowledge of the detector, the SuperK collaboration was able to extend the fiducial volume by 20\%, from the usual 22.5~kton to 27~kton in all running periods~\cite{linyan_wan_2022_6694761}, enlarging the sample size and thus its sensitivity.
We assume this improved fiducial volume in our analysis. 

%\subsubsection{Monte Carlo Simulation}
For this work, we develop a detailed Monte Carlo simulation to predict the atmospheric neutrino event rate at SuperK.
First neutrino events interacting with a water target are generated using the GENIE event generator~\cite{GENIE}.
Then, we implement software emulating the SuperK reconstruction procedures, efficiencies, and resolutions based on publicly available information.
In the end, we produce a simulation equivalent to 300 years for each phase of the experiment following the event categories defined in each case.\\

%\subsubsection{Event Classification}
The official atmospheric neutrino analysis from SuperK separates events into three categories depending on their morphology: fully contained (FC), partially contained (PC), and upward-going muons (Up-$\mu$).
In this work, we only simulate the first two as the latter has very mild impact on the sensitivity and is statistically dominated by IceCube Upgrade and ORCA.

The reconstruction is focused on the FC events as they carry the most information and are the most abundant in the energy region sensitive to the oscillation parameters, from $\SI{0.1}\GeV$ to $\SI{20}\GeV$. 
Fully contained events get categorized into Single-Ring or Multi-Ring depending on the number of rings, Sub-GeV or Multi-GeV attending to their reconstructed visible energy, and $e$-like or $\mu$-like in terms of the reconstructed ID of the most energetic ring.
In addition, depending on other reconstructed variables, events get further divided to distinguish neutrinos from antineutrinos, and charged-current from neutral-current interactions.
Single-Ring Sub-GeV samples are divided into 0-decay-electrons and $>0$-decay-electrons for $e$-like, and 0-decay-electrons, $1$-decay-electrons and $>1$-decay-electrons for $\mu$-like, where the decay-electrons are the number of delayed electrons reconstructed from the decay of charged pions and muons in the event.

\begin{figure*}
\centering
\includegraphics[width=0.49\textwidth,height=3in]{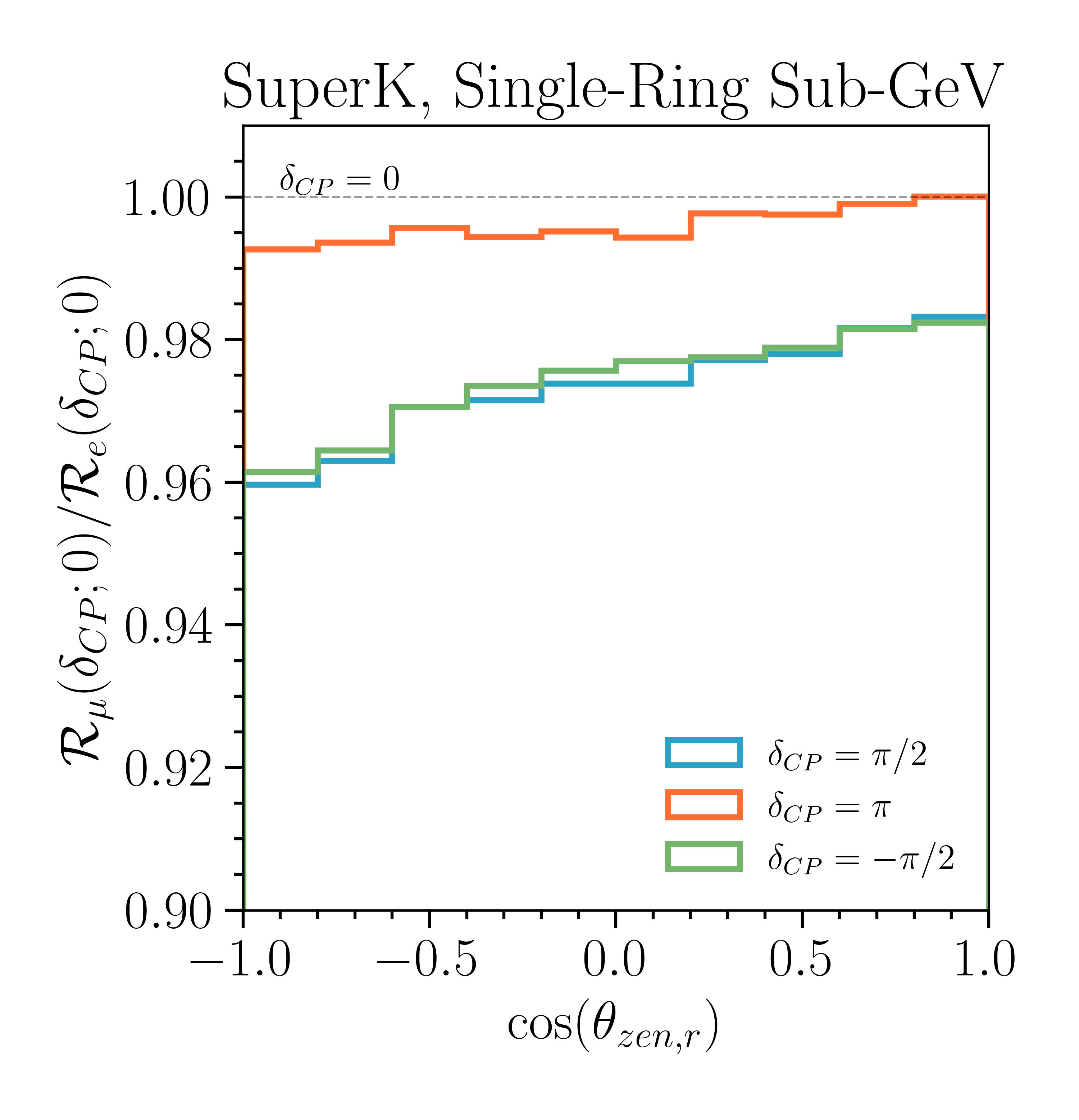}\hfill
\includegraphics[width=0.49\textwidth,height=3in]{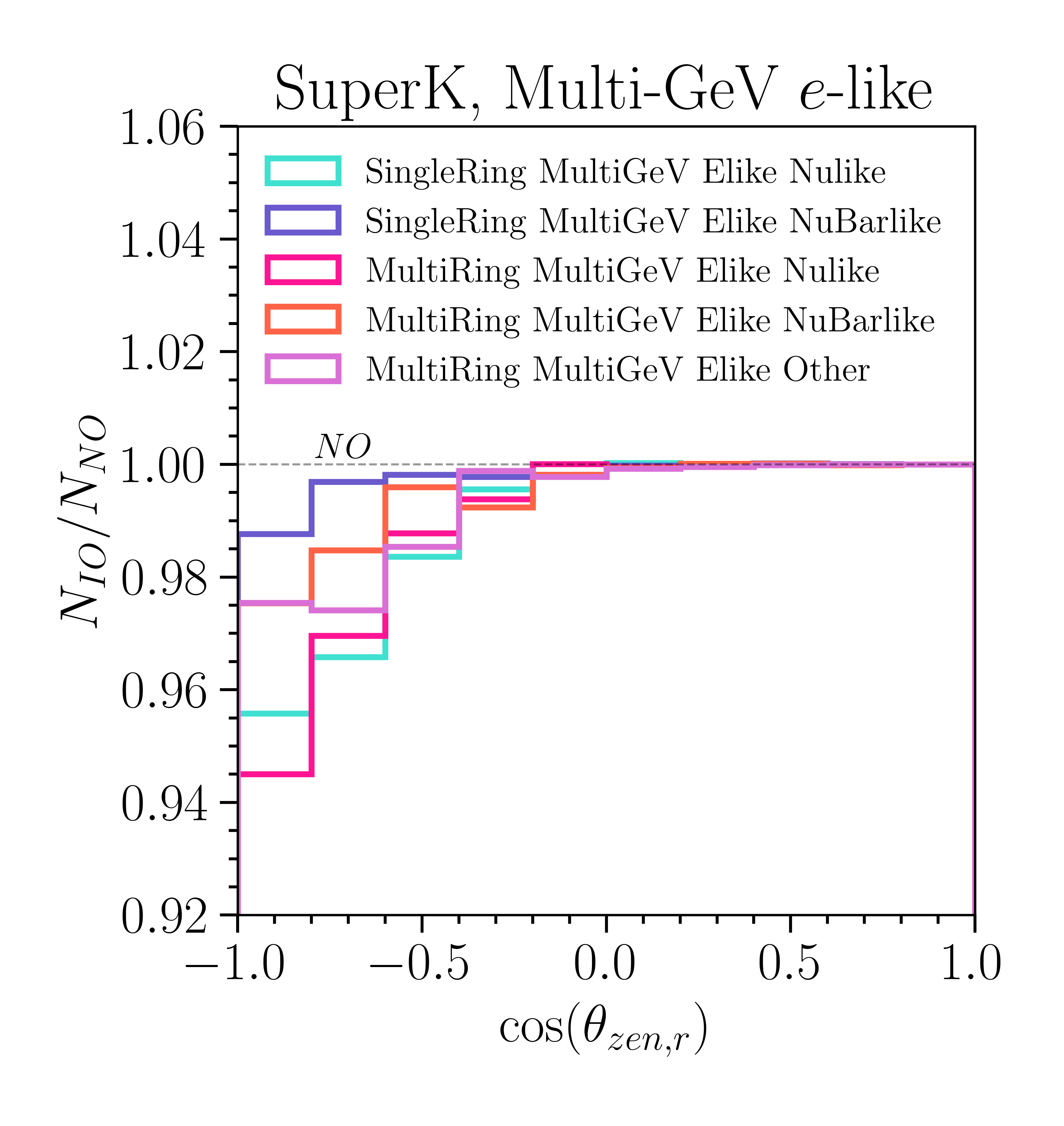}
\caption{\textbf{\textit{Impact of different values of $\delta_{CP}$ (left) and both neutrino mass orderings (right) for the most sensitive SuperK samples.}}
On the left, $\mathcal{R}_{\mu/e}(\delta_{CP};0)$ are the ratio of the number of events in Sub-GeV $\mu$-like and $e$-like samples, with 1 and 0 decay-electrons respectively, for a given value of $\delta_{CP}$ compared with the case of $\delta_{CP}=0$. 
On the right, we show the ratio between inverted ($N_{IO}$) and normal ($N_{NO}$) orderings for the number of events in SuperK Single-Ring and Multi-Ring Multi-GeV $e$-like samples.
The rest of parameters follow~\Cref{table:OscPar}.}
\label{fig:sk_dcpmo_ratio}
\end{figure*}

An additional Single-Ring Sub-GeV $\pi^0$-like sample is defined for $e$-like events passing a $\pi^0$ cut, accounting for a fraction of neutral current events producing a $\pi^0$ which decays to a pair of photons, but only one of them is reconstructed.
For these events, another sample is defined in the case that both photons are reconstructed, Sub-GeV 2-Ring $\pi^0$-like.
Single-Ring Multi-GeV $e$-like events follow the same criteria as their Sub-GeV counterparts, and no further classification is done for $\mu$-like events.
Finally, for Multi-Ring Multi-GeV $e$-like events, a two-step classification is done.
The first one tags most of the neutral-current events into the Multi-Ring Other sample.
The remaining events are divided into neutrino and antineutrino samples.

As expected from~\Cref{sec:OSC}, Sub-GeV samples are those most sensitive to the $CP$-violating phase.
The effect from different values of $\delta_{CP}$ in the most relevant samples is shown in~\Cref{fig:sk_dcpmo_ratio}.
Additionally, in~\Cref{fig:sk_dcpmo_ratio} we also show the power of Multi-GeV $e$-like samples to discern between normal and inverted orderings.
Despite their name, $\overline{\nu}$-like samples have still more neutrinos than antineutrinos; this contamination of neutrinos comes due to the modest neutrino-antineutrino separation power and the smaller $\overline{\nu}$ cross section. Therefore, the effects of different values of $\delta_{CP}$ or mass ordering scenarios are diluted. More details about the SuperK simulation can be found in~\Cref{sec:SKSim}.\\

%\subsubsection{Detector systematic uncertainties}
Detector systematics have a secondary effect on the sensitivity to the oscillation parameters, except $\delta_{CP}$, as compared with those of the flux or the neutrino-nucleon cross section.
To realistically asses the sensitivity of the experiment, we implement a subset of the most relevant SuperK detector systematic uncertainties as detailed as possible within the reach of our simulation.
Our implemented systematics follow~\cite{Abe_2018} and are summarized in~\Cref{table:SKSysts} for each of the three major detector phases aforementioned.

\begin{table*}
\begin{tabular}{l|c|c}
Systematic source & \makecell{1$\sigma$-range\\(SuperK-I to III)} & \makecell{1$\sigma$-range\\(SuperK-IV -- SKGd)} \\
\hline\hline
Energy scale	& 3\%	&	2\%	\\
FC-PC separation & 6\%  &	0.2\%	\\
FC Reduction 	& 0.3\%  & 2\%         \\
Fiducial Volume & 2\%  & 1.3\%           \\
PC Reduction 	& 3.5\%  & 1\%         \\
SubGeV2ringPi0 	& 6\%  & 6\%          \\
SubGeV1ringPi0 	& 25\%  & 15\%          \\
Multi-ring $\nu-\overline{\nu}$ & 6\%  & 3\%           \\
Multi-ring other 	& 6\%  & 4\%          \\
PC-Stop PC-Thru 	& 25\%  & 20\%          \\
$\pi^o$ ring separation & 2\%  & 2\%          \\
e-like ring separation 	& 6\%  & 2\%          \\
$\mu$ ring separation 	& 3\%  & 2\%          \\
Single-ring PID 	& 0.35\%  & 0.35\%          \\
Multi-ring PID 		& 4\%  & 4\%          \\
Decay-e tag. eff. 	& 10\%  & 10\%          \\
\end{tabular}
\caption{Summary of SuperK detector systematic uncertainties used in this work.}
\label{table:SKSysts}
\end{table*}

\subsection{Super-Kamiokande with neutron tagging}\label{sec:SKntag}
After the electronics upgrade in 2008, SuperK was able to lower the signal threshold enough to be sensitive to the $\SI{2.2}\MeV$ gamma from the de-excitation of deuterium after neutron capture on hydrogen.
Despite being very weak, this signal is reconstructed with an efficiency of $\sim$20\%.

Further, in 2018, the upgrade of the SuperK detector started to make it compatible with dissolving Gd.
At 0.2\% concentration by mass of Gd sulfate in water, 90\% of the thermal neutrons are captured by Gd with a half-life of $\sim$30~$\mu$s, emitting an $\SI{8}\MeV$ $\gamma$ cascade.
This is detected by SuperK with an efficiency of 90\%, having a final neutron tagging efficiency of $\approx$80\%.
Currently, SKGd is running at a third of the goal gadolinium concentration, which is expected to be achieved in the next years~\cite{skgd_neutrino22}.

Neutrinos produce on average fewer neutrons than antineutrinos; then in addition to the usual cut in the number of electrons from muon decays for single-ring samples, a cut is applied in the number of tagged neutrons (0 neutrons or $\ge$1 neutrons) to improve the neutrino-antineutrino separation.
This establishes new sample definitions for the analysis of SuperK-IV data.

\begin{figure*}
\centering
\includegraphics[width=0.49\textwidth,height=3in]{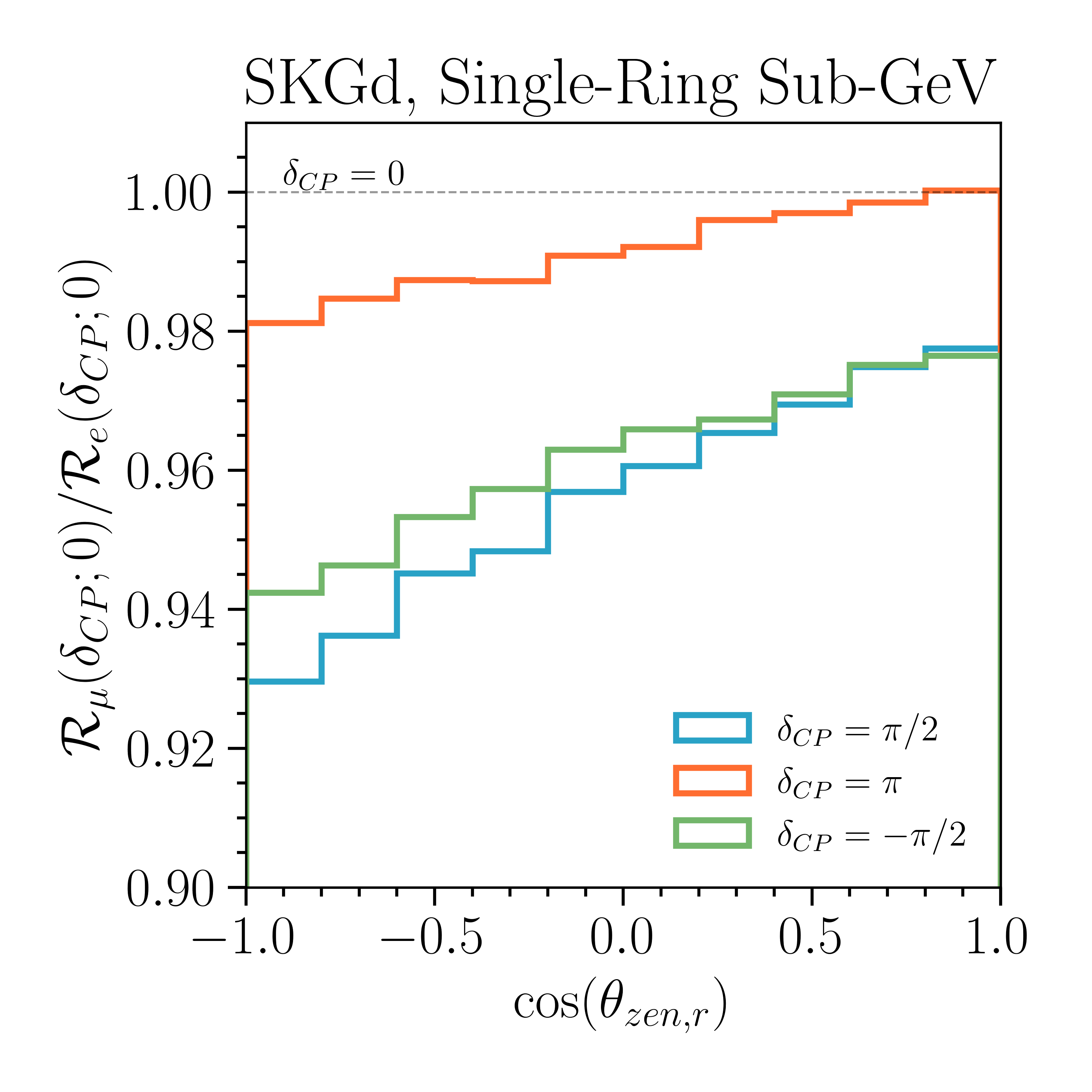}\hfill
\includegraphics[width=0.49\textwidth,height=3in]{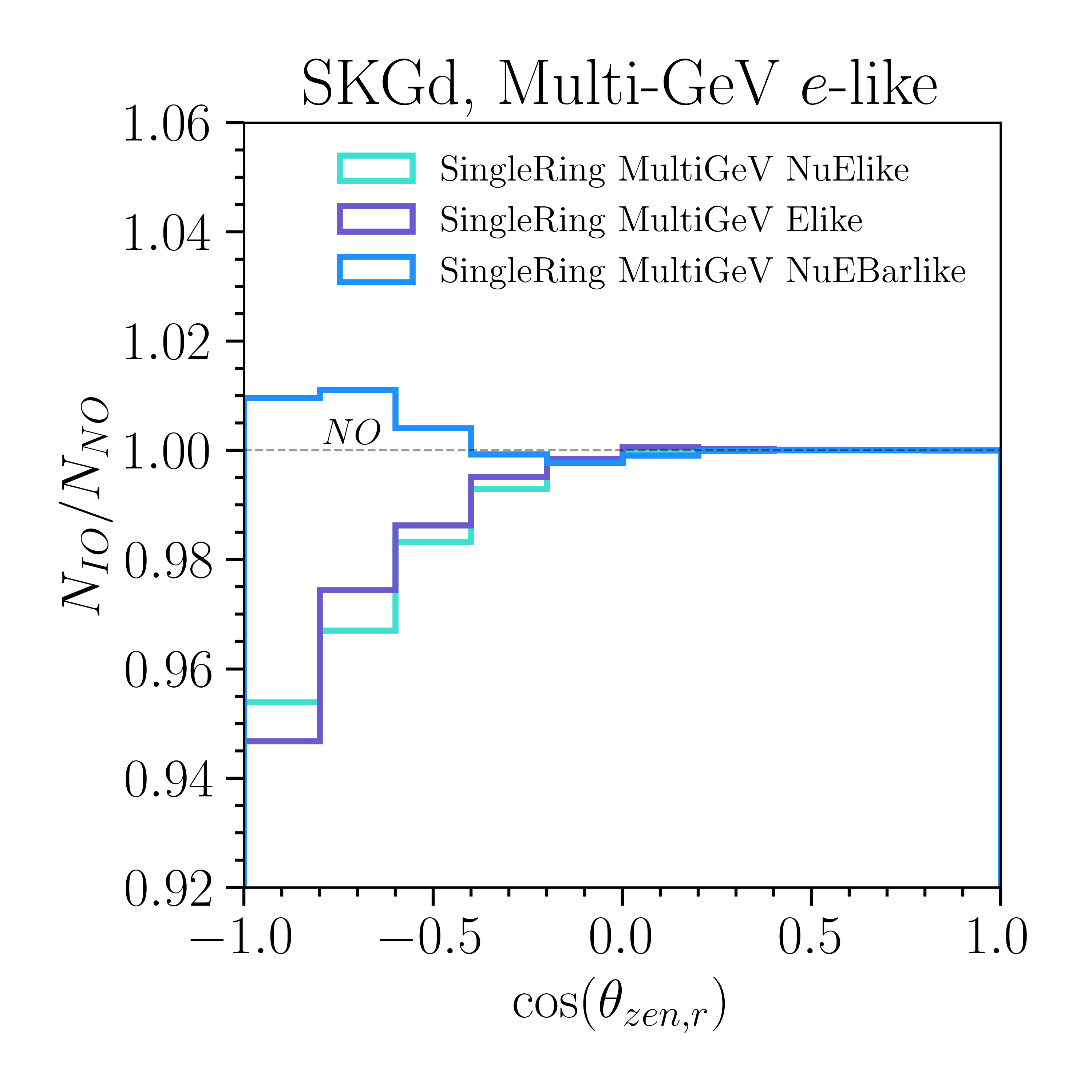}
\caption{\textbf{\textit{Impact of different values of $\delta_{CP}$ (left) and both neutrino mass orderings (right) for the most relevant SKGd samples.}}
On the left, 
 $\mathcal{R}_{\mu/e}(\delta_{CP};0)$ are the ratio of the number of events in Sub-GeV $\mu$-like and $e$-like samples, with 0 Gd-tagged neutrons, and 1 and 0 decay-electrons respectively, for a given value of $\delta_{CP}$ compared with the case of $CP$-conservation. 
 On the right, we show the ratio between inverted ($N_{IO}$) and normal ($N_{NO}$) orderings for the number of events in SKGd Single-Ring Multi-GeV $e$-like samples.
 The rest of parameters follow \Cref{table:OscPar}.}
\label{fig:skgd_dcpmo_ratio}
\end{figure*}

The SuperK atmospheric neutrino analysis includes this neutron information by defining new event samples to improve the separation between neutrinos and antineutrinos.
The previously explained Single-Ring $e$-like events with $0$ decay-electrons are divided into samples with $0$ tagged neutrons or one or more tagged neutrons, and similarly, for $\mu$-like events as described in~\cite{FernandezMenendez:2021jfk}.

This improved event classification enhances the sensitivity to those oscillation parameters behaving differently for neutrinos and antineutrinos, namely the $CP$-phase and the neutrino mass ordering. These effects are shown in~\Cref{fig:skgd_dcpmo_ratio}, for the case of Gd-neutron tagging.

\subsection{Hyper-Kamiokande}\label{sec:HK}

HyperK is a next-generation water-Cherenkov neutrino and proton-decay experiment in Japan, expected to take over the current SuperK detector in 2027.
It will be a scaled-up version of SuperK consisting of a cylinder of $\SI{74}\m$ in diameter and $\SI{60}\m$ in height, with a total mass of $\SI{258}\kton$~\cite{hk_tdr}.
The detector design is very similar to that of SuperK, with inner and outer detectors, the fiducial volume is $\SI{187}\kton$, a factor 8.4 larger than that of SuperK.
The inner detector will be instrumented with approximately 20,000 20-inch PMTs and additional $\sim$1,000 multi-PMT modules~\cite{hksnowmass_2022, jeanne_wilson_2022_6693820}.
The photo coverage will be 20\%, which is half of SuperK, but the performance is expected better due to improved photosensors.

Given the similarities between SuperK and HyperK, we used our SK-Htag simulation as the baseline Monte Carlo for HyperK.
This assumption provides a conservative estimate of the detector sensitivity as preliminary studies have already shown that neutron tagging on hydrogen will be more efficient in HyperK than in SuperK, see Ref.~\cite{HK_ntag_shota, HK_ntag_sergio}.

In the analysis, the HyperK simulated data-set is divided into the same samples as the SK-Htag, using two times the binning in zenith angle to account for the larger statistics.
Following the same logic, we conservatively assume the same detector systematic errors as in for SuperK-IV,~\Cref{table:SKSysts}.

\subsection{IceCube and the IceCube Upgrade}\label{sec:IC}

IceCube is a one $\text{km}^3$ ice-Cherenkov detector located at the South Pole~\cite{IceCube:2006tjp}, consisting of 5160 optical sensors (DOM) deployed at depths between $\SI{1450}\m$ and $\SI{2450}\m$ in the Antarctic glacier. 
The detector can observe neutrino interactions with energies beyond $\sim \SI{10}\GeV$.
The low-energy part of the atmospheric spectra is measured by IceCube-DeepCore (DeepCore), a denser sub-array of strings in the inner part of the detector of $\sim10$~Mton.
For energies around $\sim \SI{10}\GeV$, two event morphologies can be differentiated in the detector: \textit{Tracks}, which are generated by the propagation of muons through the ice, and \textit{Cascades}, produced by the propagation of electrons, taus, hadronic cascades, or electromagnetic cascades. 

As described in~\Cref{sec:OSC}, IceCube has contributed to the measurement of $\Delta m^2_{31}$ and $\sin^2\theta_{23}$. 
Using the first three years of data~\cite{IceCube:2019dyb,IceCube:2019dqi}, the sensitivity obtained is $\Delta m^2_{31} = 2.55 ^{+0.12}_{-0.11}\times 10^{-3}\text{eV}^2$ and $\sin^2\theta_{23}= 0.58^{+0.04}_{-0.13}$.
These results are systematic-limited, further evaluation of the detector response together with a larger data sample predicts a large improvement in the precision over those parameters.

\begin{figure*}[!ht]
\centering
 \includegraphics[width=0.98\textwidth]{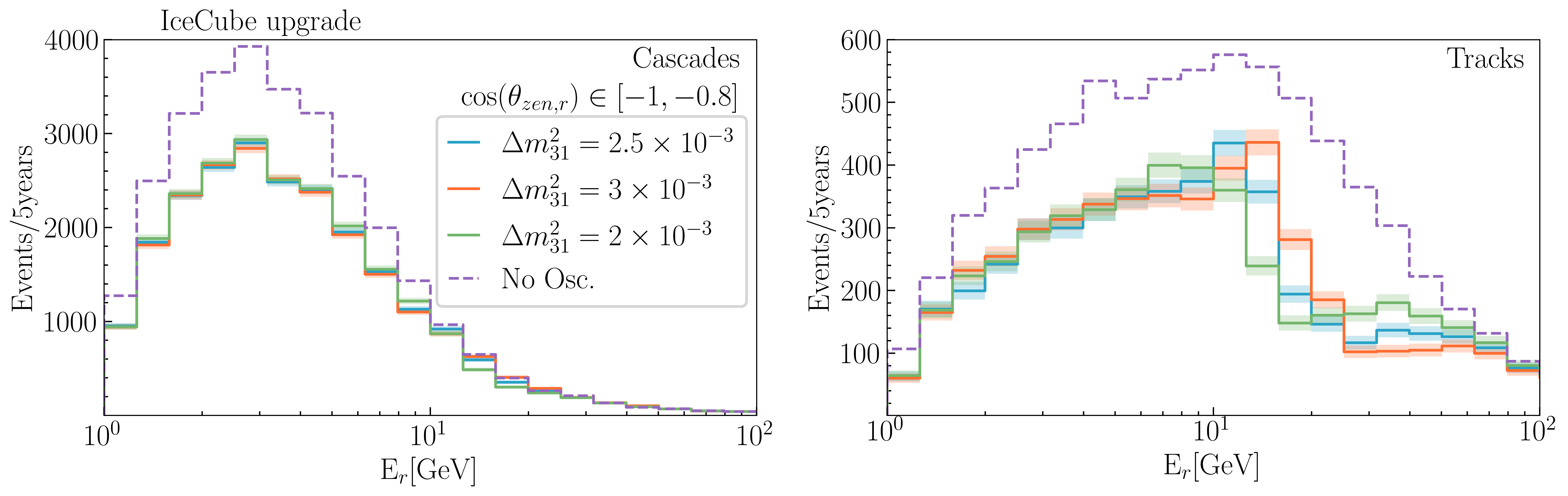}
 \includegraphics[width=0.98\textwidth]{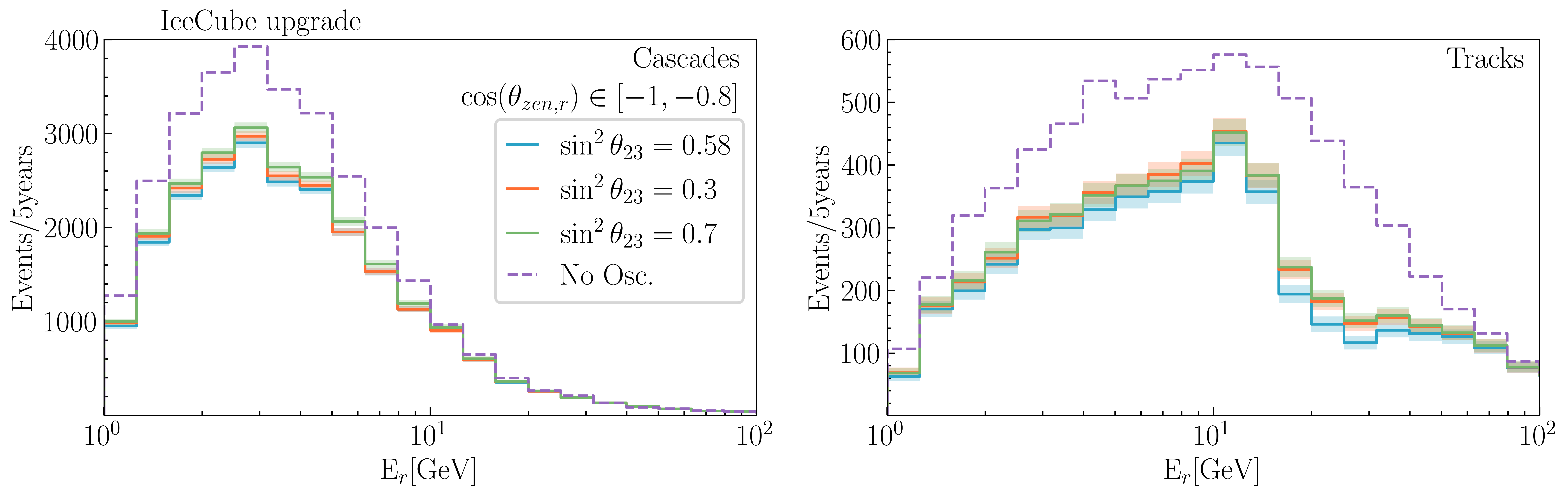}
 \includegraphics[width=0.98\textwidth]{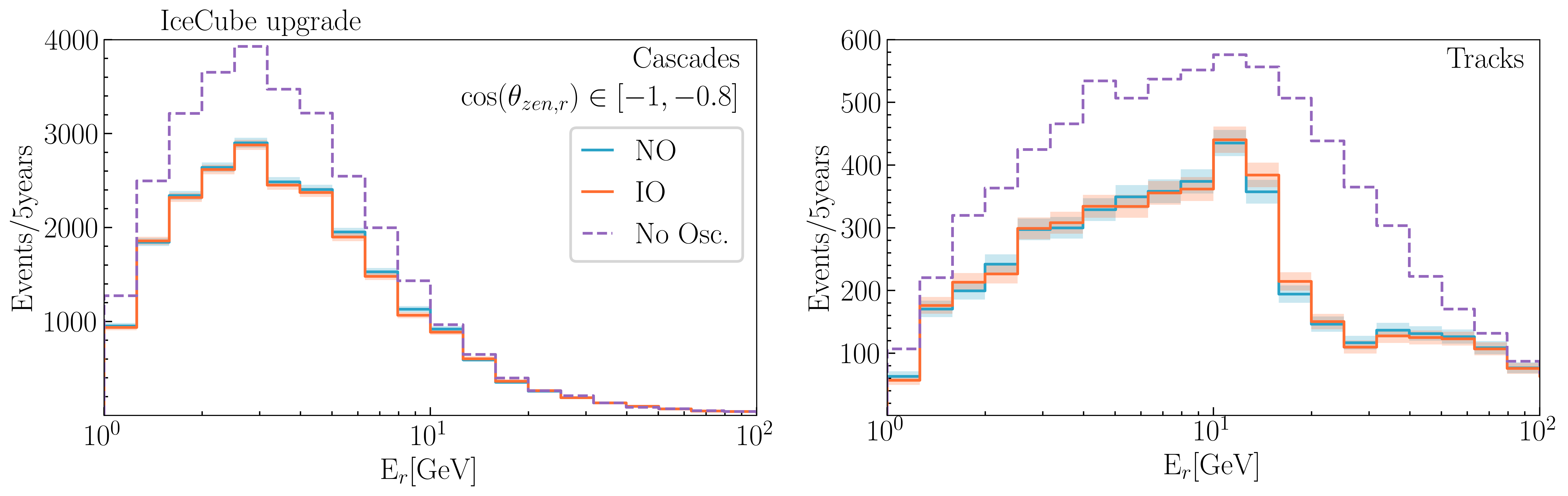}
 \includegraphics[width=0.98\textwidth]{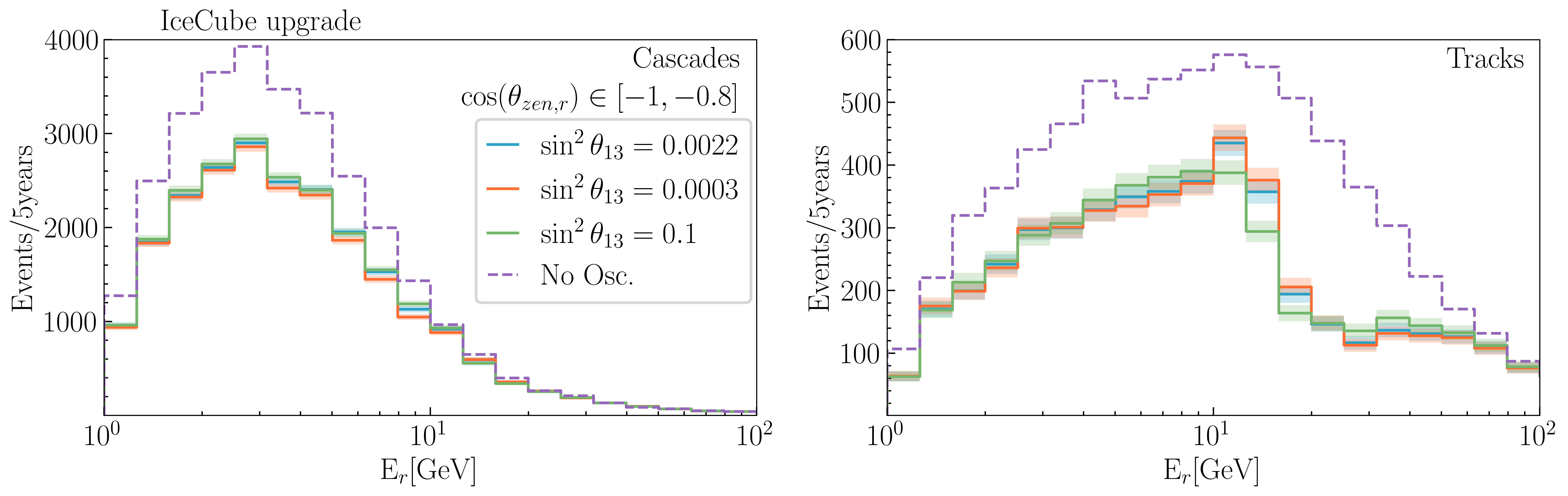}
\caption{Event distribution for different set of values of the mixing parameters for $\cos\theta \in[-1, -0.8]$ as a function of the reconstructed neutrino energy ($\text{E}_{r}$). We use the IceCube upgrade MC to evaluate the neutrino reconstructed energy~\cite{IceCube_Collaboration2020-md}. The cascade distribution is shown on the left and tracks on the right. The statistical $1\sigma$ error is shown as a band around the event distribution.}
\label{fig:ICparam}
\end{figure*}

IceCube is planning an upgrade~\cite{Ishihara:2019aao,Stuttard:2020zsj,IceCube_Collaboration2020-md} that will consist of the deployment of additional strings to increase the total volume, leading to a large sample size with an increased energy range that will extend to lower energies.
The upgrade will also increase the number of strings in the volume surrounding DeepCore, reducing the separation between the optical sensors and lowering the energy threshold to $\sim \SI{1}\GeV$.
The new range of energies will increase the total amount of events observed thanks to the soft energy spectra of the atmospheric neutrino flux, as we can see in~\Cref{fig:ICparam}.
The cascade sample dominates the low-energy sample due to the larger detector response at low energies. 

The upgraded detector will be able to accurately explore the atmospheric parameters ($\Delta m^2_{31}$, $\sin^2\theta_{23}$).
In~\Cref{fig:ICparam}, we show the impact that several values of the mixing parameters have on the event distribution for $\cos\theta\in [-1, -0.8]$.
In two upper figures, we see the impact that $\Delta m^2_{31}$ will have on the cascade (left) and track (right) distributions.
As we see from that figure, the main sensitivity comes from tracks with reconstructed energy around $\sim \SI{20}\GeV$, which coincides with the first minimum in the muon-disappearance channel. %$\sin^2\theta_{23}$ modifies the normalization of the event distribution. 
As we see in~\Cref{fig:ICparam}, tracks are sensitive to $\sin^2 2\theta_{23}$ and therefore can discriminate between maximal mixing and not, but have a very mild sensitivity to the octant.

The new range of energies accessible by the detector will bring the possibility of increasing the sensitivity over other parameters.
As mentioned in the \Cref{sec:OSC}, neutrinos are sensitive to mass ordering at the GeV due to Earth's matter effects.
In~\Cref{fig:ICparam}, we see the event distribution for both mass orderings.
The most significant difference is obtained for tracks with better angular resolution and energies below $\SI{10}\GeV$.
At the GeV scale, enhancing the effective $\tilde{\theta}_{13}$ leads to a significant effect in IceCube.
At the bottom of~\Cref{fig:ICparam}, we see how two extreme values for $\theta_{13}$ will modify the cascade and track distribution compared to the best-fit value measured in reactor experiments.
The largest sensitivity comes from tracks with energies below $\sim \SI{20}\GeV$.

In the case of the IceCube upgrade, there is still not a dedicated analysis of the most relevant detector systematics uncertainties and their impact.
For that reason, we have used the same systematics as in the search for $\nu_{\tau}$ carried out by DeepCore~\cite{IceCube:2019dqi} extended to the lower energies of the upgrade.
The detector uncertainties account for the different properties of the ice as optical absorption and scattering, the overall DOM efficiency, and the DOM response to lateral and head-on light. 
In \Cref{table:ICSysts}, we have a list of all the detector systematics used in this analysis and the $1\sigma$ range.

\begin{table}
\begin{tabular}{l|c}
Systematic source & \makecell{1$\sigma$-range\\(IC Up.)} \\
\hline\hline
Ice Absorption	& 10\% \\
Ice Scattering & 10\%  \\
Overall Optical Efficiency 	& 10\% \\
Lateral Optical Efficiency & 40\% \\
Head-on Optical Efficiency 	& free \\
Coin Fraction & 10\% \\
\end{tabular}
\caption{Summary of IceCube Upgrade detector systematic uncertainties used in this work.}
\label{table:ICSysts}
\end{table}

\subsection{KM3Net/ORCA}\label{sec:ORCA}

A new water-Cherenkov neutrino telescope is under construction in the Mediterranean sea, KM3NeT~\cite{KM3Net:2016zxf}.
This experiment will be composed of two detectors: ARCA, a cubic kilometer water Cherenkov detector that can observe high-energy neutrinos from the astrophysical sources in the northern sky, and ORCA, a megaton scale experiment that will be able to explore the neutrino properties by measuring the atmospheric neutrino flux.
In this article, we focus on the latter.

For this work, we have developed an independent Monte Carlo for ORCA based on the IceCube Upgrade simulation that takes advantage of the event simulation carried out by the IceCube collaboration.
In the simulation of ORCA, we only consider events with reconstructed energy in the range of $\SI{1.85}\GeV$ to $\SI{53}\GeV$ and keep the true neutrino variables from the IceCube simulation.
Then, the reconstructed energy, reconstructed zenith, and event morphology are computed and assigned following the distributions in~\cite{KM3NeT:2021ozk}.
The Monte Carlo event weights for ORCA are translated from those of IceCube following the ratio of effective volumes between both experiments, as shown in~\Cref{eq:Veff}. 

%Details of reconstruction and distributions can be found in Appendix.
%
\begin{equation}
w^\text{IC}_{\text{MC}} = A^{\text{IC}}_{\textit{eff}} =  V^{\text{IC}}_{\textit{eff}} \cdot \sigma \cdot \frac{1}{n_d} = w^{\text{ORCA}}_{\text{MC}} \frac{V^{\text{ORCA}}_{\textit{eff}}}{V^{\text{IC}}_{\textit{eff}}},
\label{eq:Veff}
\end{equation}

While the IceCube events are classified only into tracks and cascades, the ORCA collaboration reports a more sophisticated event classification including a third (intermediate) class.
To implement this, we reassign morphologies based on the results reported by the ORCA Collaboration in~\cite{KM3NeT:2021ozk}.
Further details of the development of the Monte Carlo simulation can be found in~\Cref{sec:ORCAMC}.

For the systematics treatment, we employ a similar set of systematics with the IceCube analysis. We use a same set of neutrino cross section-related systematics; for the detector uncertainties, we assume the same electronics behaviors including DOM efficiency rates, and we adapt the ice-related uncertainties to that of the water in the case of ORCA. 

\begin{figure*}[!ht]
\centering
 \includegraphics[width=\textwidth]{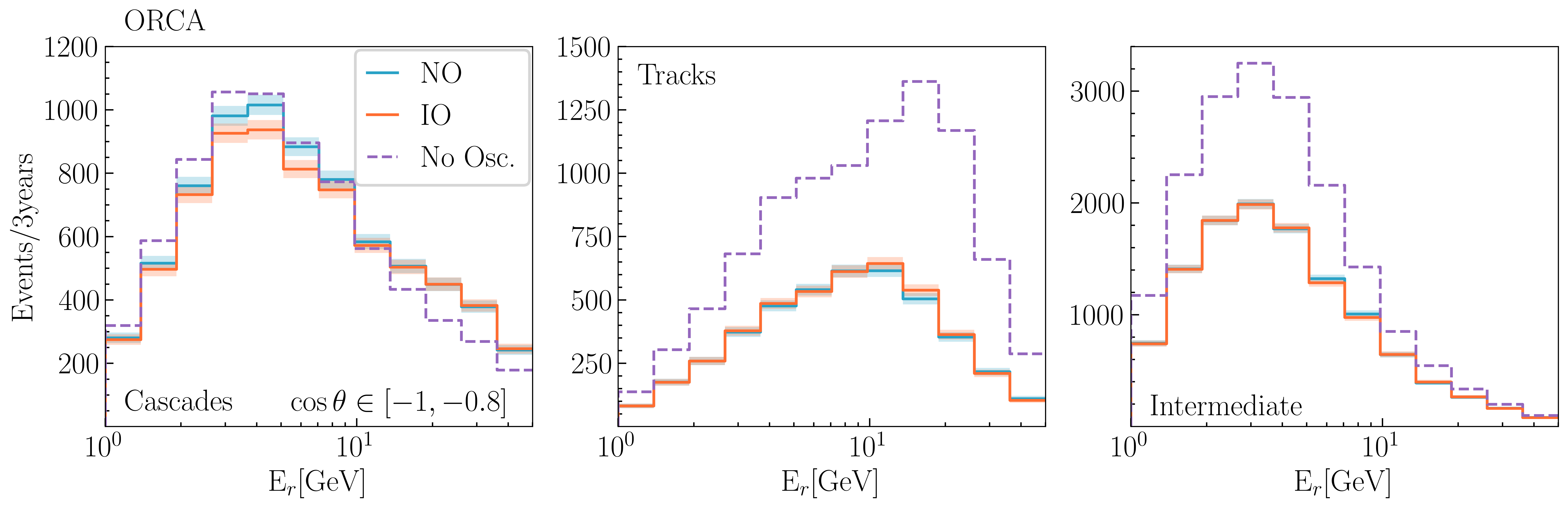}
 \includegraphics[width=\textwidth]{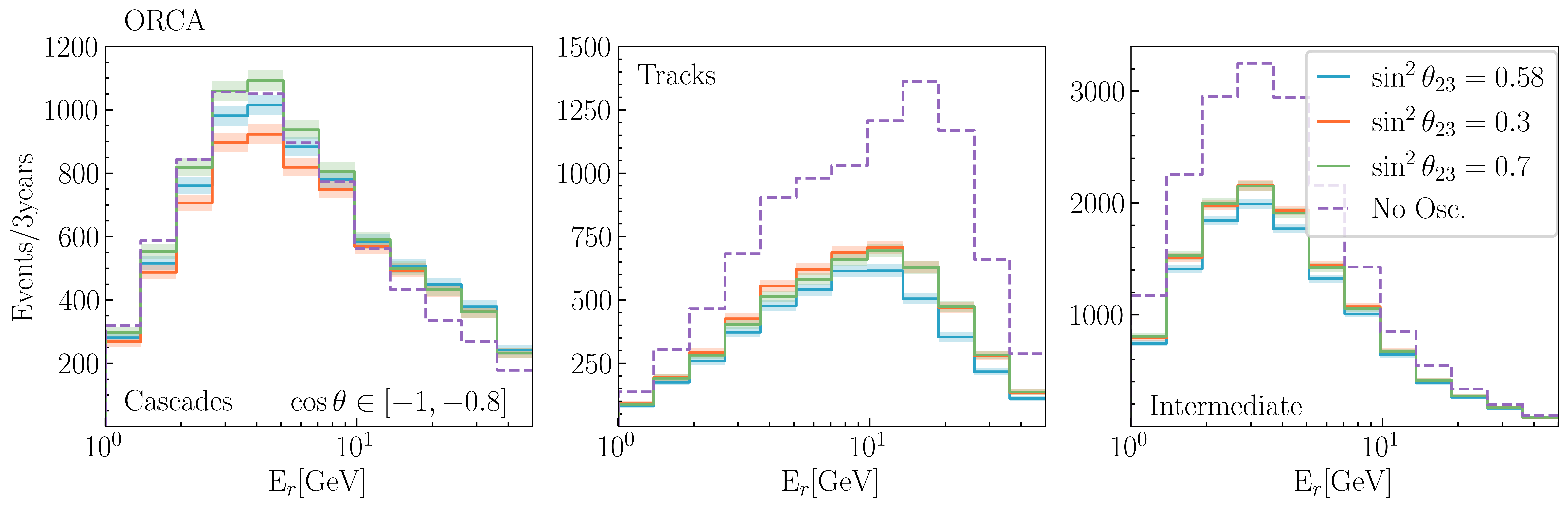}
    \caption{\textit{\textbf{Event distribution as a function of the reconstructed energy and direction for ORCA after 3 years of data taking.}} Lines correspond to event distributions (with oscillations) for track, cascade, and intermediate morphology classes respectively. }
\label{fig:ORCA_distribution_Reco}
\end{figure*}

In~\Cref{fig:ORCA_distribution_Reco}, we show the event distribution in reconstructed energy for both orderings (top) and for three different values of $\sin\theta_{23}=\{0.3, 0.58,0.7\}$.
Worth noting is that intermediate events dominate the sample for the whole energy range. 
The new event morphology improves the ``purity" in the cascade and track samples.
The neutrino mass ordering changes the number of cascades with energies around the atmospheric resonance, \Cref{fig:ORCA_distribution_Reco} (top).
Also, the normalization of the cascade sample contributes to the measurement of $\sin^2\theta_{23}$.
The same happens for tracks with energies around $\SI{5}\GeV$.
For energies above $\sim \SI{10}\GeV$, tracks are sensitive to $\sin^2 2\theta_{23}$, what helps in the separation of $\theta_{23}$ being maximal mixing or not.

Besides the uncertainties related to the flux and the cross section, the measurement carried out in ORCA will be also affected by the uncertainties in the detector response.
We consider a free normalization for each of the event morphologies used in the analysis and a $5\%$ error on the energy scale, \Cref{table:ORCASysts}. Additionally, in line with the IceCube methodology for addressing detector uncertainties, we have incorporated a set of systematic factors to consider the absorption and scattering of photons in water, as well as the response of the photomultiplier.

\begin{table}
\begin{tabular}{l|c}
Systematic source & \makecell{1$\sigma$-range\\(ORCA)} \\
\hline\hline
Energy scale & 5\% \\
Intermediate	& free \\
Tracks & free  \\
Cascades 	& free \\
Water Absorption & $10\%$\\
Water Scattering & $10\%$\\
Overall Optical Efficiency & $10\%$\\
Lateral Optical Efficiency & $40\%$\\
Head-on Optical Efficiency & $10\%$\\
\end{tabular}
\caption{Summary of ORCA detector systematic uncertainties used in this work.}
\label{table:ORCASysts}
\end{table}

\subsection{DUNE}\label{sec:DUNE}

Liquid argon time projection chamber (LArTPC) detectors have demonstrated a good capacity in the reconstruction of sub-GeV neutrinos. The energy an direction of the incoming neutrino can be reconstructed by detecting the tracks of all charged particles produced after the neutrino interaction and identifying them by their topology and energy loss. DUNE is planning to use a $20$~kton detector based on this technology to reconstruct beam neutrinos, although they can also measure the atmospheric neutrino flux. 

The reconstruction of the sub-GeV atmospheric neutrinos give us access to determine the CP-phase. 
Following~\cite{Kelly:2019itm}, the event distribution in the reconstructed energy and direction has been estimated by a simulation of the neutrino-argon interaction using an event generator.
The uncertainties considered for the outgoing protons are $10\%$ in energy and $10^{\circ}$ in the direction.
In the case of the leptons, we have used $5\%$ for the energy and $5^{\circ}$ for the direction. Due to the LArTPC capabilities in identifying low energy charged particles, the events are classify by the number of outgoing visible protons.
This allows an statistical separation between neutrinos, which dominates the fraction of the with one proton, and anti-neutrinos, which dominate the 0-proton sample.

\begin{figure}[H]
    \centering
    \includegraphics[width=0.95\textwidth]{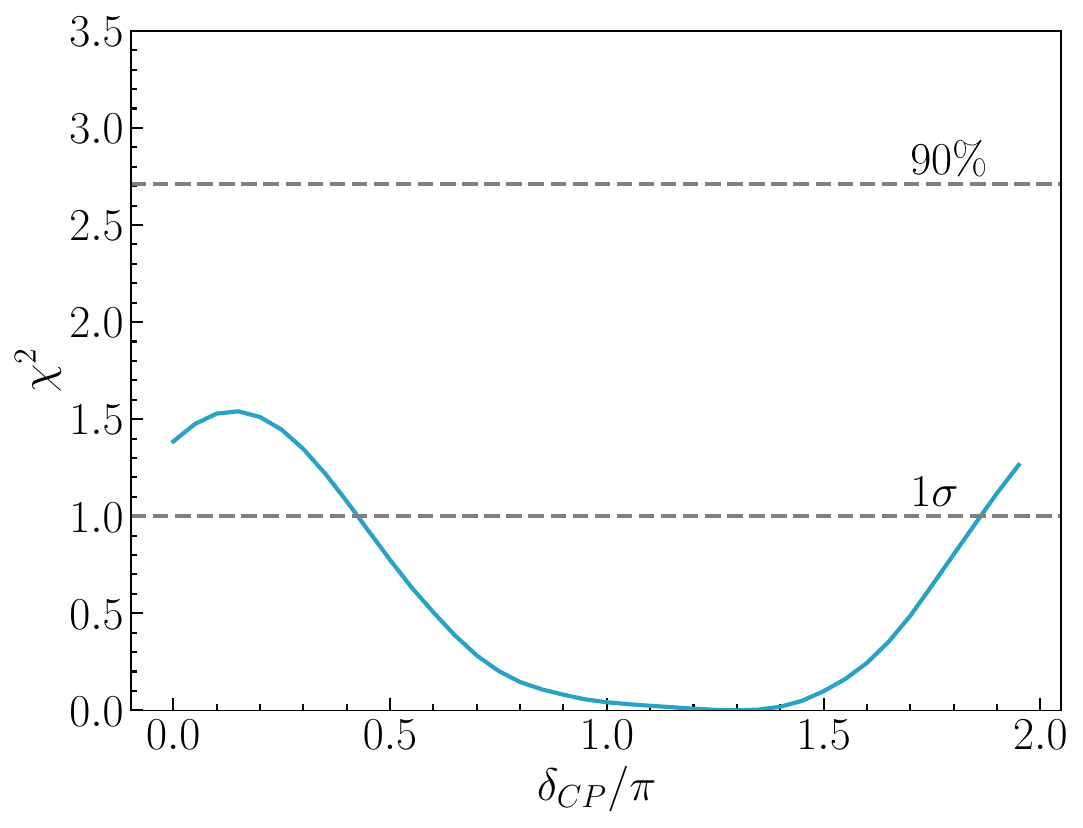}
    \caption{\textit{\textbf{DUNE sensitivity to $\delta_{CP}$}}. We consider  1~year of data taking and two modules of 10~ktonnes each.}
    \label{fig:DUNE}
\end{figure}

Considering a $20$~kton detector, taking data for 1 year, we get a sensitivity over $\delta_{CP}$ around $1.5\sigma$, ~\cref{fig:DUNE}. In this analysis, we have included only the uncertainties related to the flux and the detector response.
Another set of uncertainties related to the neutrino-argon cross section will also affect this sensitivity. The low number of events expected by the end of this decade, and the large uncertainties, makes that the DUNE measurement of the atmospheric neutrino flux cannot contribute significantly to the determination of $\delta_{CP}$ by the end of this decade.
Therefore, we have decided not included in our analysis. 

%%%%%%%%%%%%%%%%%%%%%%%%%%%%%%%%%%%%%%%%%%%%%%%
\section{Results}\label{sec:results}
In our analysis, we explore the combined sensitivity that the present and near-future generation of atmospheric neutrino experiments will have in determining the oscillation parameters for the $3\nu$ mixing scenario by the end of this decade.
Through the simulation of the different phases of SuperK, IceCube Upgrade, and ORCA, and the inclusion of all the previously discussed systematic uncertainties --- flux, cross section, and detector---, we investigate the synergies between the three experiments for the sensitivity to the $\Delta m^2_{31}$, $\theta_{23}$, $\delta_{CP}$ and $\theta_{13}$ oscillation parameters.

Unless otherwise stated, our study of the neutrino oscillation sensitivity presented here assume the values and treatment shown in \Cref{table:OscPar}.

For baselines comparable to Earth's diameter, neutrino oscillations are sizeable at $E_\nu\sim \SI{100}\GeV$ and lower.
The location of the first oscillation minimum occurs at energies around $\SI{20}\GeV$ and depends on the value of $\Delta m^2_{31}$.
The amplitude of this oscillation is modulated by $\sin^2\theta_{23}$. This energy region will be measured with large sample sizes by the IceCube Upgrade and ORCA.
SuperK also has some sensitivity to this region from the Multi-GeV Multi-Ring samples, but with smaller sample sizes.
The track sample dominates the sensitivity to $\Delta m^2_{31}$ thanks to better angular resolution.
Since the atmospheric neutrino flux is dominated by $\nu_{\mu}$, tracks are mainly sensitive to $P_{\mu\mu}$.

\begin{figure}
\centering
\includegraphics[width=\textwidth,height=3in]{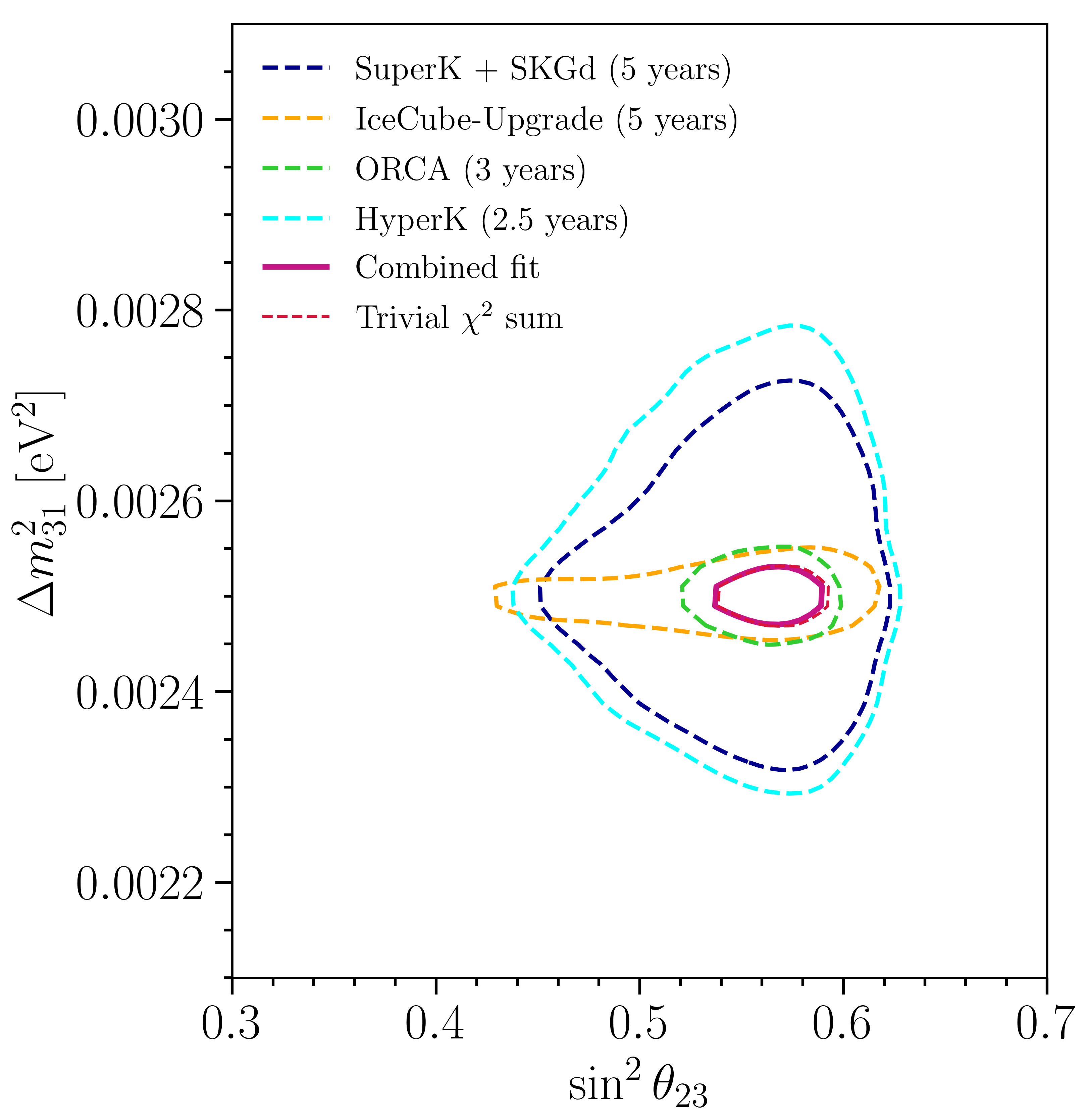}
\caption{\textbf{\textit{Two-dimensional 90\% confidence level regions for $\sin^2\theta_{23}$ and $\Delta m^2_{31}$}} for SuperK (dashed dark blue), IceCube Upgrade (dashed orange), ORCA (dashed green), HyperK (dashed cyan), and combined analysis (solid violet-red).}
\label{fig:t23vdm231}
\end{figure}

On the other hand, as seen in \Cref{sec:OSC}, the muon-disappearance channel is sensitive to $\sin^2 2\theta_{23}$, so it can only distinguish whether $\sin^2\theta_{23}$ is maximal mixing or not.
To distinguish between both octants, we need to consider the appearance channel, proportional to $\sin^2\theta_{23}$.
Water detectors have a better angular resolution compared to ice detectors due to the larger number of direct photons arriving at the PMTs.
For that reason, ORCA, SuperK and HyperK show a better precision for $\sin^2\theta_{23}$.
The combined analysis of the three experiments shows that $\Delta m^2_{31}$ can be measured at $\sim 0.6\%$ and the octant of $\sin^2\theta_{23}$ can be resolved at more than $3\sigma$ in the assumed scenario.
\Cref{fig:t23vdm231} shows that $\Delta m^2_{31}$ is dominated by IceCube Upgrade due to its better energy resolution: see~\Cref{sec:IC}.

\begin{figure}
\centering
\includegraphics[width=\textwidth,height=3.1in]{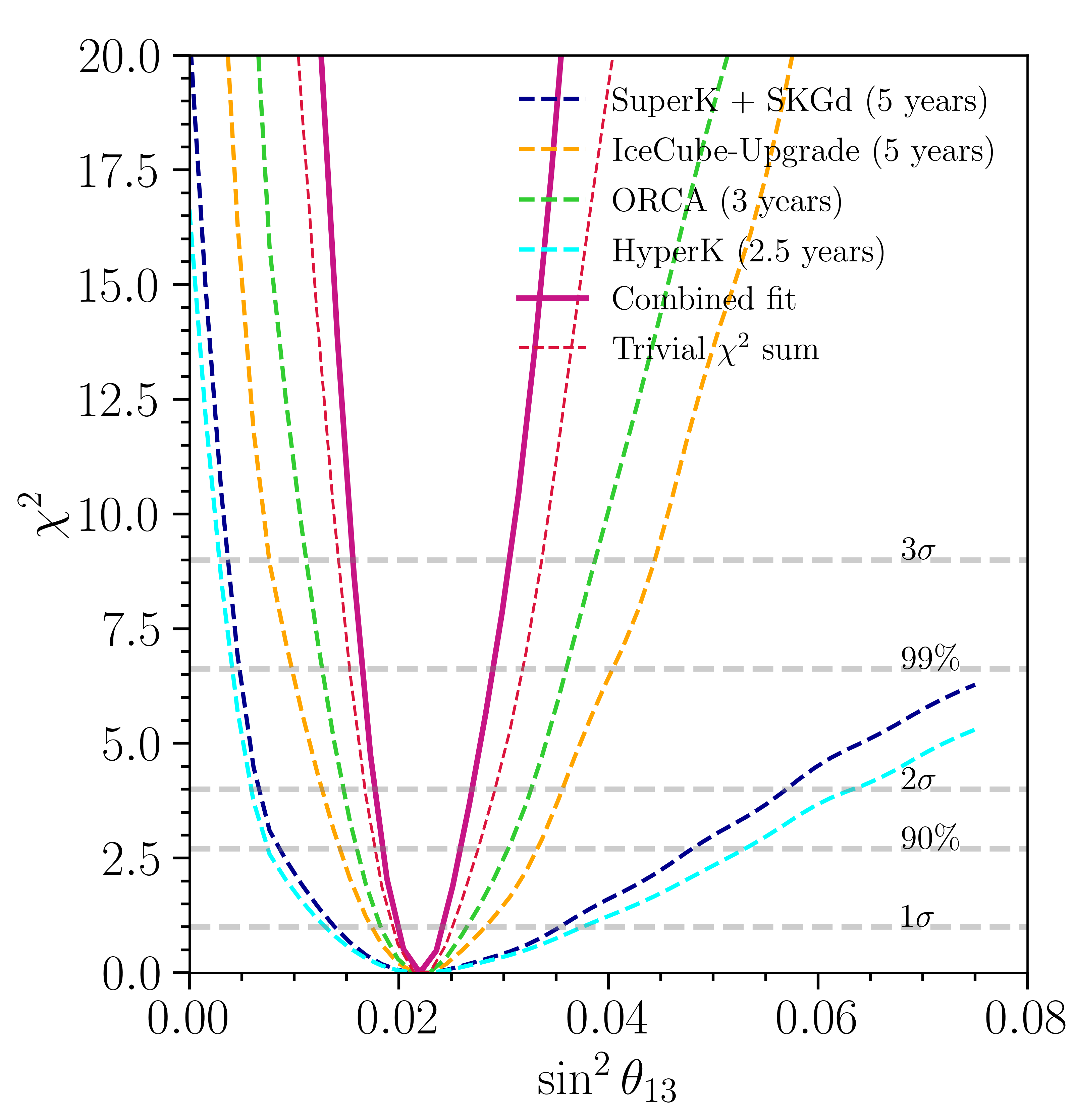}
\caption{\textbf{\textit{Sensitivty to $\sin^2\theta_{13}$}} for SuperK (dashed dark blue), IceCube Upgrade (dashed orange), ORCA (dashed green), HyperK (dashed cyan), and combined analysis (solid violet-red).}
\label{fig:t13}
\end{figure}

At the GeV scale, neutrinos crossing the mantle undergo a large flavor oscillation due to the MSW resonance at $\sim \SI{6}\GeV$.
As described in~\Cref{sec:OSC}, the resonance happens due to the effective enhancement of the $\theta_{13}$ mixing angle.
This affects neutrinos if the mass ordering is normal, and anti-neutrinos in the inverted scenario.
Although atmospheric experiments cannot differentiate between neutrinos and anti-neutrinos on an event-by-event basis, the differences in the flux and the cross section allow for statistical separation.
Since this is a very localized process in energy and direction, the events with better angular and energy resolution will contribute more to identifying it, and therefore they will contribute to the neutrino mass ordering sensitivity.
This is what we have observed in~\Cref{fig:ICparam}, where tracks show a larger modification under the mass ordering.
For the case of HyperK, SuperK and ORCA, we see in \Cref{fig:ORCA_distribution_Reco}, \Cref{fig:sk_dcpmo_ratio}, and \Cref{fig:skgd_dcpmo_ratio} that most of the sensitivity to the ordering comes from $e$-like and cascades samples in the relevant energy region.
In~\Cref{fig:ordering}, we show the sensitivity to the ordering as a function of operation time by combining the three experiments.
It would be possible to obtain a $6\sigma$ identification of the ordering after five years of SuperK with Gd in addition to its current exposure, five years of IceCube Upgrade, and three years of ORCA even with the obtained 90\% C.L. range over $\sin^2\theta_{23}$.

The effective enhancement on $\theta_{13}$ also provides the option to measure this parameter.
Although this parameter has been measured with high precision in reactor experiments~\cite{DayaBay:2018yms, RENO:2018dro, DoubleChooz:2019qbj} looking for the disappearance of $\overline{\nu}_e$ and more recently in LBL~\cite{T2K:2019bcf, Hartnell2022-do} using the appearance channel, atmospheric neutrinos bring us a complementary way to measure it via the matter effect in Earth.
Also, this will constitute the first observation of the MSW effect on Earth.
This parameter is measured by the three experiments by separate, and the combined analysis will reach an uncertainty smaller than $20\%$: see~\Cref{fig:t13}.

\begin{figure}
\centering
\includegraphics[width=\textwidth,height=3in]{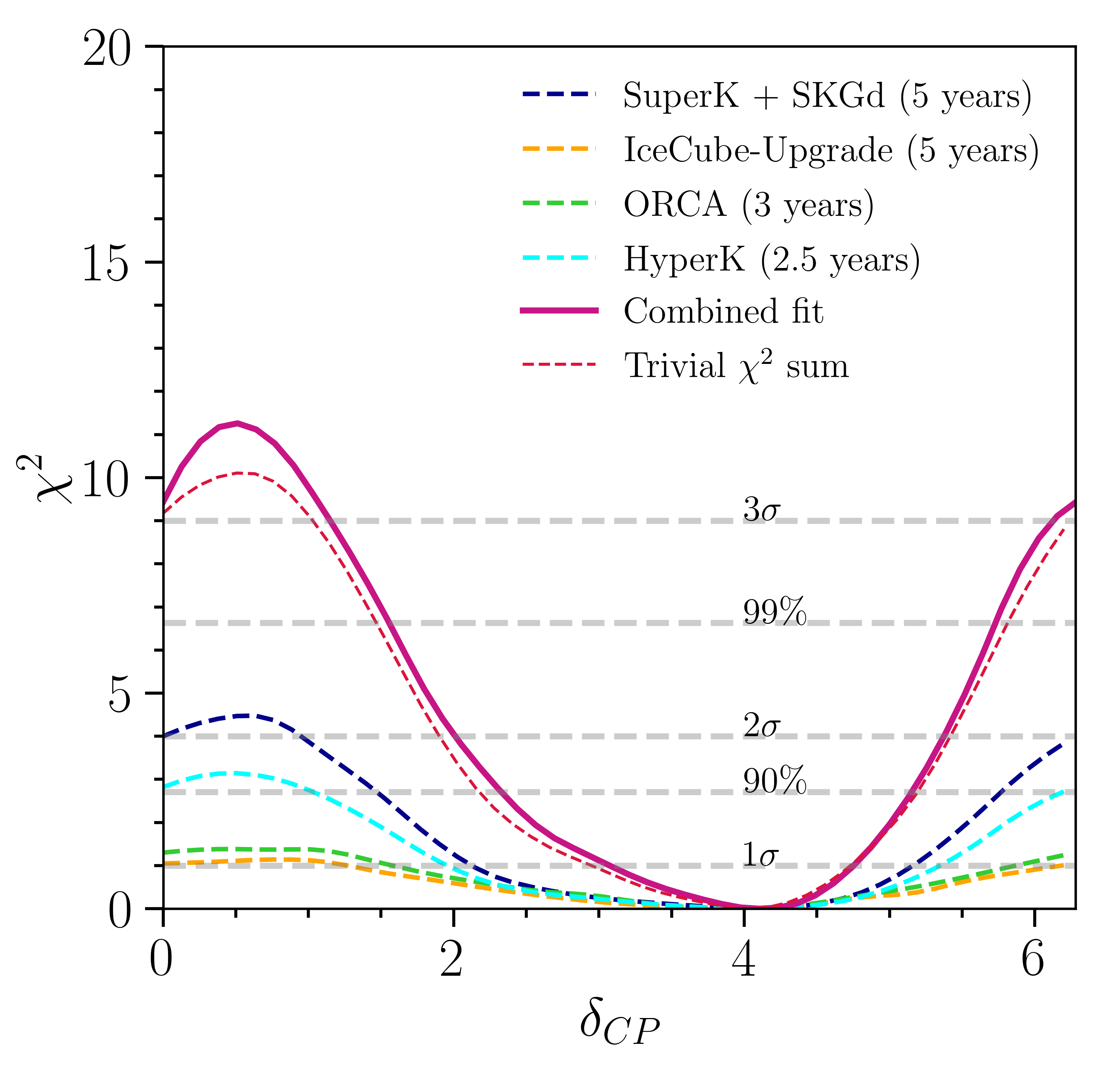}
\caption{\textbf{\textit{Sensitivty to $\delta_{CP}$}} for SuperK (dashed dark blue), IceCube Upgrade (dashed orange), ORCA (dashed green), HyperK (dashed cyan), and combined analysis (solid violet-red).}
\label{fig:dcp}
\end{figure}

Finally, the $CP$-phase is the least constrained parameter, for which almost the entire range is allowed at $3\sigma$; T2K~\cite{T2K:2021xwb} and NOvA~\cite{NOvA:2021nfi} are the only experiments have shown some sensitivity.
%and we have only a few hints from T2K~\cite{T2K:2021xwb} and NOvA~\cite{NOvA:2021nfi}.
So far, the only atmospheric measurement that shows some sensitivity over $\delta_{CP}$ comes from SuperK~\cite{linyan_wan_2022_6694761}, which excludes values around $\delta_{CP}=\pi/5$ with a significance slightly above $2\sigma$.
This parameter is the main target for the next-generation neutrino experiments.
Using the presented atmospheric experiments, the sensitivity over some parameter values can be increased up to $99\%$ C.L.: see~\Cref{fig:dcp}.
Still, the sensitivity is dominated by SuperK and HyperK, but IceCube and ORCA can get a $1\sigma$ significance thanks to their low-energy measurements. The $\delta_{CP}$ is the parameter that benefits the most from the combined analysis due to the large sample sizes from the IceCube Upgrade and ORCA constraining additionally the flux uncertainties which reduce the sensitivity to this parameter.

Previous studies such as the ones given in Refs.~\cite{Esteban:2020cvm,Capozzi:2018ubv,deSalas:2020pgw} do not consider the correlation between systematic uncertainties as we have done in this work. 
The improved methodology, provides enhanced capacity to determine $\delta_{CP}$ and $\theta_{23}$, and has lesser effect in other parameters.
Throughought this work the lines labelled `Combined Fit' and `Trivial $\chi^2$ Sum' name our analysis and a simplified uncorrelated work, respectively.
The difference between these two lines showcases the impact of the correlated systematics. 

\begin{figure}
\centering
\includegraphics[width=\textwidth,height=3in]{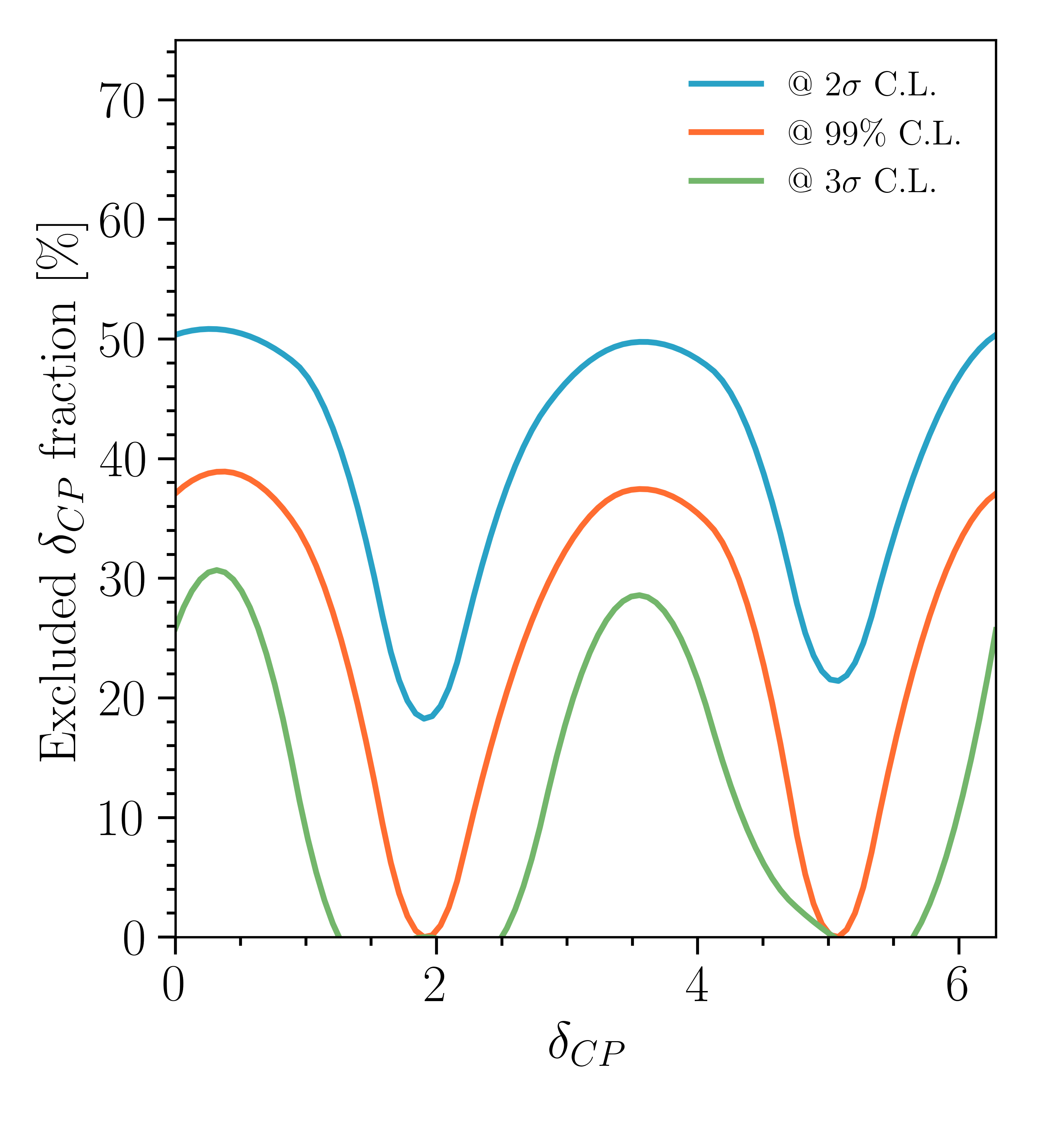}
%dcpFraction+HK_few.png}
\caption{\textbf{\textit{Excluded fraction of $\delta_{CP}$ as a function of the true value of $\delta_{CP}$.}} Lines correspond to exclusion at 2$\sigma$ (blue), 99\% (orange), and 3$\sigma$ (green)  for the combined fit assuming normal ordering and fixed $\sin^2\theta_{13}=0.022$.
$40\%$ of the parameter space can be excluded at $99\%$ confidence level for the CP-conserving scenario.
For lower confidence levels, the small deviations in the event number expectations can contribute to increase the sensitive region making the sensitivity more uniform.}
\label{fig:dcpFract}
\end{figure}

The excluded region for $\delta_{CP}$ depends on its true value because it predicts very different event distributions for electron and muon samples in SuperK and HyperK, and also, it modifies the cascade (and intermediate events) distribution in the case of IceCube Upgrade (ORCA).
Since almost the entire space is allowed, we have explored what fraction of $\delta_{CP}$ can be excluded at some confidence levels and for different true values of $\delta_{CP}$: see~\Cref{fig:dcpFract}.
The most favorable scenario would correspond to $\delta_{CP}=0$ or $\pi$, where $60\%$ of the whole space can be excluded at $90\%$ C.L. and $30\%$ of the space is explored with a significantly larger than $3\sigma$.

In~\Cref{table:1sigma}, we summarize the projected 1$\sigma$ regions for all four oscillation parameters considered free in this combined analysis and with the true values from~\Cref{table:OscPar}.

\begin{table}[ht]
\begin{tabular}{l|c}
Parameter & $1\sigma$ range \\
\hline\hline
%\\[-0.8em]
$\sin^2\theta_{13}$ & [$0.0199$, $0.0242$]  \\
%\\[-0.8em]
$\sin^2\theta_{23}$ &  [$0.554$, $0.578$]  \\
%\\[-0.8em]
$\delta_{CP}$       &  [$3.12$, $4.74$] \\
%\\[-0.8em]
$\Delta m^2_{31}$ [eV$^2$]   &  [$0.002487$, $0.002514$] \\
% $\Delta m^2_{31}$ ($10^{-3}$eV$^2$)   &  [$2.484$, $2.516$] \\
\end{tabular}
\caption{$1\sigma$ sensitivity range from the combined analysis of the atmospheric neutrino experiments.}
\label{table:1sigma}
\end{table}

%%%%%%%%%%%%%%%%%%%%%%%%%%%%%%%%%%%%%%%%%%%%%%%
\section{Discussion and Outlook}\label{sec:disc}
\begin{figure}
\includegraphics[width=0.9\textwidth, height=5.cm]{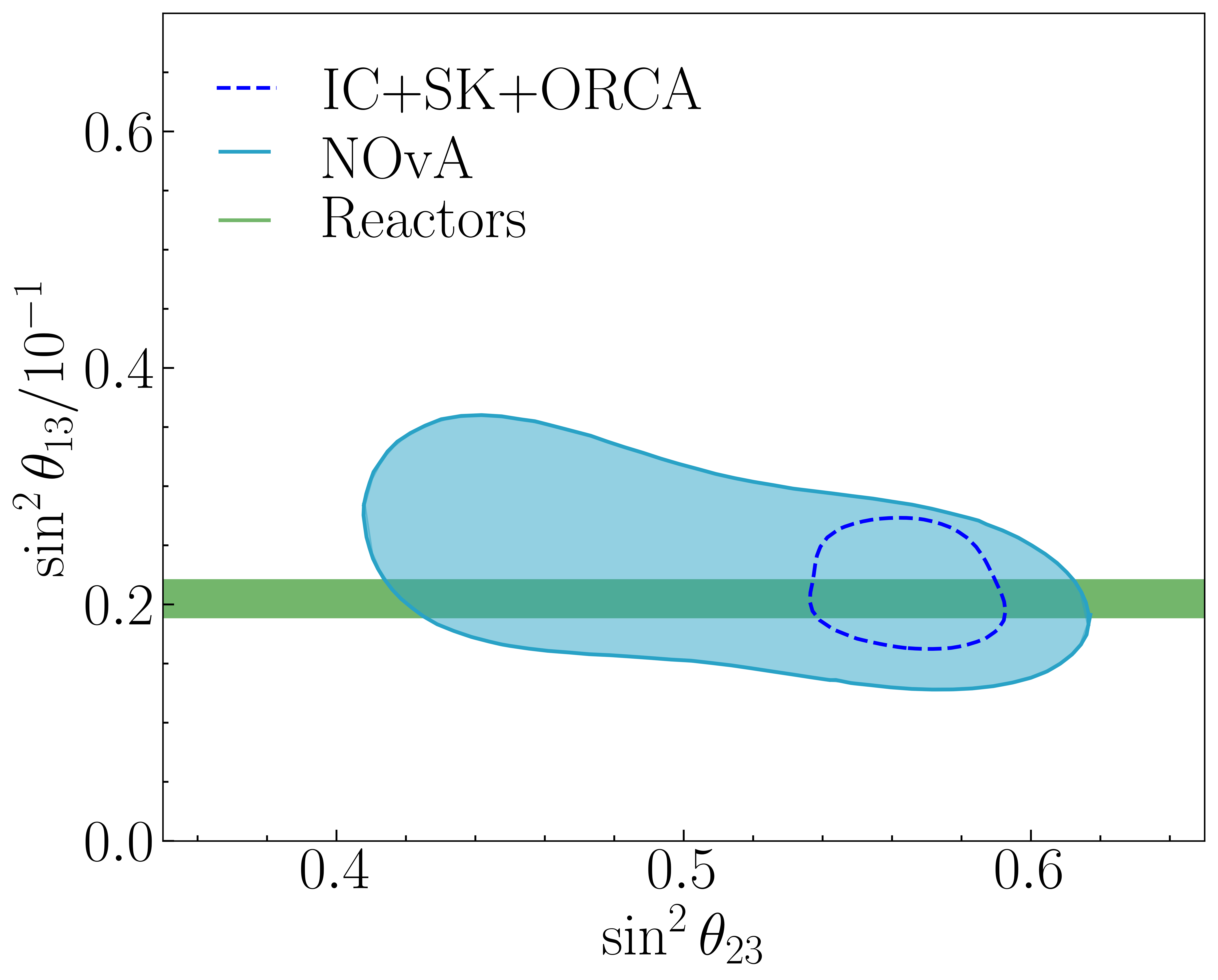}
\includegraphics[width=0.9\textwidth, height=5.cm]{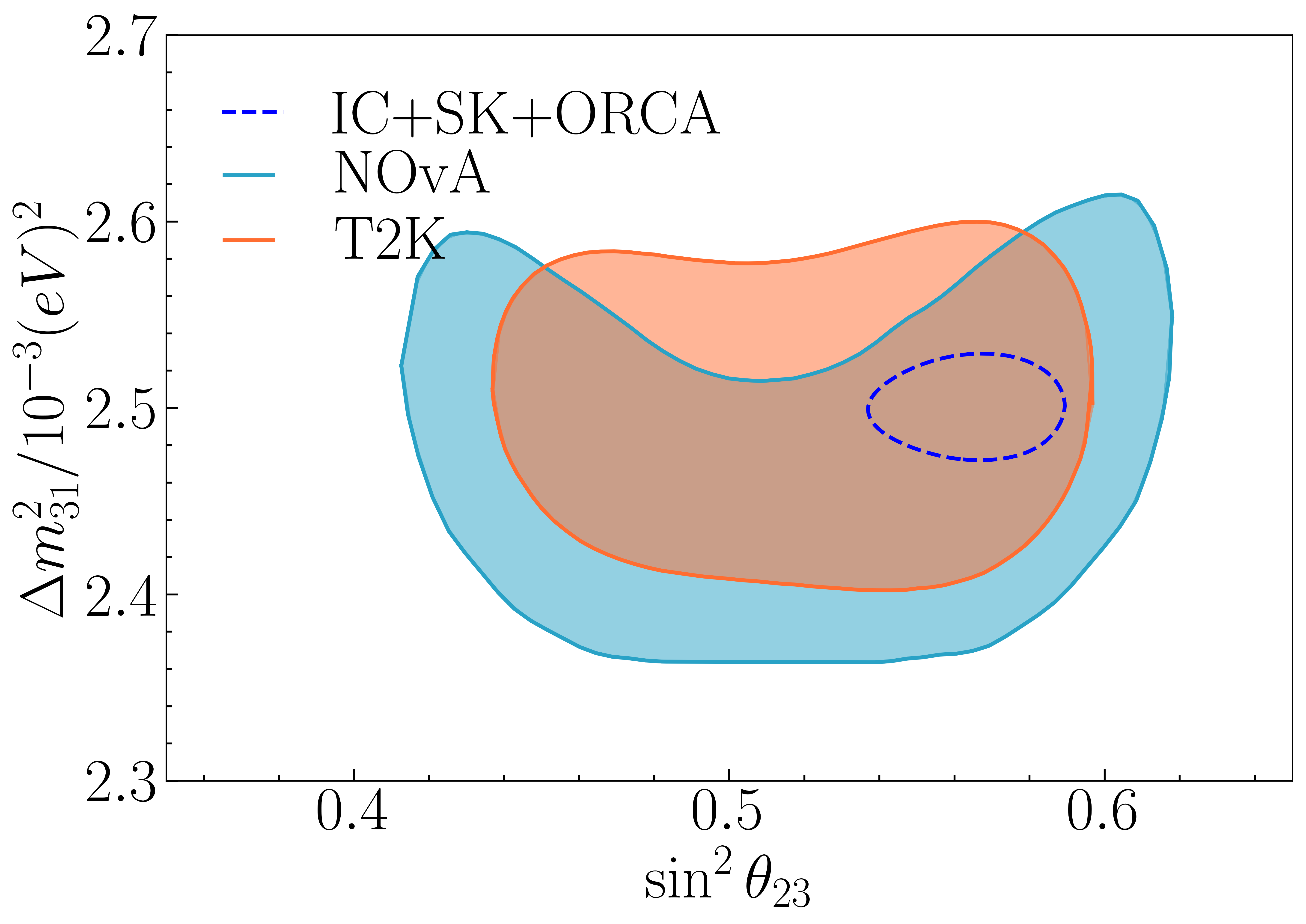}
\includegraphics[width=0.9\textwidth, height=5.cm]{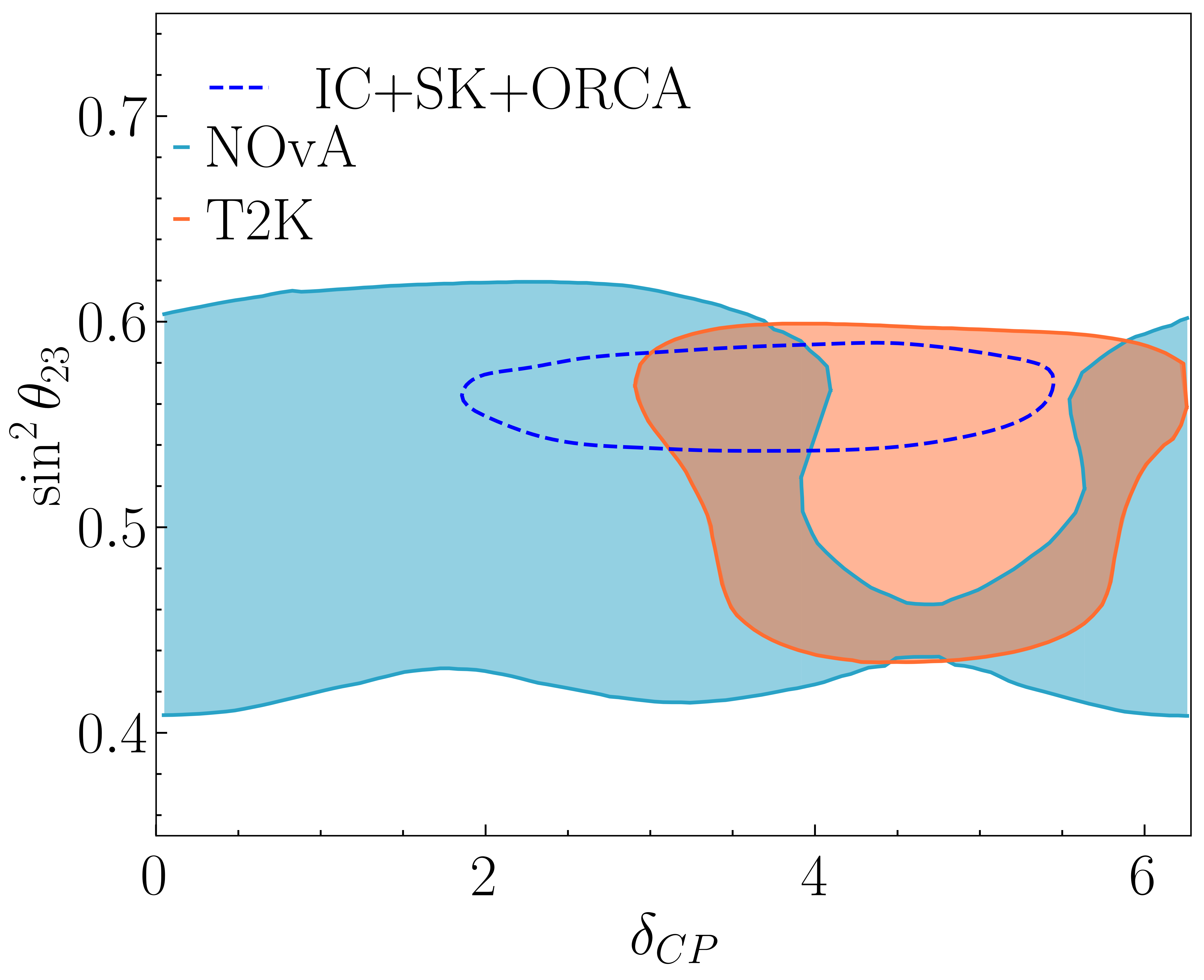}
\centering
\caption{\textbf{\textit{Present sensitivity to the 3$\nu$ mixing parameters}} Comparison between the latest results from reactor and LBL measurements, and the future prediction from the combine analysis of SuperK, IceCube upgrade and ORCA.}
\label{fig:status}
\end{figure}

Atmospheric neutrinos have played a very important role in determining the oscillation parameters since they were observed by SuperK~\cite{Super-Kamiokande:1998kpq}.
As highlighted in this article, the large range of energies and baselines covered by the flux crossing Earth leads to a great sensitivity in the determination of the atmospheric parameters ($\Delta m^2_{31}$ and $\theta_{23}$).
The matter effect at the GeV scale on neutrinos crossing the core and the mantle also brings the possibility of measuring the mass ordering and the so-called reactor angle ($\theta_{13}$).
Finally, at the GeV scale and below, the deviations in the appearance channel and statistical differentiation between neutrinos and anti-neutrinos allow the constraint of the $CP$-violation phase.

The results obtained in our analysis indicate that soon, atmospheric neutrinos can provide a complementary role in constraining the oscillation parameters and improving, in some cases, the precision obtained by long-baseline experiments.
For instance, in the case of mass ordering, the combined analysis of LBL has a $2\sigma$ preference for inverted ordering due to the tension in $\delta_{CP}$~\cite{Kelly:2020fkv}, while the latest results from atmospheric neutrinos show a 
preference for normal ordering with a significance larger than $2\sigma$~\cite{linyan_wan_2022_6694761}.
The results presented in this study demonstrate that atmospheric neutrinos alone are poised to achieve a significance exceeding $6\sigma$. 
This suggests that these parameters will be measured with remarkable precision by the conclusion of this decade, as illustrated in~\Cref{fig:ordering}.
%The results obtained in this work show that atmospheric neutrinos alone will be able to reach a significance of more than $6\sigma$, see~\Cref{fig:ordering}. 

\begin{figure*}[!ht]
     \centering
         \includegraphics[width=7cm, height=7cm]{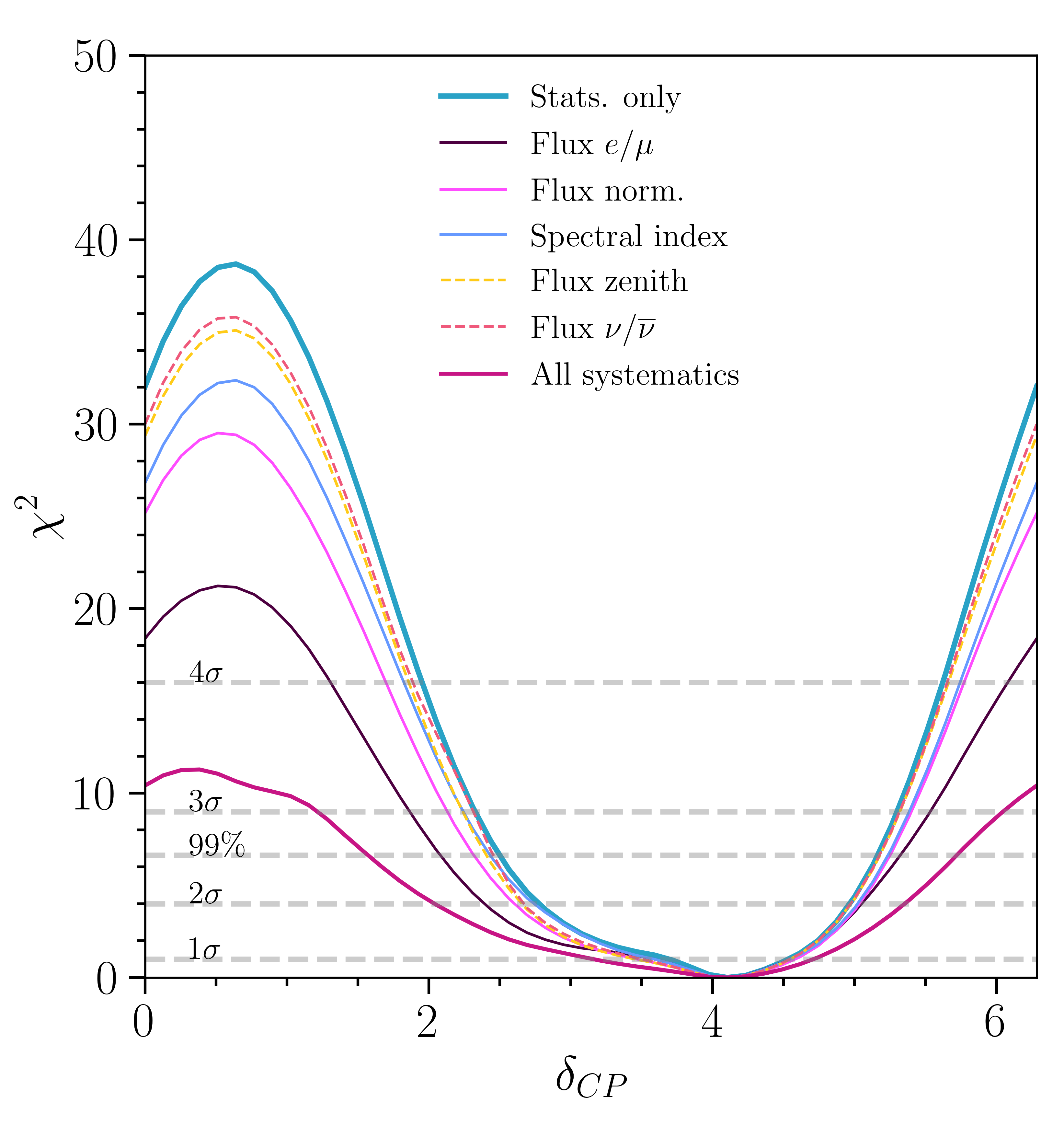}
     \hfill
         \includegraphics[width=7cm, height=7cm]{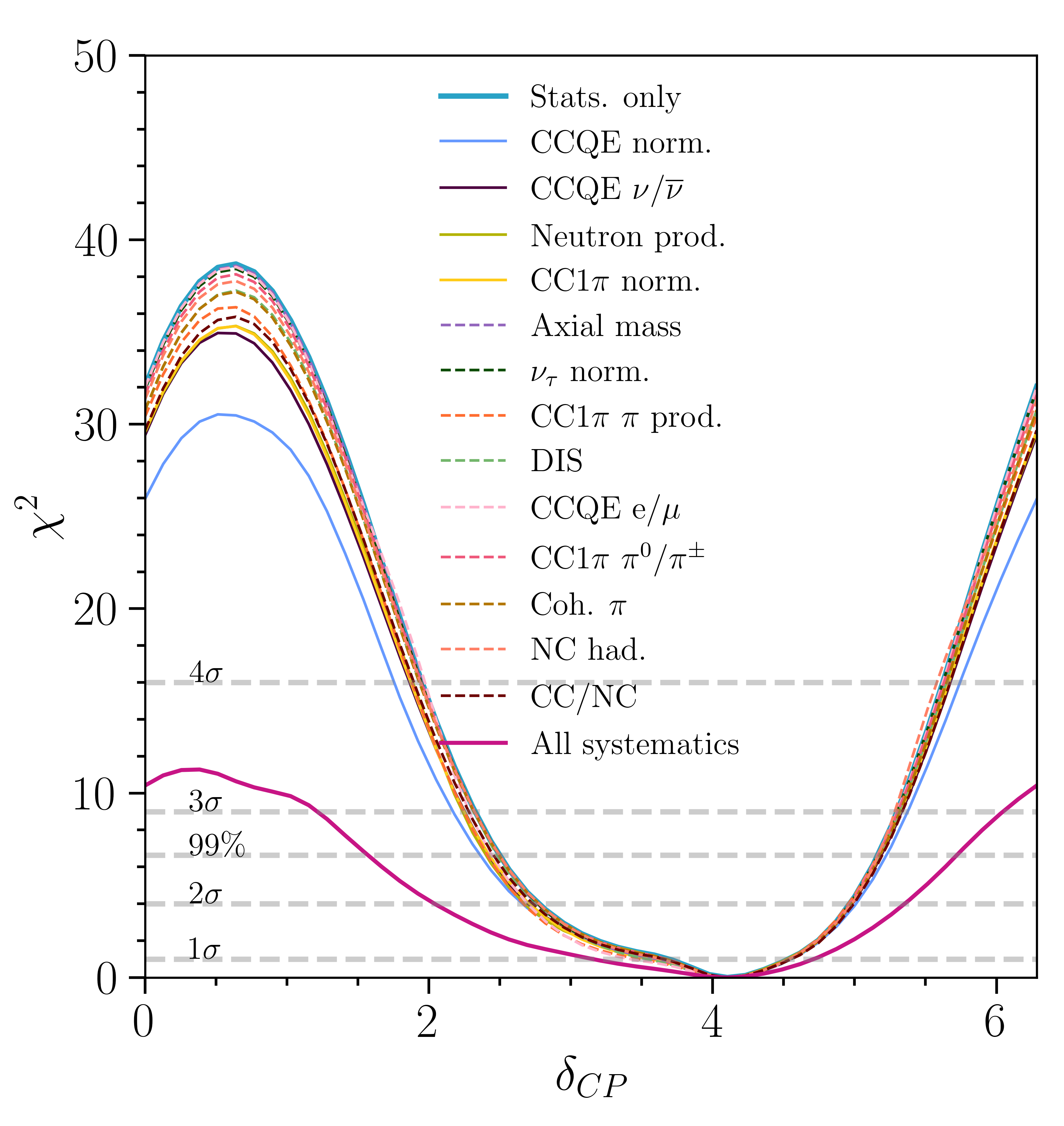}
     \hfill
         \includegraphics[width=7cm, height=7cm]{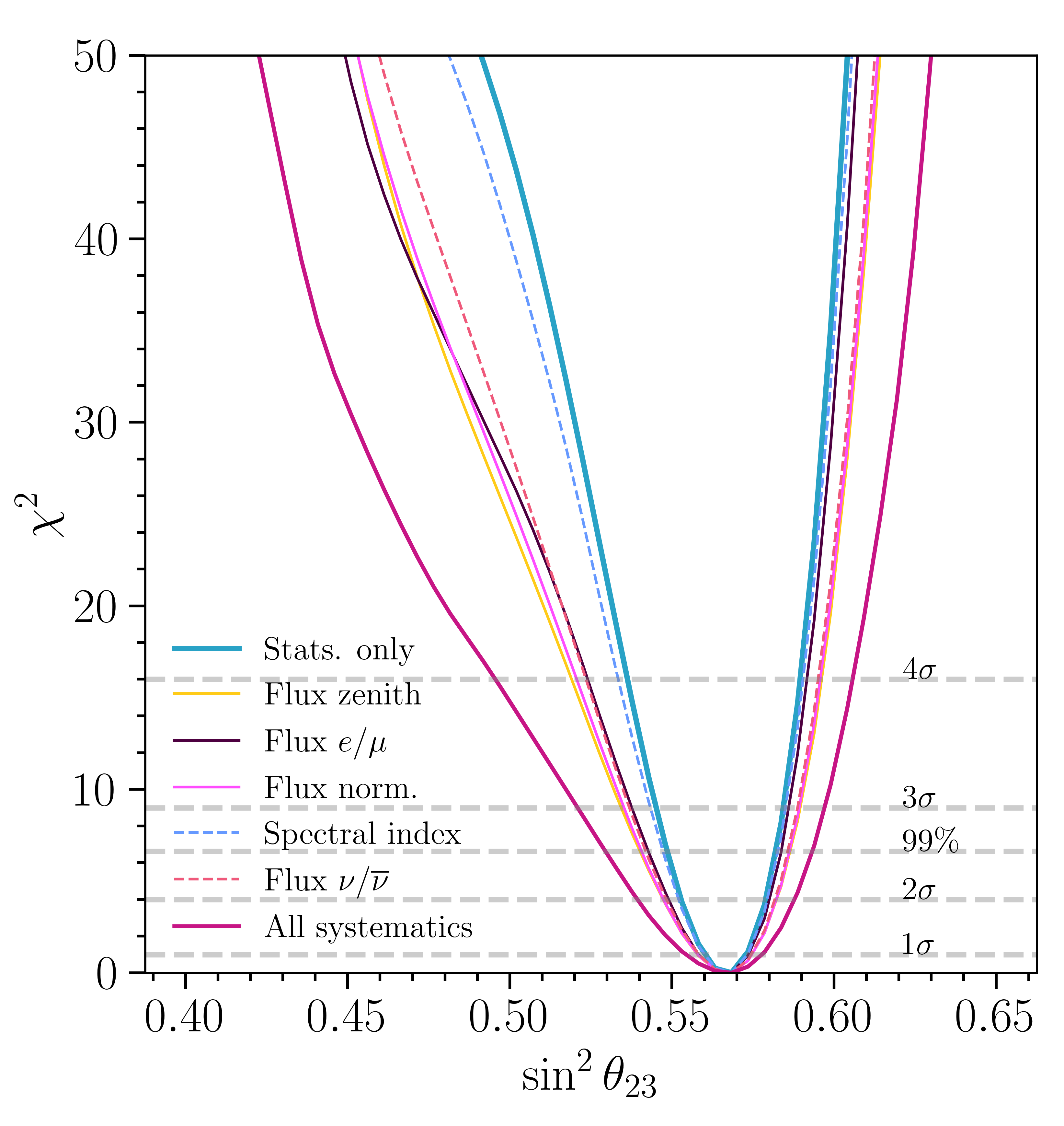}
     \hfill
         \includegraphics[width=7cm, height=7cm]{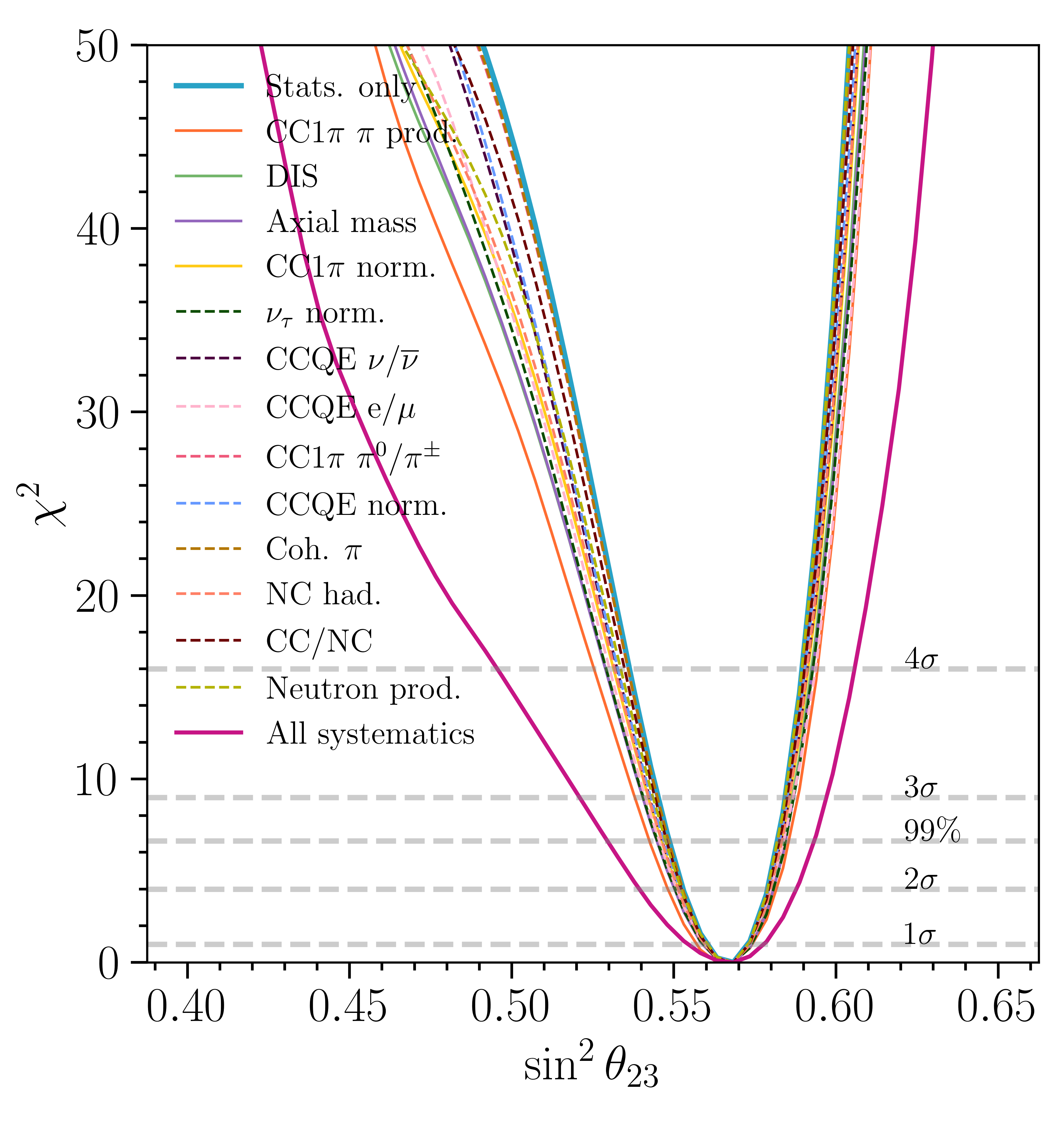}
     \hfill
         \includegraphics[width=7cm, height=7cm]{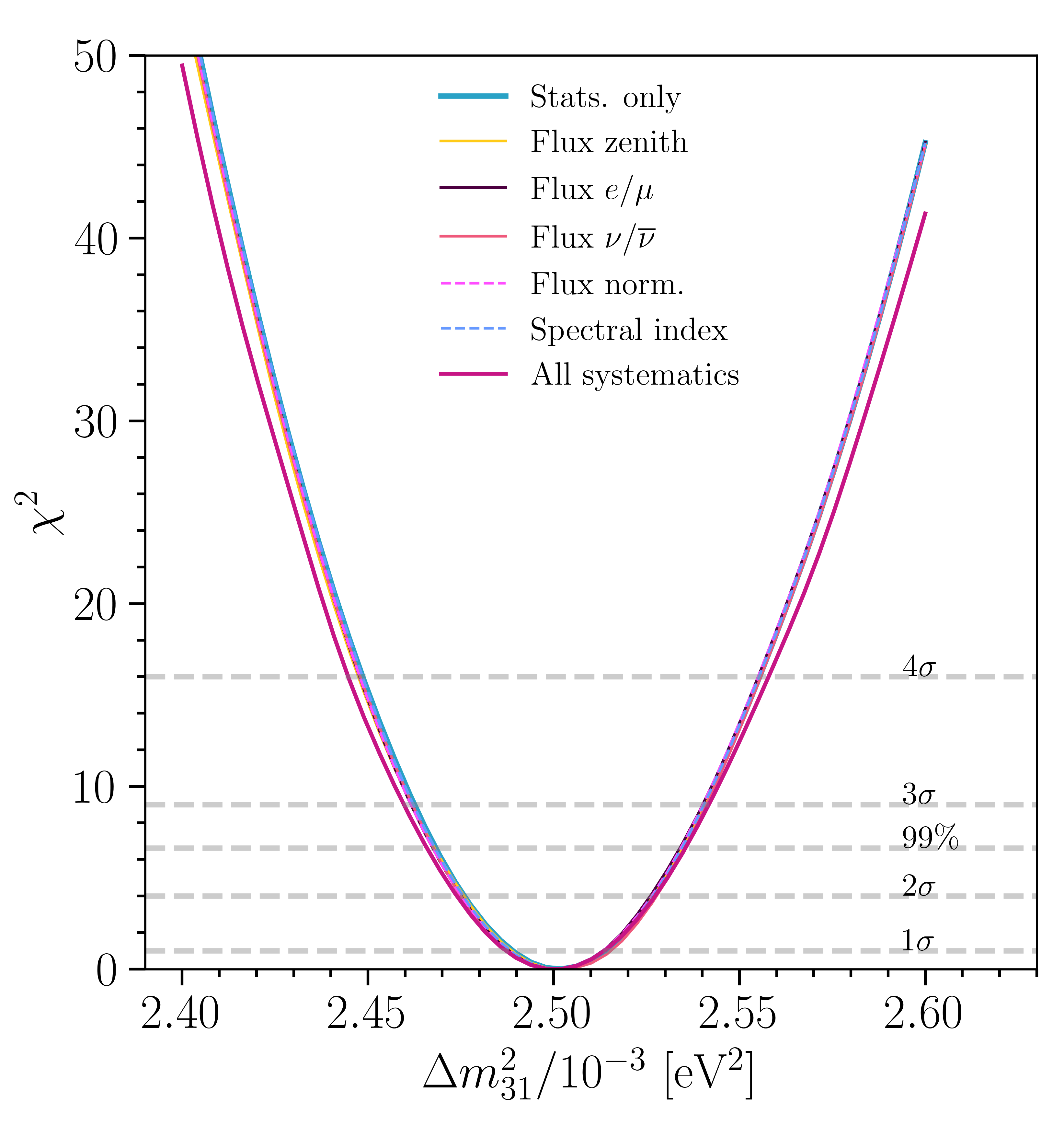}
     \hfill
         \includegraphics[width=7cm, height=7cm]{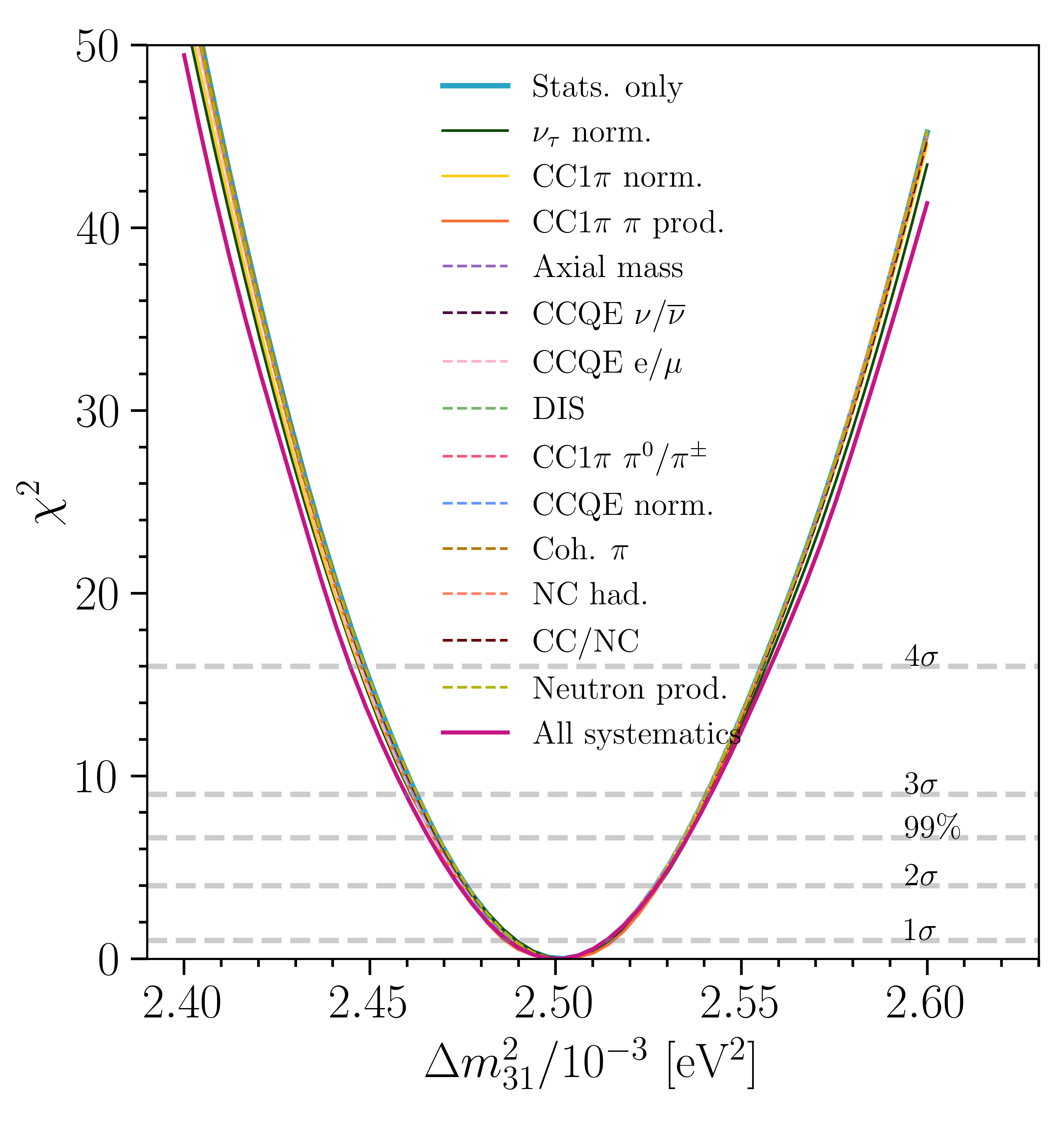}
        \caption{\textbf{\textit{From top to bottom, the impact of each atmospheric neutrino flux (left) and cross section (right) systematic uncertainties on the sensitivity to $\delta_{CP}$, $\sin^2\theta_{23}$ and $\Delta m^2_{31}$.}}
        For clarity, solid lines represent the most relevant systematics, while dashed lines represent the rest.
        }
        \label{fig:flux+xsec_syst}
\end{figure*}

The measurement of the atmospheric parameters is currently dominated by LBL experiments.
In the case of $\Delta m^2_{31}$, the current sensitivity is dominated by T2K~\cite{Bronner2022-ey} and NOvA~\cite{Hartnell2022-do}, and it is known with $\sim 1.1\%$~\cite{Esteban:2020cvm} according to latest global analyses.
The sensitivity to this parameter can be largely improved up to $\sim 0.5\%$ using just atmospheric neutrinos.
The latest results on $\sin^2\theta_{23}$ indicate a preference for the upper-octant at a significance smaller than $1\sigma$, and maximal mixing is excluded with less than $2\sigma$ significance.
If we assume the current best-fit value of the global analyses, atmospheric neutrinos can achieve a $3\sigma$ sensitivity over maximal mixing, ruling out the lower octant with a higher significance than the LBL experiments, as shown in~\Cref{fig:status}. These results predict that atmospheric neutrinos will rule out the wrong octant of $\theta_{23}$ with a large significance by the end of the decade.
Most of the sensitivity over those two parameters comes from the region above $\sim \SI{10}\GeV$ dominated by the large sample size of IceCube Upgrade and ORCA measurements and, to a lesser extent, SuperK and HyperK.
In that region, the event distribution is dominated by $\nu_{\mu}$ that interact via DIS.
Therefore, flux and cross section systematic uncertainties have a mild effect on the sensitivity as shown in~\Cref{fig:flux+xsec_syst}.

Until now, the $CP$-violation phase has been explored with a low significance and, only a small region around $\delta_{CP}=\pi/2$ is excluded at more than $3\sigma$~\cite{Esteban:2020cvm}.
By the end of the decade, a significant fraction of the parameter space will be excluded by atmospheric neutrinos, as shown in~\Cref{fig:dcpFract}.
That will also help in resolving the actual tension between T2K and NOvA in the determination of $\delta_{CP}$ and $\sin^2\theta_{23}$~\cite{Kelly:2020fkv} thanks to the possibility to differentiate between the two octants of $\sin^2\theta_{23}$, as shown in~\Cref{fig:status}.

\begin{figure}[H]
     \centering
         \includegraphics[width=7cm, height=7cm]{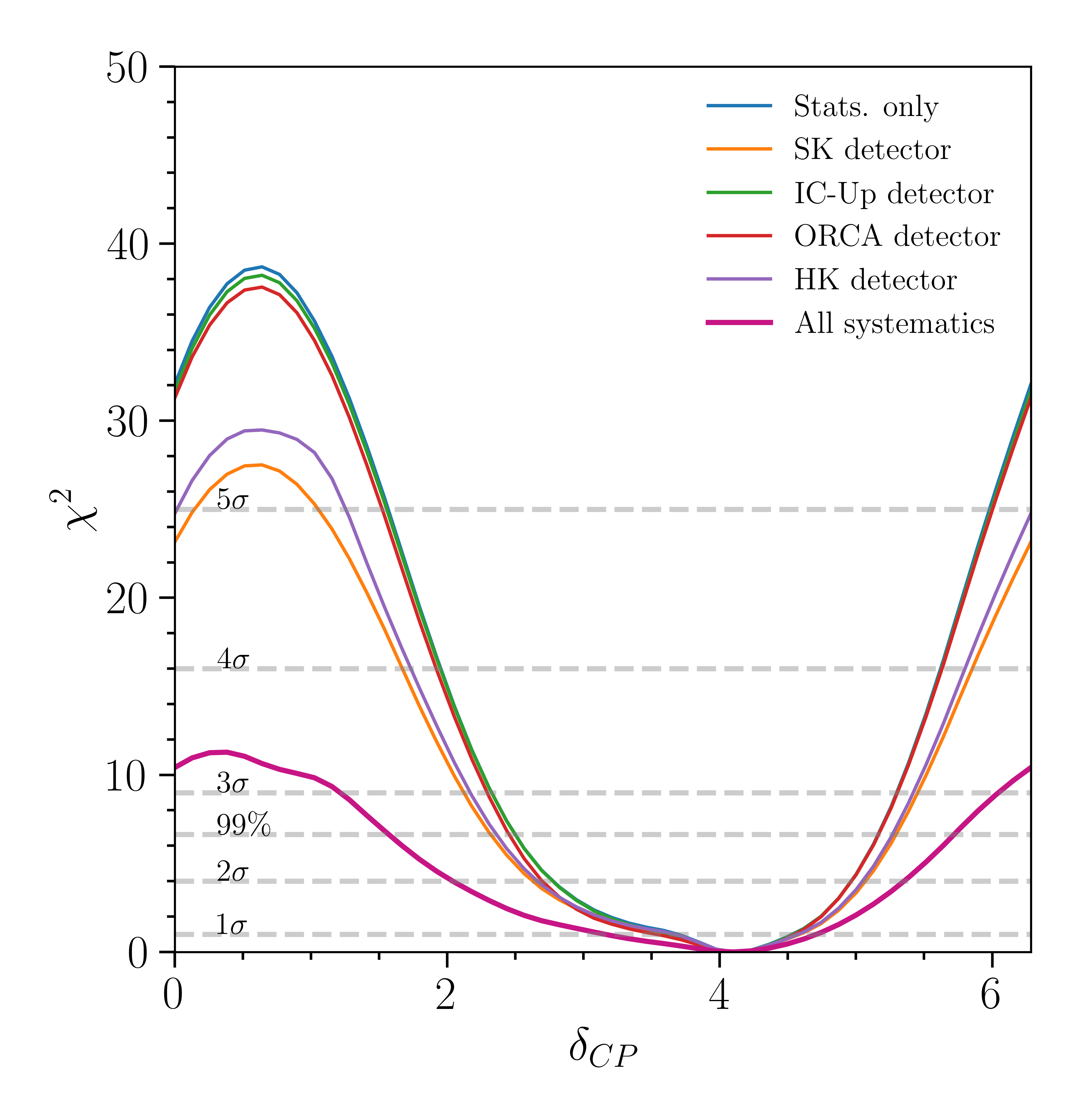}
     \hfill
         \includegraphics[width=7cm, height=7cm]{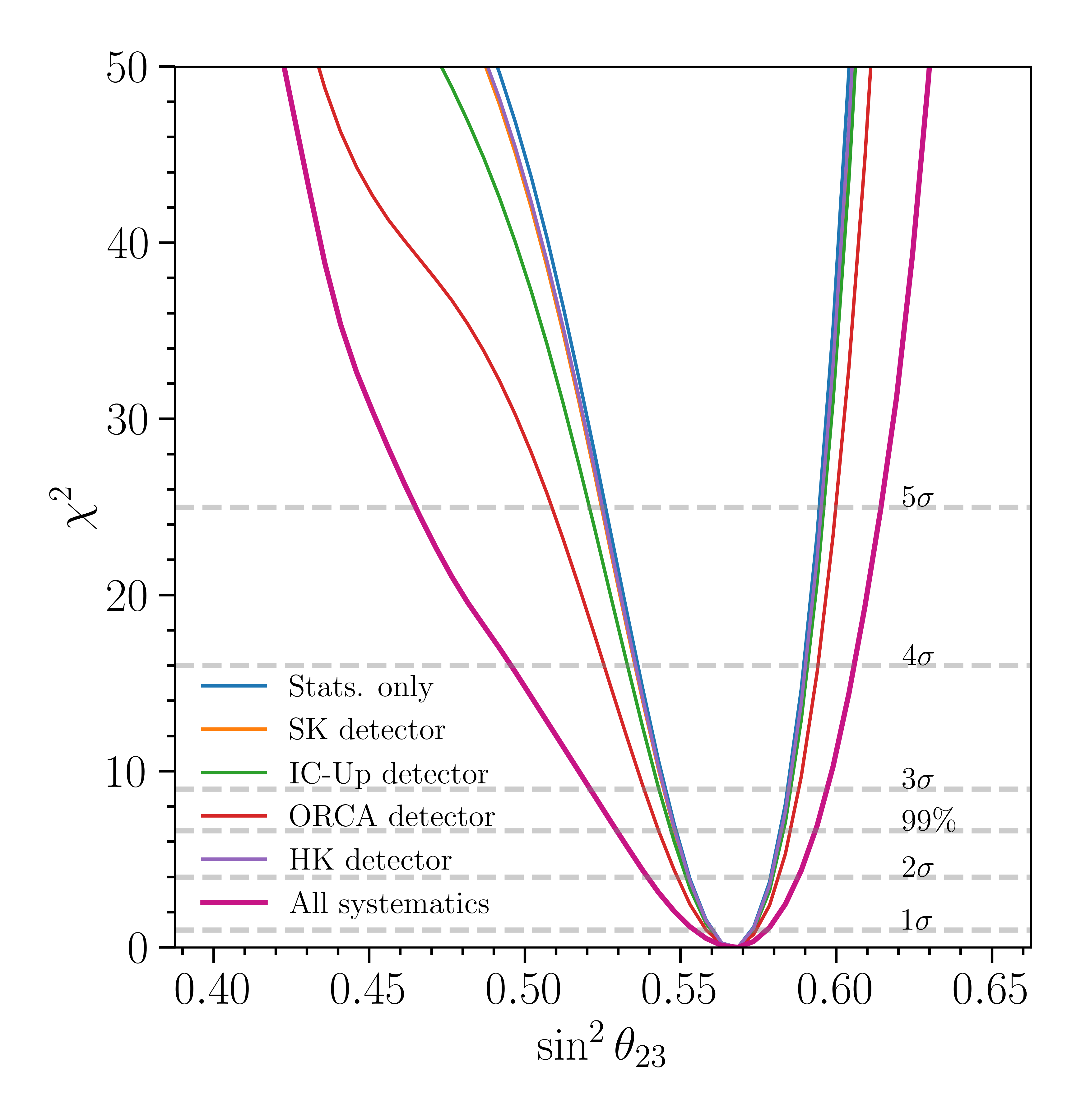}
     \hfill
         \includegraphics[width=7cm, height=7cm]{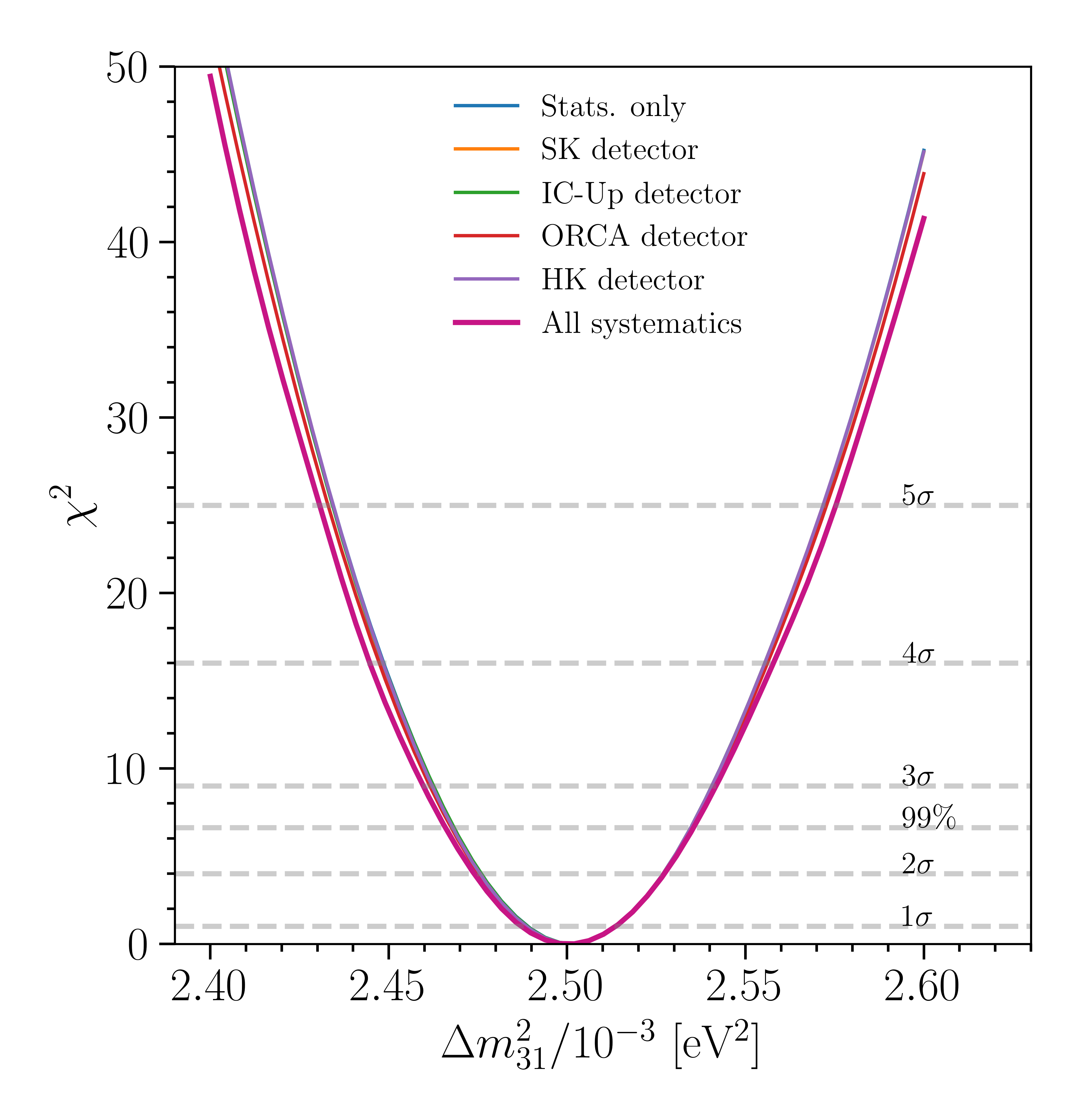}

        \caption{\textbf{\textit{From top to bottom, impact of each detector's systematic uncertainties on the sensitivity to $\delta_{CP}$, $\sin^2\theta_{23}$ and $\Delta m^2_{31}$.}}}
        \label{fig:det_syst}
\end{figure}

We undertake a systematic study to examine how potential future improvements in various systematic uncertainties will enhance sensitivity regarding that specific parameter.
The sensitivity to the $CP$-phase is heavily impacted by the uncertainties of the flux, detector and, to a lesser extent, cross section, see~\Cref{fig:flux+xsec_syst}.
The flavor ratio and normalization below $\SI{1}\GeV$ of the flux, and those related to the CCQE cross section, are the uncertainties reducing the sensitivity to this parameter. Detector systematics have a similar quantitative impact in the sensitivity as those of the flux.
An independent and complementary to long-baseline experiments measurement of $\delta_{CP}$ is of utmost importance to boost the precise picture of the 3-flavor neutrino mixing.
Ancillary measurements of the low-energy cosmic-ray flux and hadron production in proton and $^4_2$He scattering with nitrogen and oxygen nuclei, as well as development of more complete models, would narrow these uncertainties with a significant enhancement in the sensitivity to $\delta_{CP}$ from atmospheric neutrinos.
Further, as discussed in~\Cref{sec:xsec}, the current and projected experiments for measuring the neutrino cross section below $\SI{1}\GeV$ needed for next-generation accelerator neutrino experiments will provide valuable input for improving the measurement of the $CP$-phase with atmospheric neutrinos.

Finally, the measurement of $\theta_{13}$ is dominated by reactor experiments~\cite{DayaBay:2018yms, RENO:2018dro, DoubleChooz:2019qbj}, but lately, LBL experiments have measured this parameter with a considerable precision~\cite{T2K:2019bcf, Hartnell2022-do}.
We have explored the possibility of measuring $\theta_{13}$ in the atmospheric neutrino flux using the matter effects at the GeV scale.
The results indicate that by the end of the decade, it will be possible to reach a $20\%$ precision with the atmospheric neutrino flux, see~\cref{fig:t13}.
Although the precision is not comparable to the reactor measurements, it will certainly be on the same order as LBL measurement, \cref{fig:ICparam}. Exploring the mixing parameters at different energy scales might be a convenient way to search for new physics~\cite{Babu:1993qv,Chankowski:1993tx,Feruglio:2017rjo,Criado:2018thu,Novichkov:2019sqv,Babu:2021cxe}. 

Besides, the resulting scenario also enables atmospheric neutrinos to play a prominent role in the measurement of $\nu_\tau$ cross section.
Tau neutrinos are not expected from the unoscillated atmospheric flux, but they are measured in the detectors due to neutrino oscillations and strongly depend on the associated oscillation parameters.
This way, such a combined analysis will provide very valuable input for the lower end of charged-current $\nu_\tau$ cross section, see the discussion in Refs.~\cite{Martinez-Soler:2021sir,Denton:2021rsa}.

%%%%%%%%%%%%%%%%%%%%%%%%%%%%%%%%%%%%%%%%%%%%%%%%
\section{Conclusion}\label{sec:concl}
In this article, we explored the sensitivity of current and soon-to-operate water(ice)-Cherenkov atmospheric neutrino detectors ---namely, IceCube Upgrade, ORCA, SuperK, and HyperK --- to determine neutrino oscillation parameters.
To simulate these experiments we have developed dedicated Monte Carlo simulations and reproduce with good fidelity their experimental results. 
We incorporate more than 80 sources of systematic uncertainties and treat the correlated uncertainties between experiments, namely those associated to the common cross section and flux.
Through a comprehensive study, we motivate a combined data-fit from these experiments by showing the few-percent level precision that it would provide to the measurement of the remaining oscillation parameters --- in particular, $\theta_{23}$, $\Delta m^2_{31}$ --- and the neutrino mass ordering, as well as providing a constraint on the value of the $CP$-phase independent from long-baseline neutrino experiments.
Furthermore, the tools and the analysis presented in this work comprise the crucial first step to properly include atmospheric neutrinos in global fits.

Additionally, we identify the synergies among experiments and the common systematic uncertainties diminishing the sensitivity for each parameter.
We have an extended discussion of the sources of these uncertainties; in particular, regarding the flux and cross section inputs. 
It is worth noting that while our analysis uses conservative estimations of the flux and cross section uncertainties, both  are expected to be improved through new measurements and further theoretical developments.
In other words, this is a motivation for further ancillary measurements and methods, since an improved set of aforementioned estimations will greatly benefit the sensitivity of the analyzed experiments, and of particular relevance for the sensitivity to $\delta_{CP}$.

On the detector side, our analysis is also conservative. 
The reconstructions used for all the experiments considered in this work use traditional reconstruction techniques.
These traditional techniques have been shown to underperform compared to new machine-learning-enhanced reconstruction methods, which show greater accuracy and improved execution time, see, e.g.,~\cite{Abbasi:2021ryj, perioML}.
Additionally, in the case of the IceCube Upgrade, the reconstruction used in this work does not take full advantage of the next-generation sensors which have improved light collection.
Improvements are also expected for ORCA as the detector development proceeds.
In summary, the detector systematics assumed in this work can be taken as a conservative baseline that is expected to improve as this decade unfolds.

Finally, we emphasize that the results from a combined fit of atmospheric neutrinos would provide a very valuable input for the next-generation neutrino physics program towards the precise measurement of the $CP$-violating phase in the lepton sector.
This combined analysis nurtures itself from more than 40 years of global expertise and measurements of neutrino-water interactions paving the road for next-generation neutrino-water experiments such as Hyper-Kamiokande and IceCube-Gen2.
 
\acknowledgements{}

We thank Ali Kheirandish, Matheus Hostert, Pedro Machado, Michele Maltoni, Antoine Kouchner, Francis Halzen, Sergio Palomares-Ruiz, Alfonso Garcia-Soto, Luis Labarga, and Roger Wendell for useful discussions.
CAA will like to dedicate this work to E. Fernandini, who was an inspirational educator.
CAA, IMS, and MJ are supported by the Faculty of Arts and Sciences of Harvard University. PF is supported by the Department of Physics of the University of Liverpool and by the Donostia International Physics Center.
Additionally, CAA and IMS are supported by the Alfred P. Sloan Foundation.
We thank Jean DeMerit for carefully proofreading this manuscript.

\newpage

\bibliography{atmos}

\begin{thebibliography}{227}
\bibitem [{a_f()}]{a_footnote}%
  \BibitemOpen
  \href@noop {} {}\bibinfo {note} {If neutrinos are Majorana particles, two
  additional phases are present in this matrix, but they are not observable via
  flavor oscillation, thus we will ignore them in the rest of the
  discussion.}\BibitemShut {Stop}%
\bibitem [{d_f()}]{d_footnote}%
  \BibitemOpen
  \href@noop {} {}\bibinfo {note} {If the long-baseline measurement is combined
  with SuperK, the preference for NO rises to $\Delta\chi^2 \sim
  7$}\BibitemShut {NoStop}%
\bibitem [{e_f()}]{e_footnote}%
  \BibitemOpen
  \href@noop {} {}\bibinfo {note} {At higher energies, the $CP$-term is largely
  suppressed due to $\Delta_{21}$ that goes to zero, so that $\delta_{CP}$ has
  a larger impact in the neutrino evolution at sub-GeV energies and for
  Earth-size baselines.}\BibitemShut {Stop}%
\bibitem [{f_f()}]{f_footnote}%
  \BibitemOpen
  \href@noop {} {}\bibinfo {note} {The atmospheric experiments can use the
  electron and muon event samples to search for the $CP$-violation
  effects.}\BibitemShut {Stop}%
\end{thebibliography}

%%%%%% SUPPLEMENTAL MATERIAL STARTS HERE

\clearpage
\newpage

\onecolumngrid
\appendix

\ifx \standalonesupplemental\undefined
\setcounter{page}{1}

%\newcounter{SIfig}
%\setcounter{SIfig}{1}
\setcounter{figure}{0}

\setcounter{table}{0}
\setcounter{equation}{0}
\fi
\renewcommand{\thepage}{Supplemental Methods and Tables -- S\arabic{page}}

\renewcommand{\figurename}{SUPPL. FIG.}

\renewcommand{\tablename}{SUPPL. TABLE}
\renewcommand{\theequation}{A\arabic{equation}}

\newcounter{SIfig}
\renewcommand{\theSIfig}{SUPPL. FIG. \arabic{SIfig} }

%%%%%%%%%%%%%%%%%%%%%%%%%%%%%%%%%%%%%%%%%%%%%%%
\section{Super-Kamiokande and Hyper-Kamiokande Simulations}\label{sec:SKSim}
% This is trying if cref works with appendix \cref{sec:SKSim} \ref{fig:sktopo}\\
As stated in the text, for this analysis we assume that HyperK will perform very similar to the SuperK detector.
The simulation for each of the SuperK experiment phases is based on the output of version \textit{v2.12} of the GENIE generator \cite{andreopoulos2015genie}.
The obtained true information of the neutrino interaction in water is then translated into the relevant reconstructed variables for analysis using the effective response of the SuperK detector.
Given the complexity of the SuperK reconstructions and the variety of event topologies, the emulation of the SuperK detector is done on an event-by-event basis similar to reconstruction software. For that purpose, many distributions and efficiencies were taken from different sources to account for the different performances of the detector depending on its phase \cite{PhysRevLett.107.241801, Abe_2018}.

First, the true neutrino information is used to select all the particles that may emit a detectable signal by SuperK, that is photons or charged particles above the Cherenkov threshold by more than 30~MeV in kinetic energy. Further, we use PYTHIA package \cite{Sj_strand_2006, Sj_strand_2015} to enable the decay of unstable particles within the SuperK event time window. This is of special relevance for heavy mesons produced in neutrino interactions above 1~GeV, the hadronic decays of $\tau$ leptons, and the decay of neutral pions produced by sub-GeV neutrinos. The resulting detectable particles from the decay are added to the previous particles.
This list of final state particles gives a first estimate of the number of rings that the event will have, true Cherenkov rings.\\

\begin{figure}[h!]
\centering
\includegraphics[width=0.55\textwidth,height=3in]{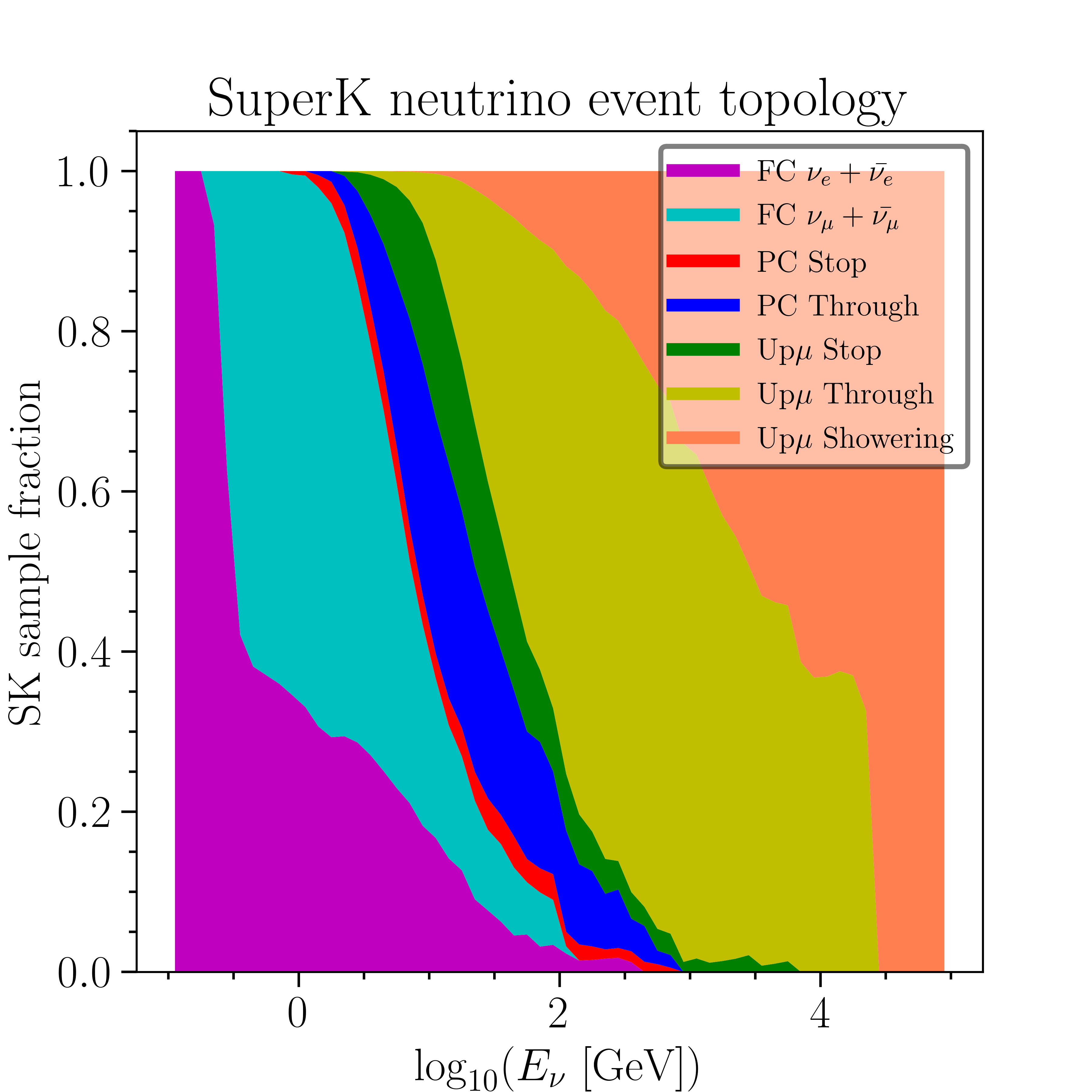}
\caption{\textbf{\textit{Probability of a neutrino event being classified into the major topology categories in SuperK as a function of the true neutrino energy.}}}
%\refstepcounter{SIfig}
\refstepcounter{SIfig}\label{fig:sktopo}
\end{figure}

The relevant information and kinematics of these particles are then fed to the event reconstructor. Making use of the digitized distributions, efficiencies, and resolutions available from \cite{Irvine2014DevelopmentON, Jiang2019StudyOT, Pik}, we compute the reconstructed momenta, directions, and particle IDs (e-like or $\mu$-like) for each of them. If there is only one detectable ring in the final state, the reconstruction process end there.\\
In the case of more than one particle able to produce a detectable ring, the closeness between the reconstructed directions is evaluated, if the angle between them is smaller than 70$^o$ those rings would be identified as a single ring by SuperK fitter (APFit) and, therefore, their signals merged into a single ring. In these cases, the resulting merged ring will have fuzzy edges, and therefore, always assumed to be $e$-like. In the end, the total visible energy, reconstructed direction, and particle ID of the most energetic ring are computed and used to classify each event according to the SuperK sample definition.

Naturally, given the complexity of SuperK's reconstruction and classification, an implementation of this sort has important limitations. For illustration purposes, we will briefly describe two show-cases to overcome these issues:
\begin{itemize}
\item{Event topologies:} We use the event rate table and spectrum from~\cite{Abe_2018} to produce~\ref{fig:sktopo}. This way, we overcome our limitations to class and event as fully-contained, partially-contained, or upward-going muon in the absence of the actual detector geometry. To mimic the relevance of SuperK geometry in event classification, we randomly assign one of these topologies based on the true neutrino energy and before any reconstruction takes place.
\item{Secondary interactions:} The GENIE output does not take into account all the secondary interactions that may happen to the produced particles with water and to which the detector is sensitive. The most relevant products from these secondary interactions are electrons from $\mu$ decays and neutrons, as they are used for event classification. For them, we rely on the available distributions of their multiplicity from various interactions, flavor channels and samples, in \cite{FernandezMenendez:2017ccn}.
\end{itemize}

\begin{figure*}
\centering
\includegraphics[width=\textwidth,height=0.75\textwidth]{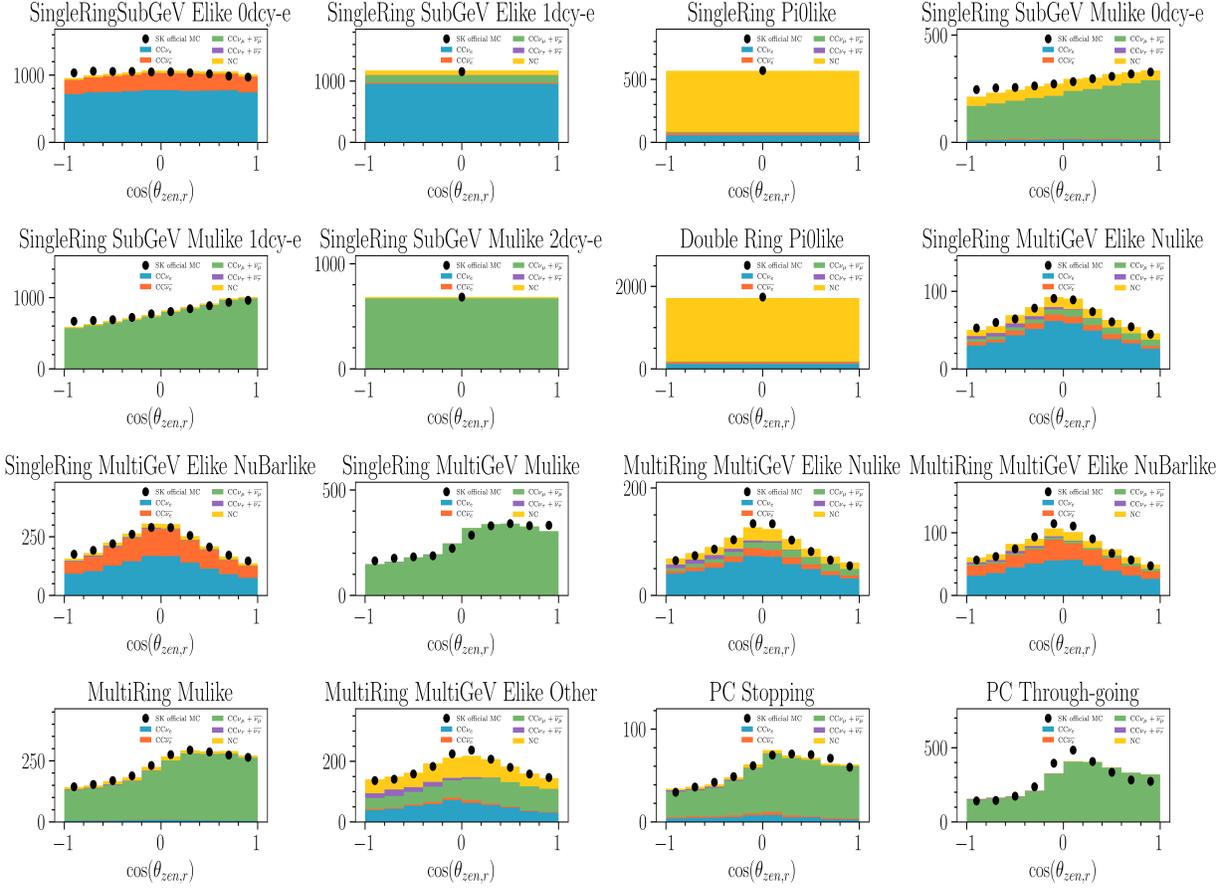}
\caption{\textbf{\textit{Reconstructed cosine zenith event distribution for each SuperK event category.}} Event rates divided by neutrino species (colored histograms), and compared to the results shown in \cite{Abe_2018} (black dots). The oscillation parameters assumed are normal ordering, $\sin^2\theta_{23}=0.587$, $\Delta m^2_{31}=2.5\cdot10^{-3}~eV^2$, $\delta_{CP}=4.18$, $\sin^2_{13}=0.018$.}
\refstepcounter{SIfig}\label{fig:SKzenith}
\end{figure*}

Once the reconstruction is completed for all the MC events, we tune the weights of the events so they reproduce the official SuperK event rates tabulated in \cite{Abe_2018}.\\

\begin{figure*}[h!]
    \centering
    \subfloat[][Reconstructed cosine zentih event distributions for Single-Ring Multi-GeV $e$-like samples with 0 decay-electrons and 0 (left) and $>$0 H-tagged neutrons. The same exposure and oscillation parameters as in \ref{fig:SKzenith} are assumed.]{
    \includegraphics[width=0.8\textwidth]{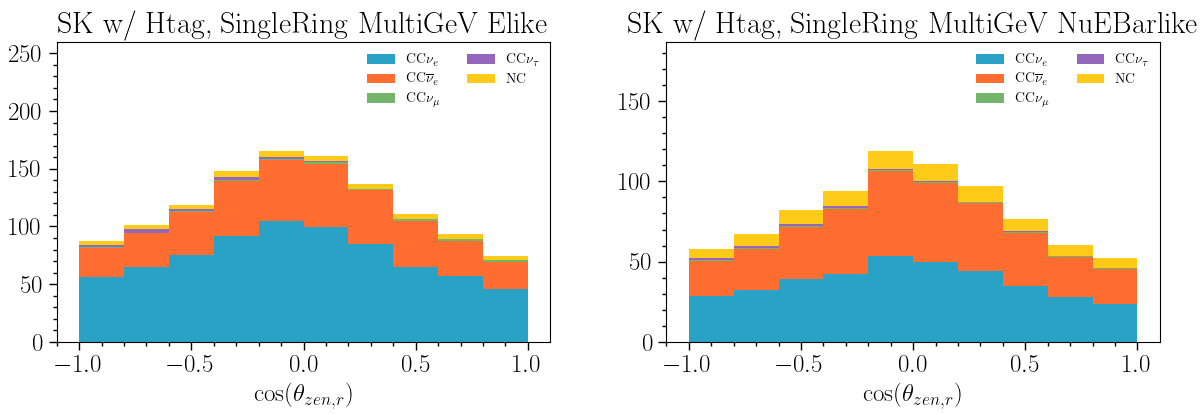}}\\
    \subfloat[][Reconstructed cosine zentih event distributions for Single-Ring Multi-GeV $e$-like samples with 0 decay-electrons and 0 (left) and $>$0 Gd-tagged neutrons. The same exposure and oscillation parameters as in \ref{fig:SKzenith} are assumed.]{
    \includegraphics[width=0.8\textwidth]{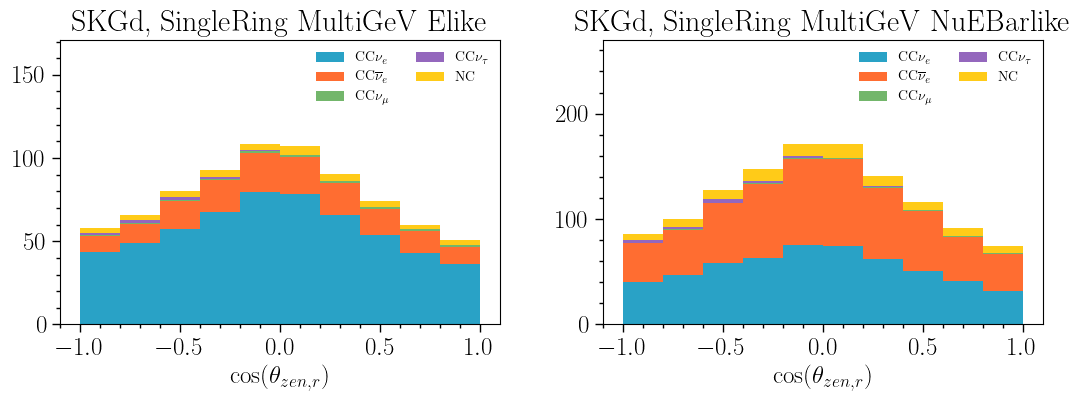}}\\
\caption{\textbf{\textit{Improvement in separation between neutrinos and antineutrinos from neutron tagging in SuperK.}}}
\refstepcounter{SIfig}\label{fig:SK_gd+htagging}
\end{figure*}

\begin{figure}[h!]
\centering
\includegraphics[width=0.45\textwidth,height=2.6in]{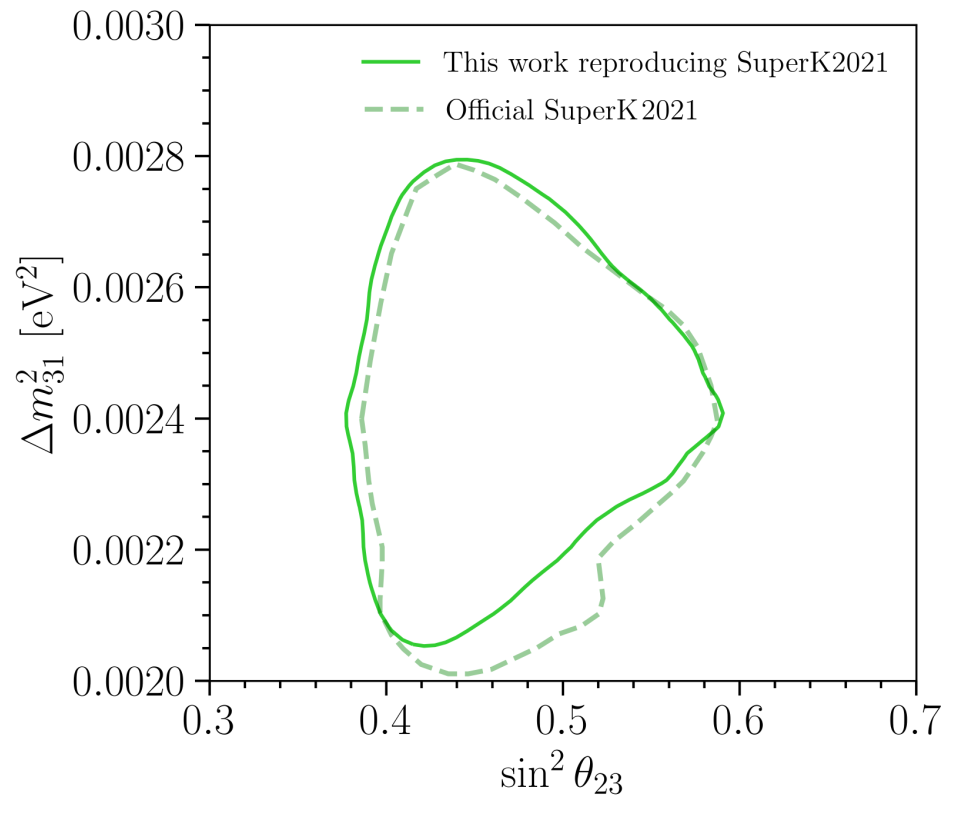}\hfill
\includegraphics[width=0.45\textwidth,height=2.6in]{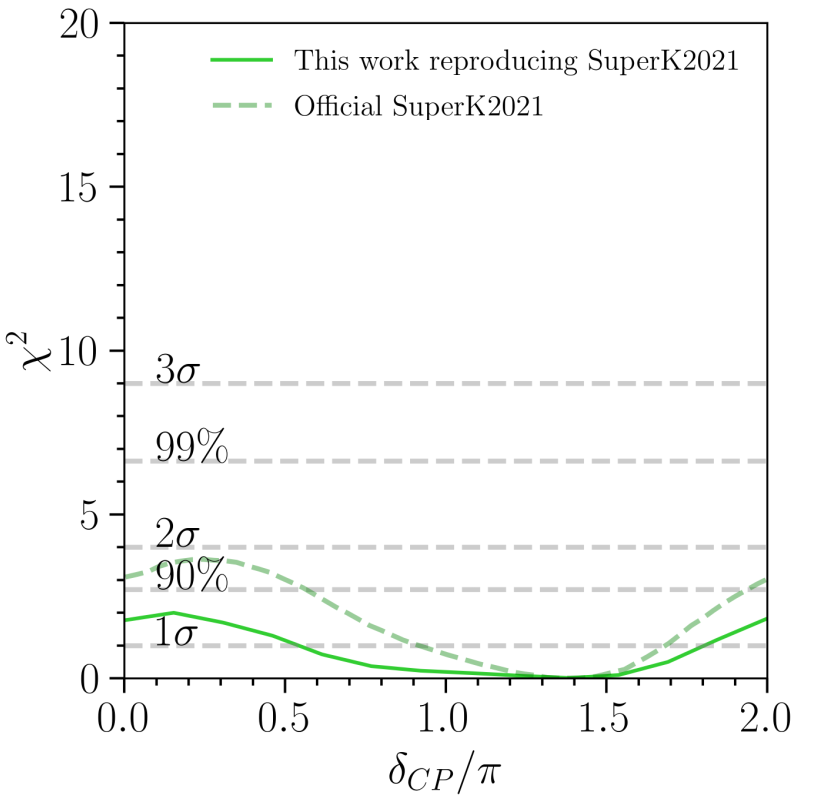}
\caption{\textbf{\textit{Comparison of 90\% confidence level contour for $\sin^2\theta_{23}$ and $\Delta m^2_{31}$, and sensitivity to $\delta_{CP}$ parameters obtained from this work's and SuperK official 2021 analyses~\cite{FernandezMenendez:2021jfk}.}} The former assumes the SuperK 2021 analysis, exposure and best fit values.}
\refstepcounter{SIfig}\label{fig:skrepro}
\end{figure}

To cross-check the validity of the developed SuperK atmospheric neutrino simulation, in~\ref{fig:skrepro} we compare the obtained sensitivity for $\sin^2\theta_{23}$, $\Delta m^2_{31}$ and $\delta_{CP}$ with the official one reported by the SuperK collaboration in 2021, \cite{FernandezMenendez:2021jfk}. This comparison is particularly relevant as it employs the usual event samples for SuperK-I to SuperK-III, and implements new ones based on the number of tagged neutrons on hydrogen for SuperK-IV. 
It should be noted that this comparison is done between the fit SuperK data and an Asimov set from our work assuming the corresponding best fit values. It can be seen that for the senstivity of the $CP$-phase, we obtain a conservative estimate as compared with the SuperK data; similar behavior was reported by the SuperK collaboration in \cite{linyan_wan_2022_6694761}.

\begin{figure}
\centering
\hfill
\includegraphics[width=0.45\textwidth]{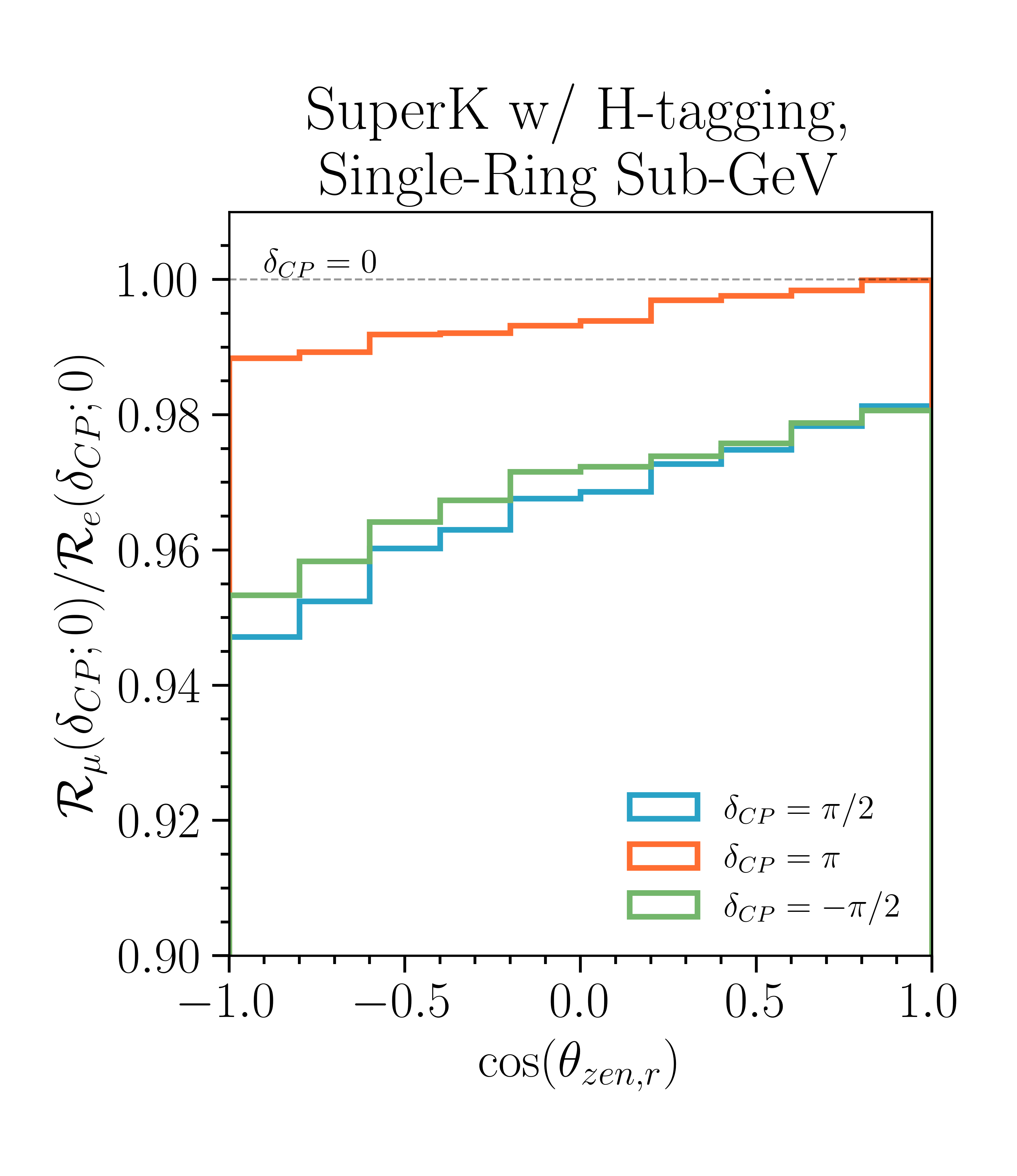}\hfill
\includegraphics[width=0.45\textwidth]{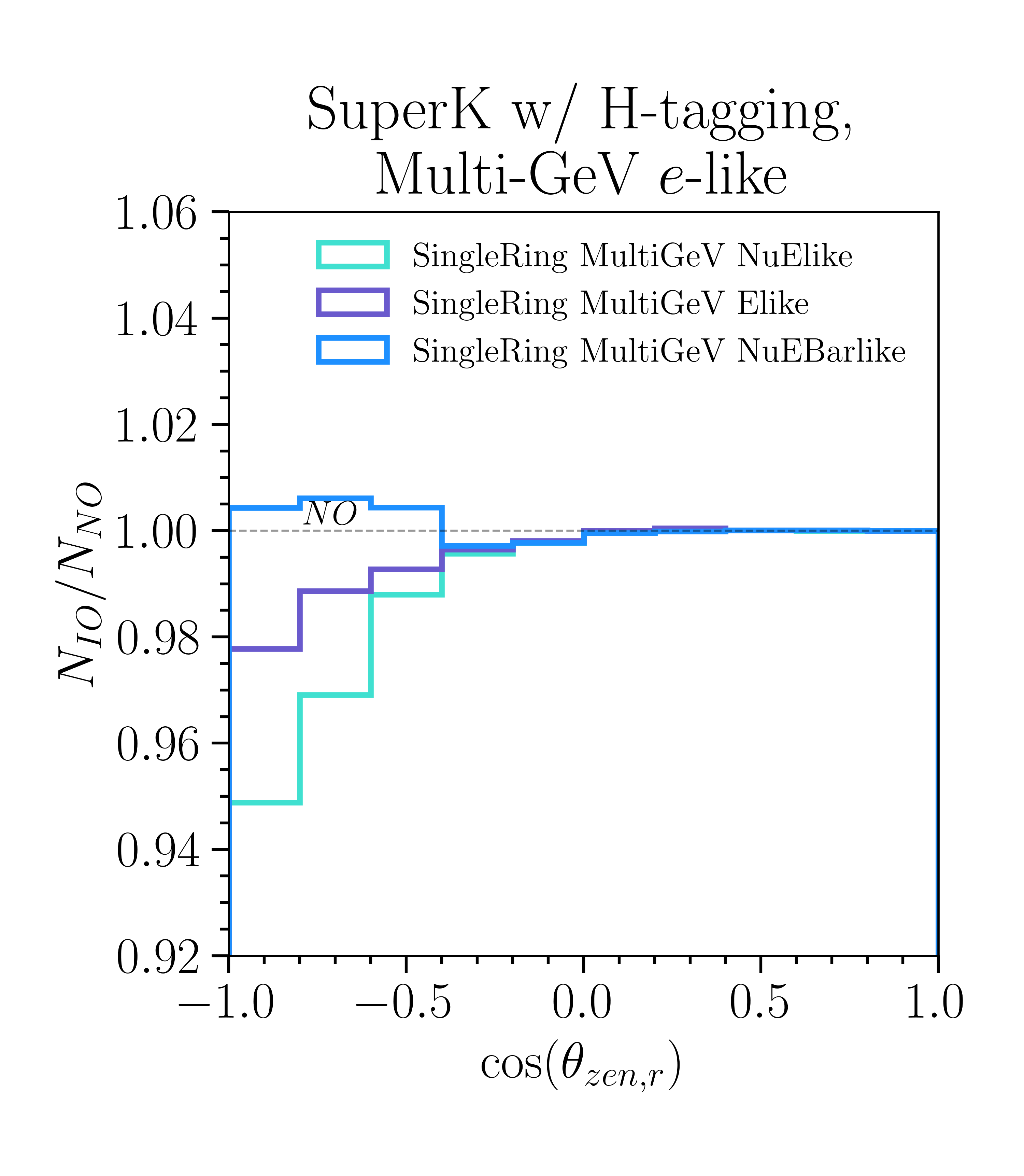}
\hfill
\caption{\textbf{\textit{In the left, the double ratio in the number of events for SuperK with H-neutron tagging Sub-GeV $\mu$-like and $e$-like samples assuming various values of $\delta_{CP}$ compared to $CP$-conservation. In the right, the ratio between inverted and normal orderings for the number of events in SuperK with H-neutron tagging single ring Multi-GeV $e$-like samples.}}}
\refstepcounter{SIfig}\label{fig:skh_dcp+mo}
\end{figure}

For the sake of completeness, we show in \ref{fig:skh_dcp+mo} the analog plots to \Cref{fig:skgd_dcpmo_ratio} for the case of H-neutron tagging.

%%%%%%%%%%%%%%%%%%%%%%%%%%%%%%%%%%%%%%%%%%%%%%%
\section{Neutrino telescopes simulation}\label{sec:ORCAMC}
\begin{figure*}[hbt!]
\centering
 \includegraphics[width=0.45\textwidth]{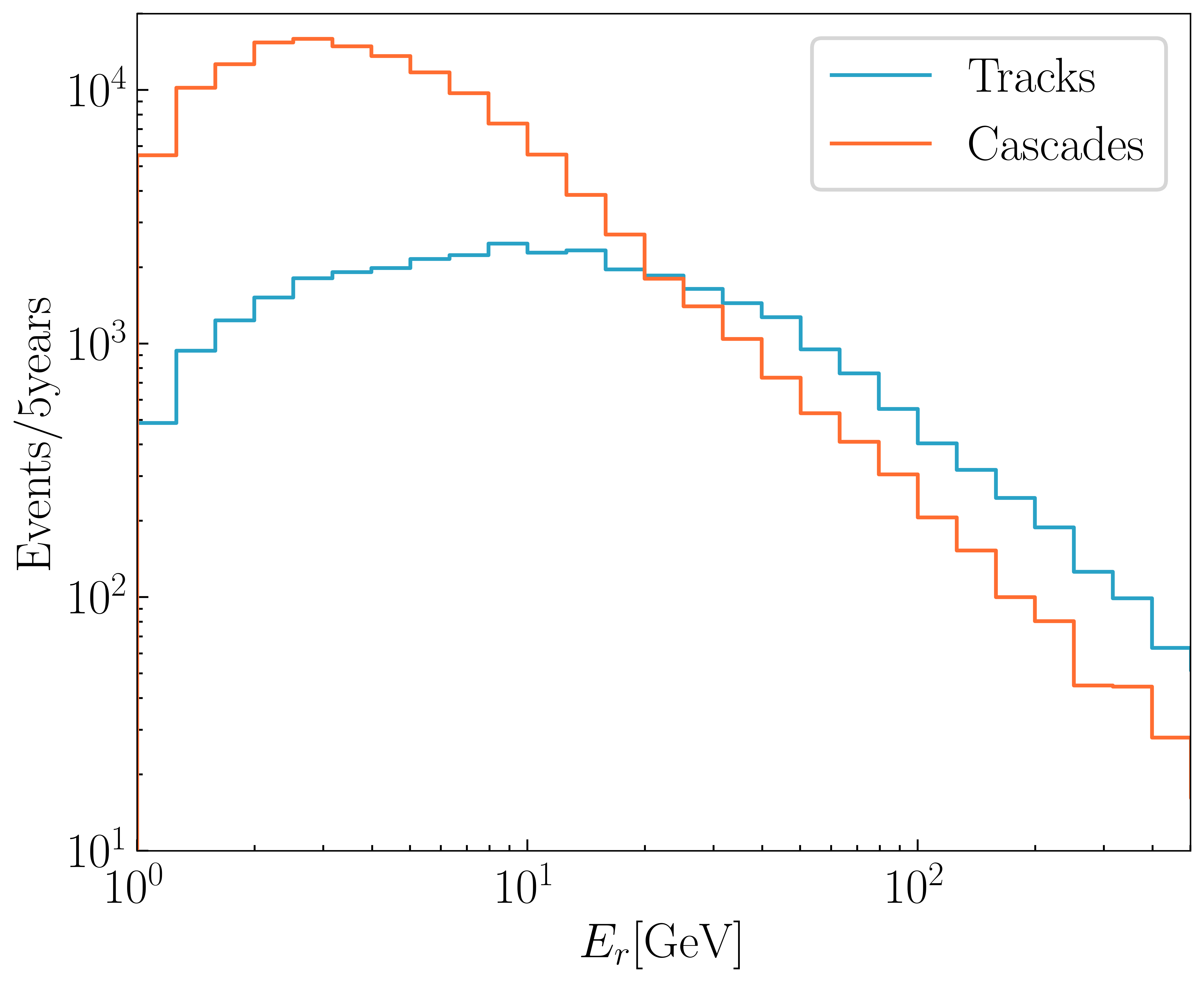}
 \hfill
 \includegraphics[width=0.45\textwidth]{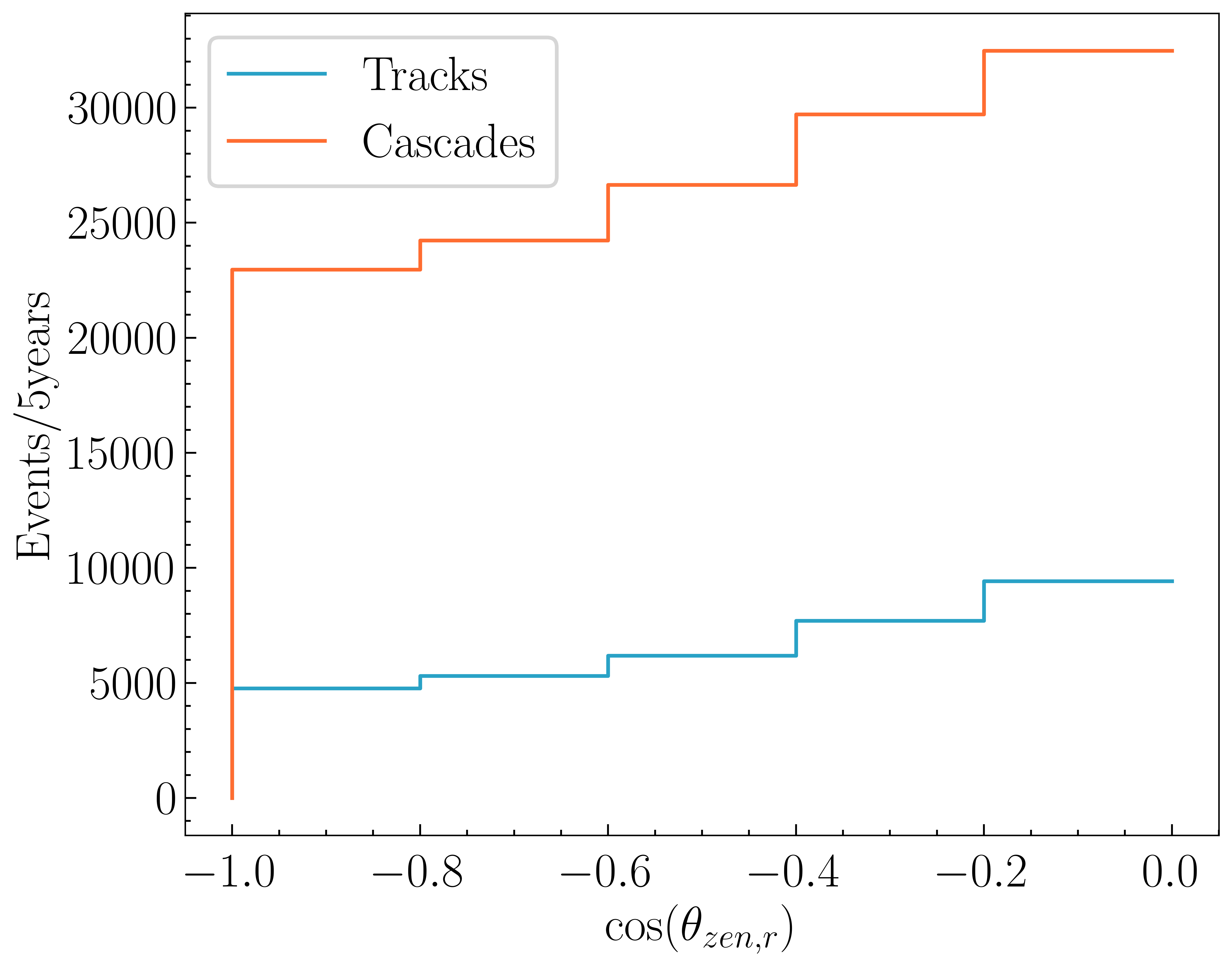}
\caption{\textbf{\textit{Event distribution as a function of the reconstructed energy and direction for IceCube Upgrade after 5~years of data taking.}}}
\refstepcounter{SIfig}\label{fig:supp_ICEvent}
\end{figure*}

\begin{figure}
\centering
 \includegraphics[width=\textwidth]{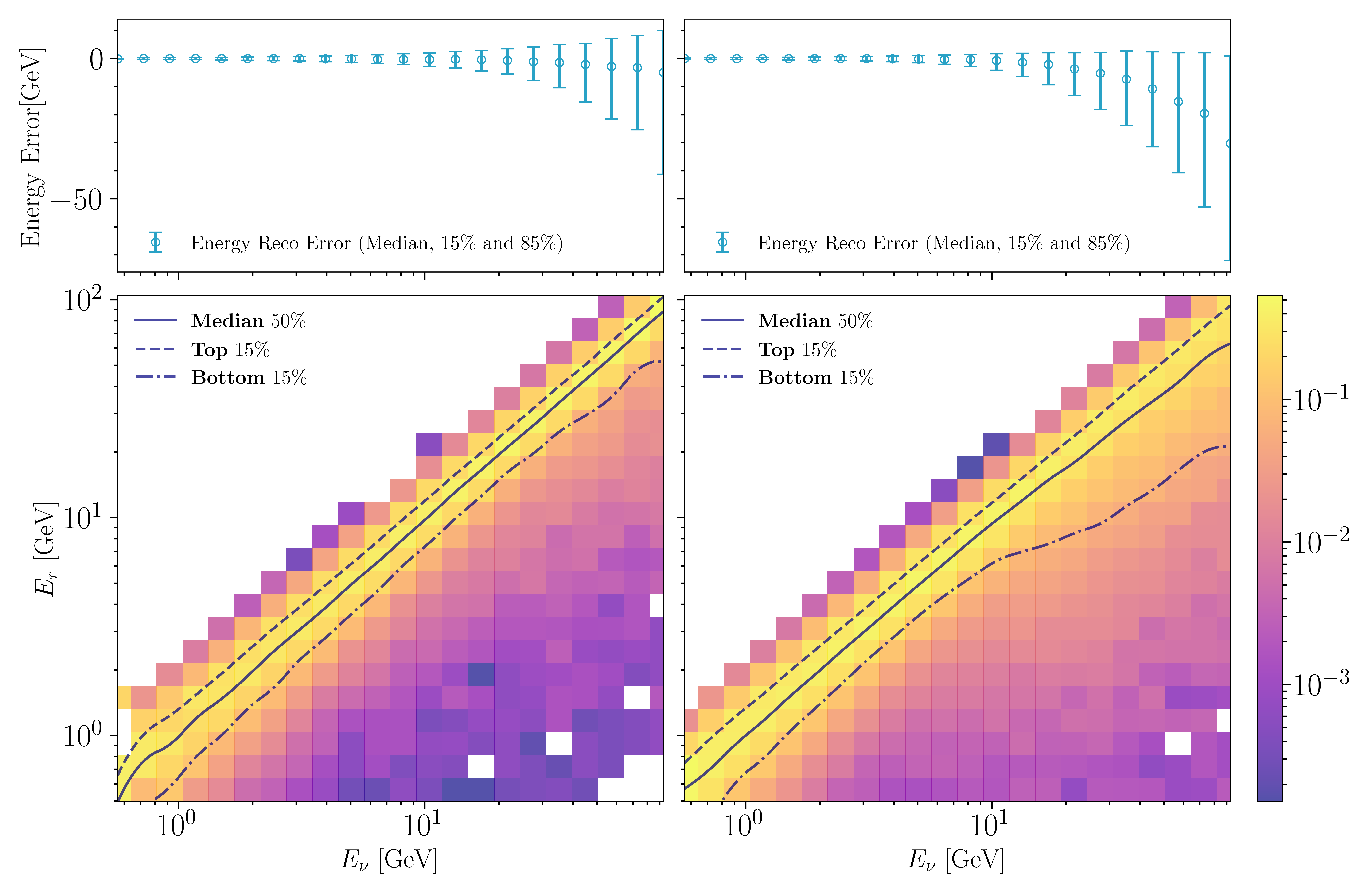}
    \caption{\textit{\textbf{Energy reconstruction resolution for track and cascade class events for the IceCube Upgrade Monte Carlo simulation.}} Plots are column-normalized, the lines correspond to the median, top and bottom 15\% in each true energy bin.}
\refstepcounter{SIfig}\label{fig:IC_distribution_Reco}
\end{figure}

This work uses the public release of Monte Carlo (MC) simulation for IceCube Upgrade DeepCore~\cite{IceCube_Collaboration2020-md}. This release considers the detector response for neutrino energies ranging from below 1~GeV to 1~TeV, and zenith angles from $0$ to $\pi$. In our analysis, we use a 2d histogram in reconstructed energy and cosine of zenith, with 20 bins in log-scale between $E_{r}=1$~GeV and $E_{r}=100$~GeV and 10 bins between $\cos(\theta_{zen,r})=-1$ and $\cos(\theta_{zen,r})=1$.
%includes MC events with true energy ranging from sub-GeV to 100~GeV, and with true zenith angle ranging from $0$ to $2 \pi$, but 
%we only choose to analyze upgoing events with $-1 \leq \cos \theta_{zen\, \nu} \geq 0$. Each event has a simulated reconstructed energy as well as reconstructed azimuth and zenith angles. 
The event distribution as a function of reconstructed energy and cosine zenith angle is shown in~\ref{fig:supp_ICEvent}. To obtain the expected event distribution for each value of the oscillation parameters, we sum over all the MC events the product of the effective area, the neutrino flux and the oscillation probability. We also include a factor $2\pi$ to account for the azimuthal distribution of the flux and running time of the experiment, that in the case of IceCube upgrade is 5~years. The reconstructed events are separated into 2 morphological classes: tracks and cascades, depending whether the event has a visible track or not. The energy resolution for each of these two classes is shown in~\ref{fig:IC_distribution_Reco}, and it is based on the performance of the standard DOMs used by the collaboration. For the upgrade it is planned to use new optical sensors called D-Egg and mDOM~\cite{Classen:2017sng,Makino:2019qmn,IceCube:2021xht,IceCube-Gen2:2021wka}, that will improve the energy and neutrino directional reconstruction. A new MC that consider the capabilities of those new sensors it is needed.
%The event distribution 
%The Monte Carlo weight of each event is the effective area for this event multiplied by $4\pi$ to account for the solid angle integral.

To reproduce the expected ORCA response, we have developed a simulation based of the IceCube Upgrade MC.
Based on the information provided by ORCA collaboration~\cite{KM3NeT:2021ozk}, we consider events with reconstructed energy between $1.85$~GeV and $50$~GeV. 

\begin{figure}[t]
\centering
 \includegraphics[width=0.9\textwidth]{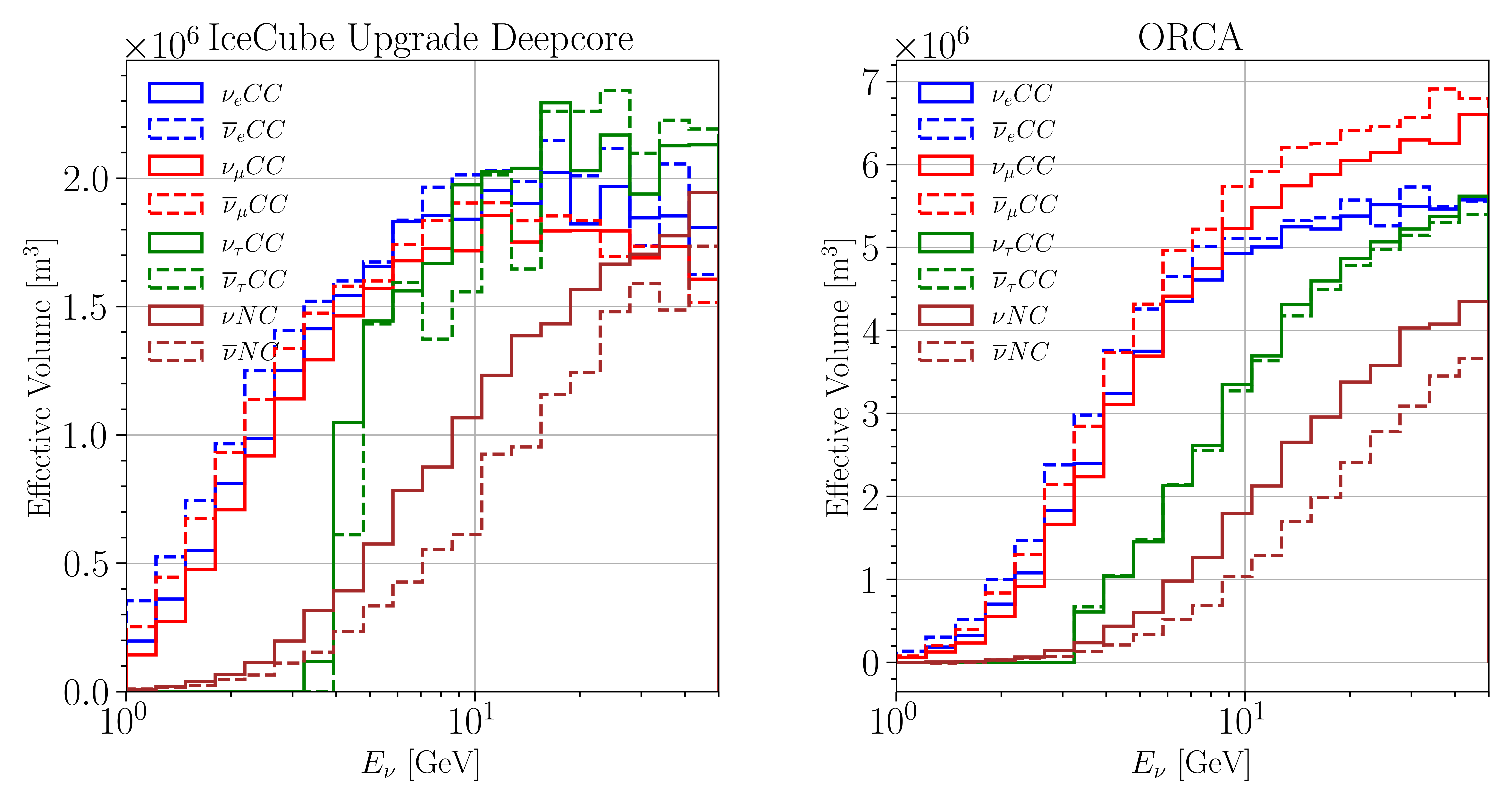}
    \caption{\textit{\textbf{Effective Volumes as functions of true neutrino energy for IceCube Upgrade and ORCA, respectively.}}}
\refstepcounter{SIfig}\label{fig:Eff_Vol}
\end{figure}

\begin{figure*}[ht!]
\centering
 \includegraphics[width=\textwidth]{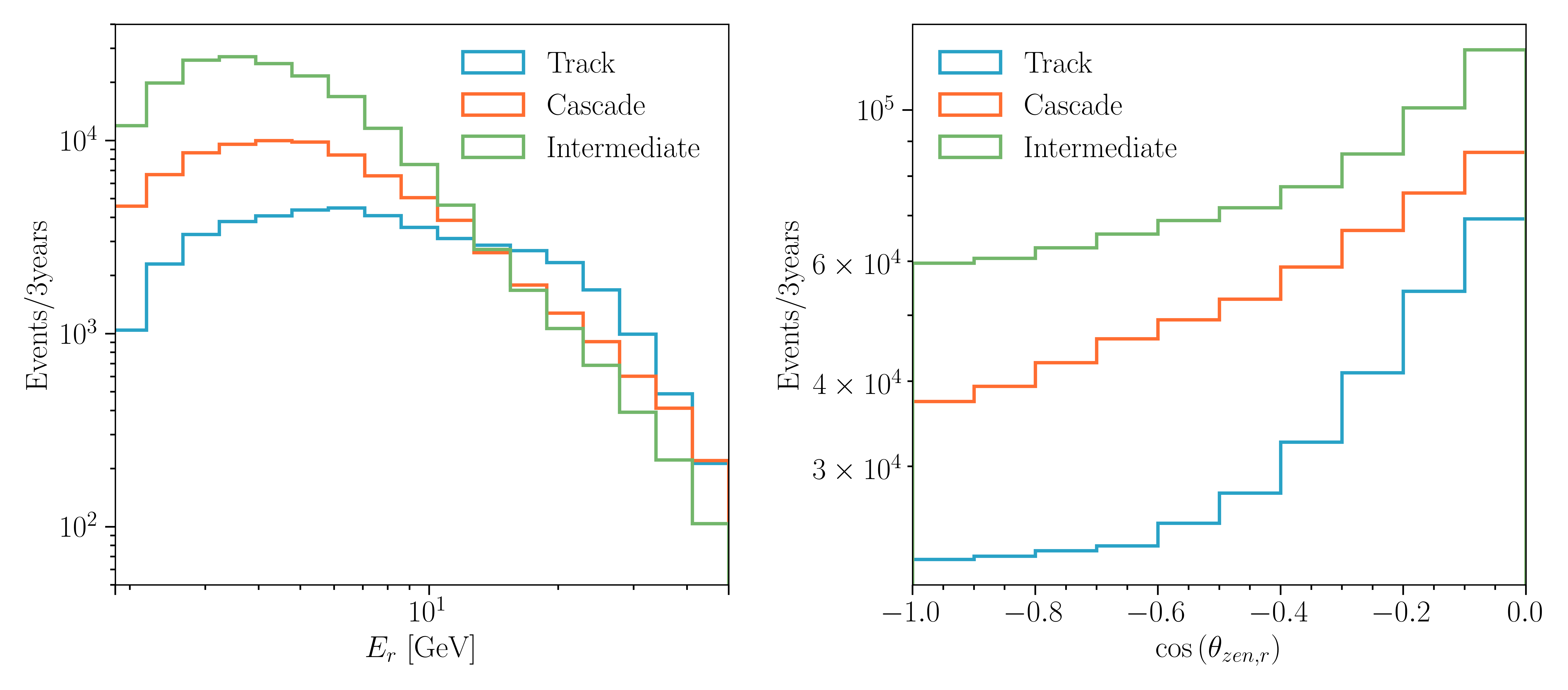}
    \caption{\textit{\textbf{Event distribution as a function of the reconstructed energy and direction for ORCA after 3 years of data taking.}} Oscillation parameters are chosen based on NuFit5.0 best fit parameters. Lines correspond to event distributions (with oscillations) for track, cascade, and intermediate morphology classes respectively. }
\refstepcounter{SIfig}\label{fig:supp_ORCA_distribution_Reco}
\end{figure*}

\begin{figure*}
\centering
 \includegraphics[width=\textwidth]{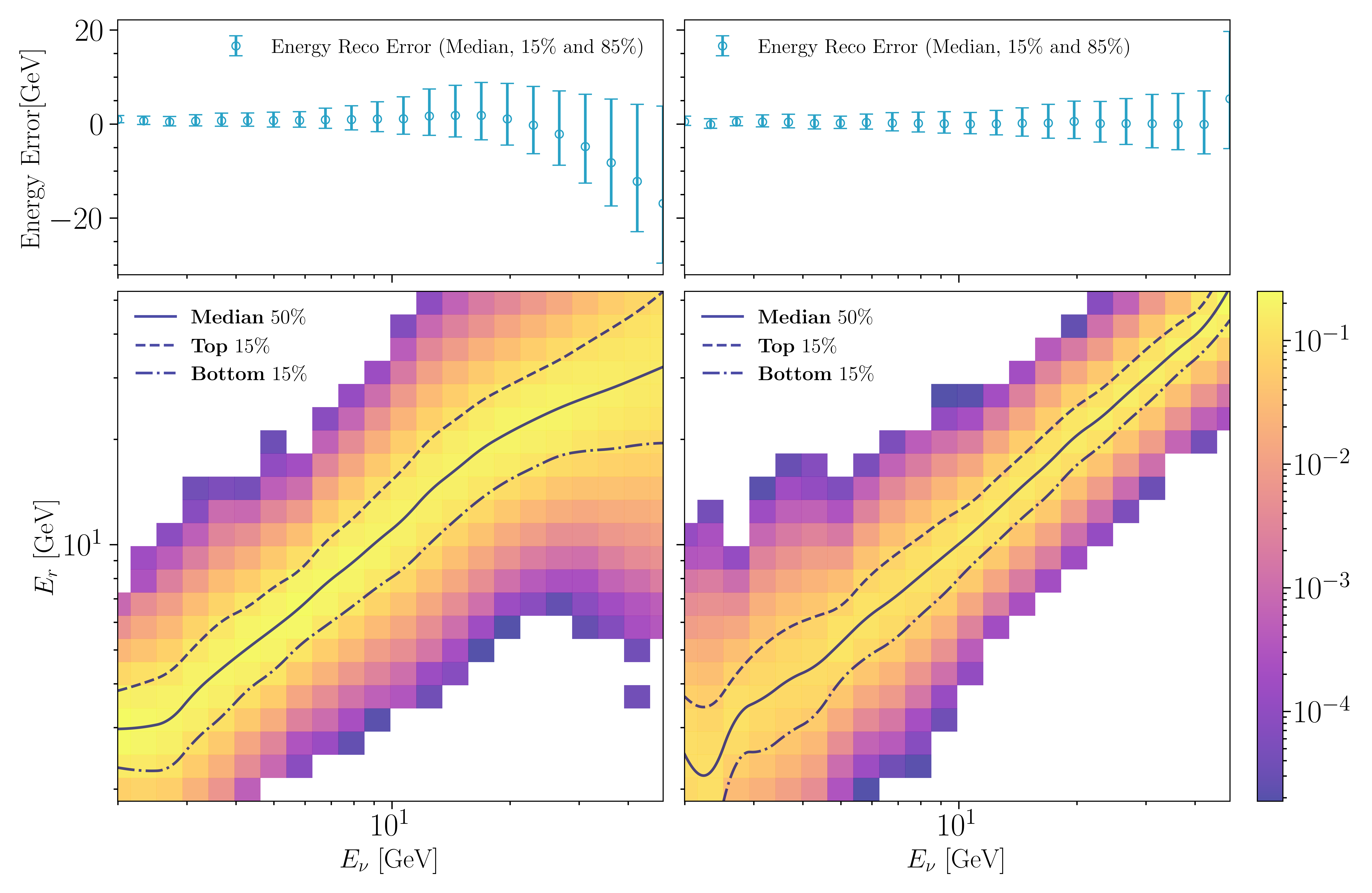}
    \caption{\textit{\textbf{Energy reconstruction resolution for track and cascade class events for the ORCA Monte Carlo simulation this work developed.}} Plots are column-normalized, the lines correspond to the median, top and bottom 15\% in each true energy bin.}
\refstepcounter{SIfig}\label{fig:Energy_Reco}
\end{figure*}

\begin{figure}
\centering
\includegraphics[width=0.7\textwidth]{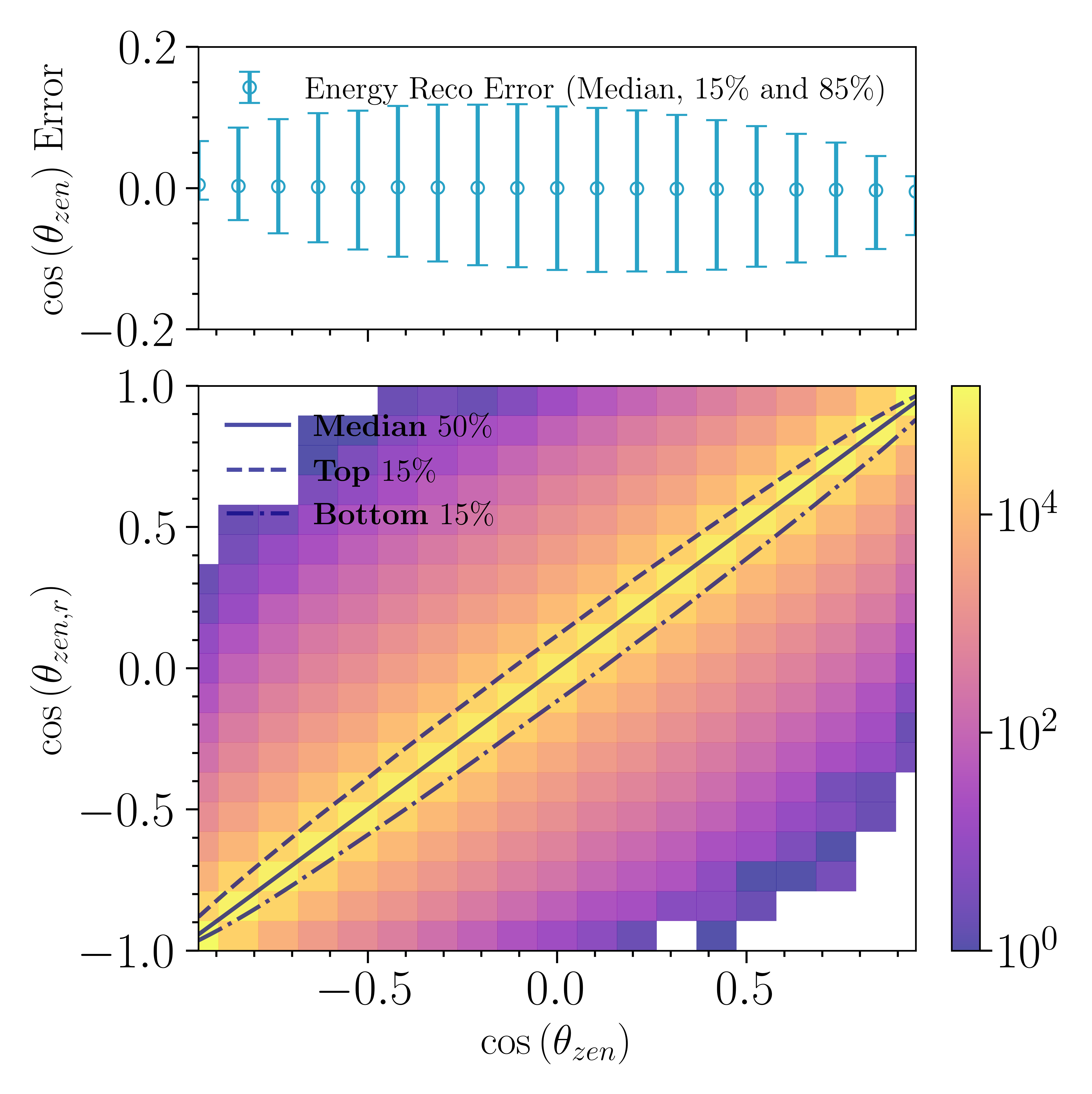}
    \caption{\textit{\textbf{Zenith angle reconstruction resolution for the ORCA Monte Carlo simulation this work developed.}} Plots are column-normalized, the lines correspond to the median, top and bottom 15\% in each true energy bin.}
\refstepcounter{SIfig}\label{fig:Zenith_Reco}
\end{figure}

Keeping the true information about these events, we randomly assign the reconstructed information including energy and zenith angle to reproduce the corresponding distributions given in~\cite{KM3NeT:2021ozk}. In~\ref{fig:Energy_Reco} we show the relation between the true neutrino energy ($E_{\nu}$) and the reconstructed energy ($E_{r}$) for cascades (left) and tracks (right), for the intermediate events we consider an average of the previous distributions. In~\ref{fig:Zenith_Reco} we show the relation between the true neutrino zenith and the reconstructed direction. We then calculate Monte Carlo weights for the events based on the effective volume information. To avoid artifacts in the sensitivity due to under-sampling, we duplicate each event 15 times weighting each copy by $1 / 15$ of the original MC weight. 
%This is done to enlarge the MC statistics and avoid artifacts in the sensitivity due to under-sampling that will occur in the later steps.
To estimate the event distribution, we proceed in a similar way as in the IceCube analisys. The event distribution as a function of $E_{r}$ and $\cos(\theta_{zen,r})$ is shown in~\ref{fig:supp_ORCA_distribution_Reco}. The comparison between IceCube Upgrade DeepCore and ORCA effective volumes is shown in~\ref{fig:Eff_Vol}. For both experiments, we observe an enhancement of the number of events in the horizontal direction that is larger for tracks-like events. The origin of such enhancement is due to the atmospheric flux, thanks to the longer path of mesons before reaching the Earth on those directions, and the larger response of the detector to low energy cascades.

\begin{figure}
\centering
\includegraphics[width=0.55\textwidth]{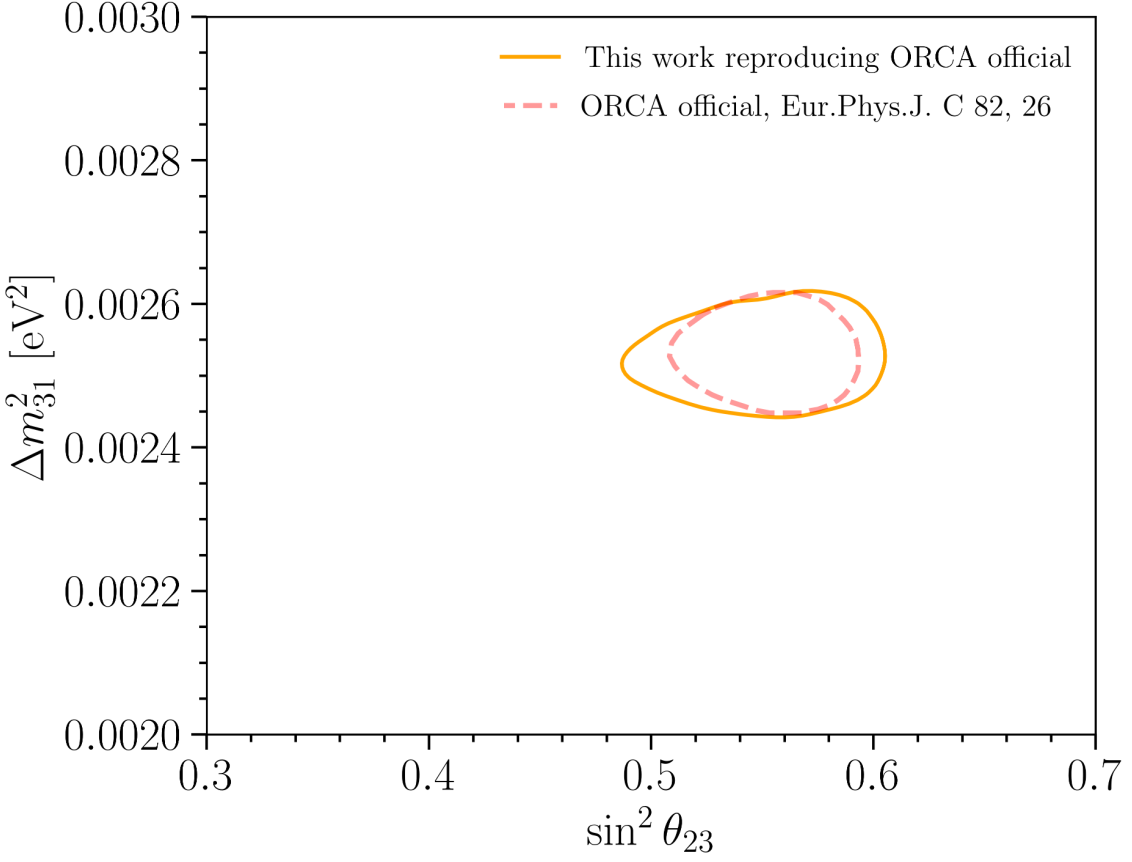}
\caption{\textbf{\textit{90\% confidence level contour obtained from this work's and ORCA official sensitivty studies.}}}
\refstepcounter{SIfig}\label{fig:orcarepro}
\end{figure}

As we noted before in this paper, ORCA analysis is based in three different morphologies: tracks, cascades and intermediate events. In our simulation, the fraction of events that contribute to each morphology follows the expectation from ORCA collaboration for the different flavors and energies. 
%Reconstructed event morphology classes in this work are assigned to reproduce the fractions of different morphology classes at different energies for different current and flavor neutrino and antineutrinos given in the ORCA paper~\cite{KM3NeT:2021ozk}. 
Due to lack of information on the kinematical properties of each and morphology in ORCA, %misidentification and energy and $y$ values, 
%in this work we avoid assigning a reconstructed class of tracks in the IceCube Upgrade simulation to identified cascades and vice versa while generating the ORCA MC from the IC Upgrade MC prior; instead we only assigns intermediate class to event morphogy when such a situation appears. 
we made a random distribution of each MC events between the three classes, avoiding the assignment of an initial track into a cascade and vice versa.
For some bins, this results in a reduced fraction of tracks and cascades and an increased fraction of the intermediate events. This almost only happens for $\nu_e$ charged current events at higher energy bins, where there isn't a lot of statistics. We checked that the final result of the ORCA simulation is not dramatically affected by this. In~\ref{fig:orcarepro}, we show the good agreement obtained between our simulation and the estimated sensitivity published by ORCA collaboration. 

\section{Variation of sensitivity with the octant of $\theta_{23}$}\label{sec:TH23}

\begin{figure}
\includegraphics[width=0.31\textwidth]{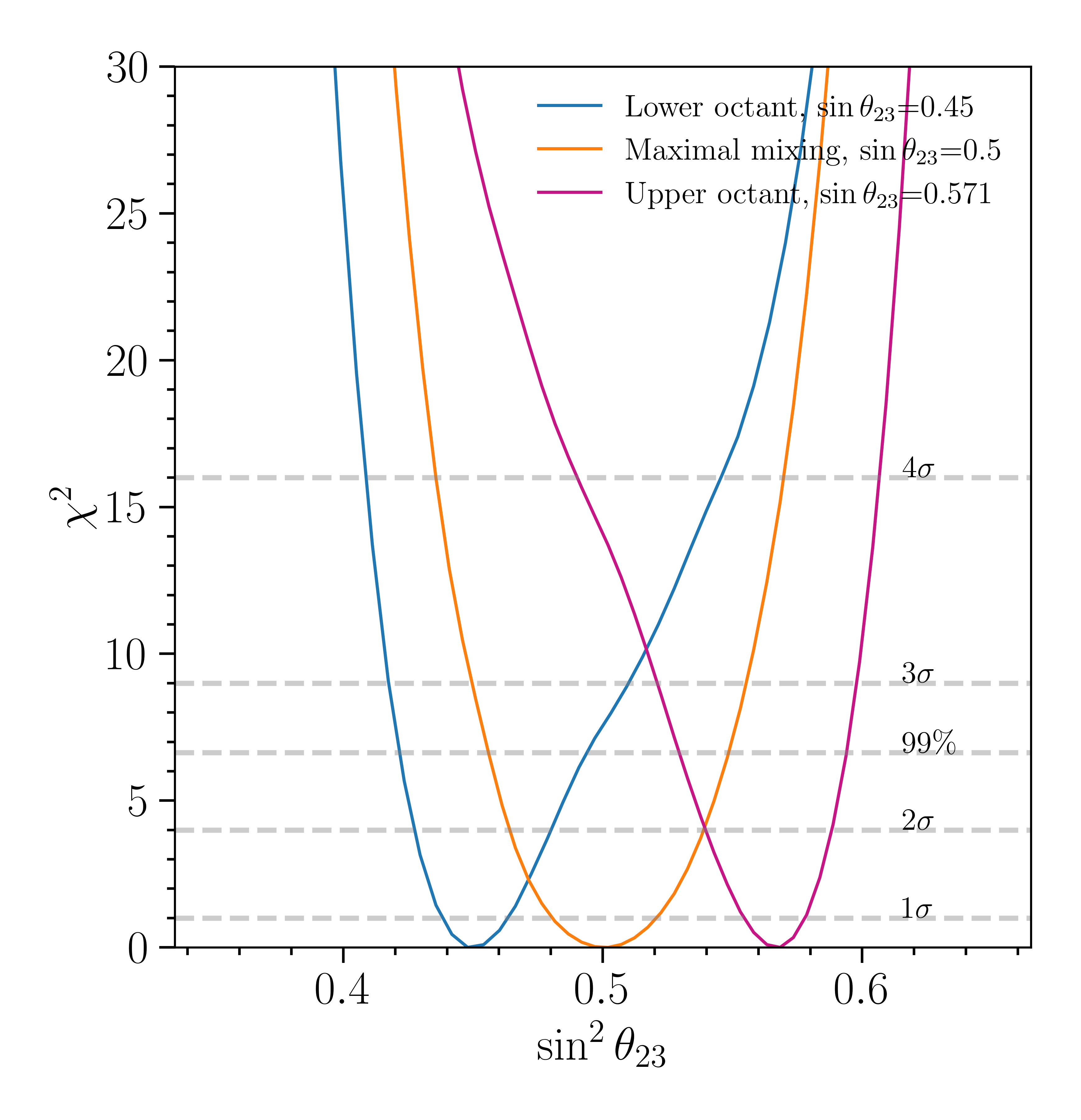}
\includegraphics[width=0.31\textwidth]{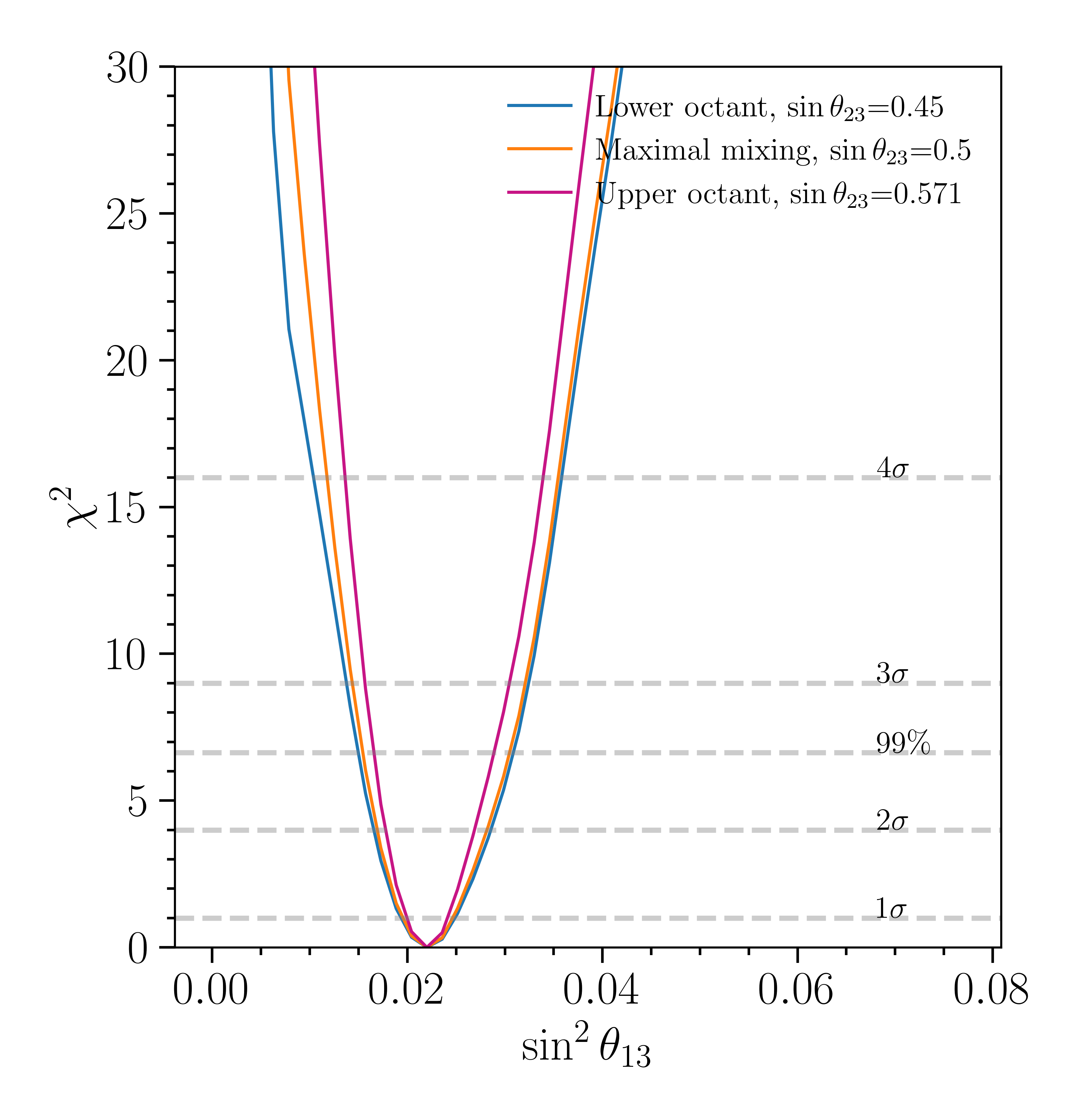}
\includegraphics[width=0.31\textwidth]{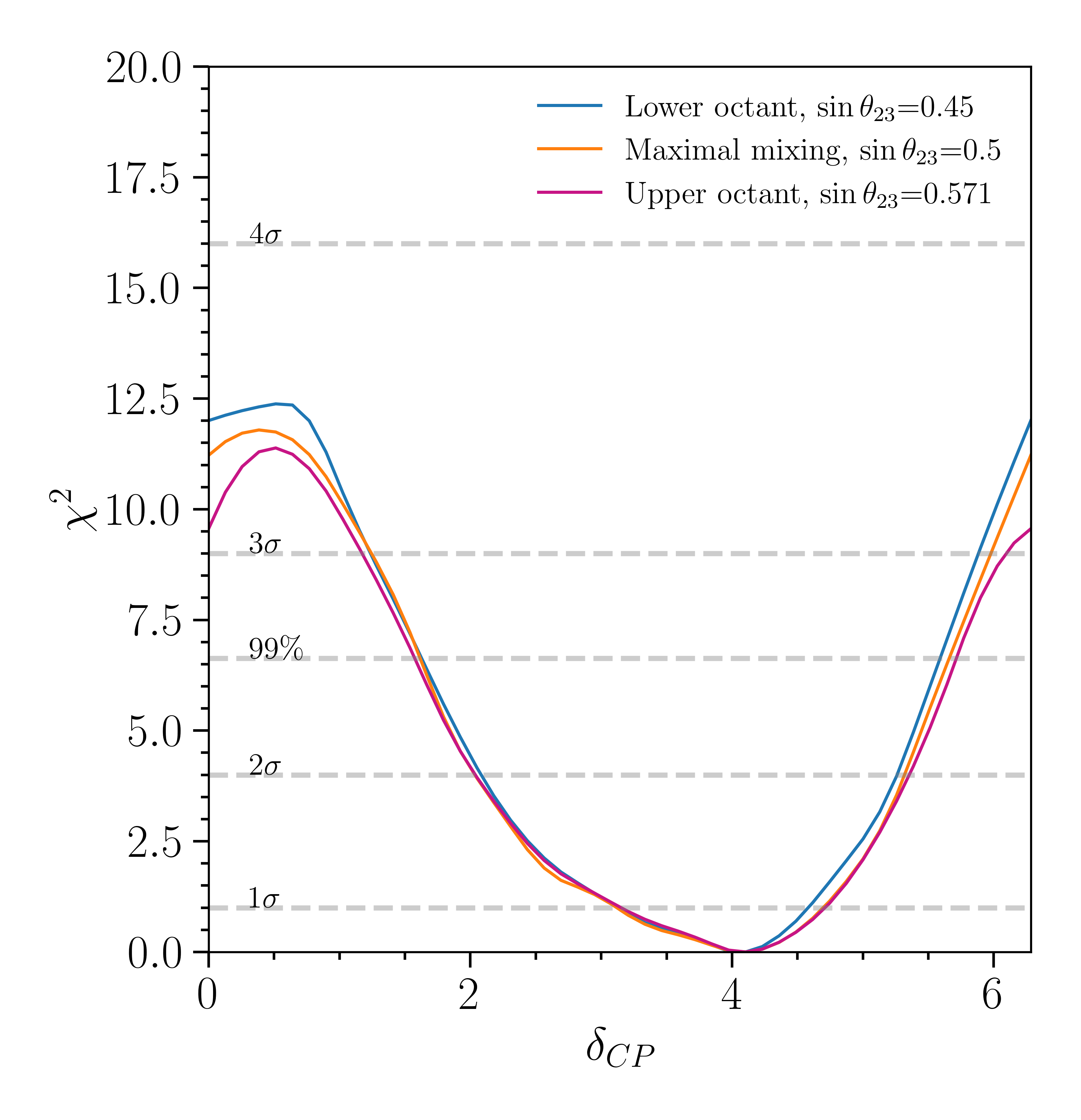}
\centering
\caption{\textbf{\textit{Sensitivity for different values of $\sin^2\theta_{23}$.}}}
\label{fig:octant}
\end{figure}

The sensitivity obtained in this work relies on the best-fit obtained from a global analysis~\cite{Esteban:2020cvm}.
Still, the sensitivity may change if we consider a different benchmark scenario.
In the case of the solar parameters ($\Delta m^2_{21}$ and $\sin\theta_{12}$) or $\Delta m^2_{31}$ and $\sin\theta_{13}$, the present data has shrunk the uncertainty to a few perfect.
For $\sin\theta_{23}$ a $20\%$ of parameter space it is still allowed at $3\sigma$, and for $\delta_{CP}$ allowed range increases to $\sim 80\%$. In the case of the CP-phase, we already explored how the sensitivity changes for different true values of $\delta_{CP}$ \Cref{sec:results}.
In this section, we focus on $\sin\theta_{23}$ considering two other benchmark values, maximal mixing ($\sin^2\theta_{23} = 0.5$) and lower octant ($\sin^2\theta_{23} = 0.45$).
The results are shown in \Cref{fig:octant}.
The impact of the octant of $\sin^2\theta_{23}$ on $\Delta m^2_{31}$ and $\delta_{CP}$ is negligible.
To resolve the octant of $\sin^2\theta_{23}$, for both scenarios (lower and upper octant) it is possible to exclude the wrong octant at $3\sigma$.

\end{document}